\documentclass[preprient2]{aa}
\usepackage{graphicx}
\usepackage[breaklinks,colorlinks,citecolor=blue,linkcolor=blue]{hyperref}
\usepackage{txfonts}
\usepackage{lscape}
\usepackage{epstopdf}
\usepackage{CJK}
\usepackage{float}
\usepackage[absolute]{textpos}
\usepackage{amsmath}
\usepackage{mathrsfs}

\newcommand{\HI}{\ion{H}{I}}
\newcommand{\hctn}{HC$_3$N}

\begin{document}
\begin{CJK}{UTF8}{gbsn}

\title{A FAST Survey of HINSA in PGCCs Guided by HC$_3$N}

\author{
Xunchuan Liu (刘训川)  \inst{1,2,3}\and 
Yuefang Wu\inst{1,2}\and 
Chao Zhang \inst{1,4} \and Ningyu Tang \inst{5,6} \and
Tie Liu \inst{3} \and 
 Ke Wang \inst{2} \and
 Di Li \inst{6,7} \and
 Lei Qian \inst{6} \and
Sheng-Li Qin \inst{4} \and
Jarken Esimbek \inst{8}  \and
Junzhi Wang \inst{3} \and 
 Jinghua Yuan \inst{6}   \and
 Fengwei Xu  \inst{1,2} \and 
 Lixia Yuan \inst{9}
}
\institute{
    Department of Astronomy, School of Physics, Peking University, 100871 Beijing, China\\
    \email{liuxunchuan@qq.com}\\
    \email{ywu@pku.edu.cn}
    \and
    Kavli Institute for Astronomy and Astrophysics, Peking University, 100871 Beijing, China
    \and
    Shanghai Astronomical Observatory, Chinese Academy of Sciences, Shanghai 200030, China 
    \and
    Department of Astronomy, Yunnan University, Kunming, 650091, China
    \and
    Department of Physics, Anhui Normal University, Wuhu, Anhui 241002, China
    \and   
    National Astronomical Observatories, Chinese Academy of Sciences, Beijing 100101, China 
    \and
    NAOC-UKZN Computational Astrophysics Centre, University of KwaZulu-Natal, Durban 4000, South Africa
    \and
Xinjiang Astronomical Observatory, Chinese Academy of Sciences, 830011, Urumqi, China
\and
Purple Mountain Observatory, Qinghai Station, 817000, Delingha, China
}
\abstract{
Using the Five-hundred-meter Aperture Spherical radio Telescope (FAST), we search  for \ion{H}{I} narrow-line self-absorption (HINSA) features
in  twelve Planck Galactic cold clumps (PGCCs), one starless core L1521B and four star forming sources.
Eight of the 12 PGCCs have emission of $J$=2-1 of cyanoacetylene (HC$_3$N). 
With an improved HINSA extraction method more
robust for  weaker and blended features with high velocity resolution, the detection rates
of HINSA in PGCCCs are high, at 92\% overall (11/12) and 87\% (7/8) among sources with HC$_3$N $J$=2-1 emissions.
Combining the data of molecular spectra and Planck continuum maps,
we studied the morphologies, abundances and excitations of \ion{H}{I}, CO and HC$_3$N in PGCCs.
The distribution of HINSA is similar to that of CO emission.
HINSA tends to be not detected in regions associated with warm dust and background ionizing radiation,
as well as regions associated with stellar objects.
The abundances of \ion{H}{I} in PGCCs are approximately $3\times 10^{-4}$, and vary within a factor of $\sim$3.
The non-thermal velocity dispersions traced by C$^{18}$O $J$=1-0 and HINSA are consistent with each other (0.1--0.4 km s$^{-1}$),
larger than those of HC$_3$N ($\sim$0.1 km s$^{-1}$).
Carbon chain molecule abundant PGCCs provide a good sample to study HINSA.
}


\maketitle

\section{Introduction}
The detection of  \HI~with the 21 cm line ($^{2}S_{1/2}$, $F$=1-0; $f_0=1420.4058$ MHz) 
is a milestone in interstellar matter in our Galaxy \citep{1951Natur.168..356E}. 
Immediately after, astronomers tried to investigate star formations in atomic
clouds, only to realize that the atomic clouds have too high a temperature and too low a density
to form stars. In 1960s, molecular clouds were identified to be the birth place of stars. Since
hydrogen is the major component of molecular clouds, the syntheses of molecular hydrogen (H$_2$) from H atoms is
critical for star formation. Measurement of [H]/[H$_2$] is very important for understanding the
chemical states of targeted sources \citep[e.g.][]{1962Icar....1...13C,1963ApJ...138..393G,1983ApJS...53..591D}. 
However, both of the observations of \ion{H}{I} and H$_2$ in molecular clouds are very difficult.  H$_2$ 
is an non-polarized molecule lacking permanent electric dipole moment, and thus cannot be detected in microwave band.
Meanwhile, the crowded velocity components contributed by background and foreground sources
make it also hard to identify \HI~emission associated with any exact cloud component in the
Galaxy \citep{2003ApJ...585..823L}. 

\HI~Narrow-line Self-Absorption (HINSA) was demonstrated to be  a good tracer of molecular
clouds \citep[][and references therein]{2003ApJ...585..823L}. 
\ion{H}{I} Self-Absorption (HISA)  occurs if cold atomic hydrogen 
is in front of a warmer emission background \citep[e.g.][]{1974AJ.....79..527K,1979A&AS...35..129B,2020A&A...634A.139W}.
HISA is common but diverse in gas outside the solar circle.
The majority of HISA features have no obvious $^{12}$CO emission counterparts \citep{2000ApJ...540..851G,
2010ASPC..438..111G}.
\citet{2005ApJ...626..887K} searched for HISA features within the Southern Galactic Plane
Survey (SGPS), finding \ion{H}{I} number densities of a few cm$^{-3}$.
The origins of HISA can be varied, including the
cold neutral medium with low temperatures   \citep[15--35 K,][]{2001ApJ...551L.105H}
and the ``missing link'' clouds in status between atomic clouds and molecular clouds \citep{2005ApJ...626..887K},
as well as the cold \ion{H}{I} component within the molecular clouds \citep{1977A&A....54..933W}.
As a special case of HISA, HINSA was defined as self-absorption feature with corresponding CO emission and line width 
comparable or smaller than that of CO \citep{2003ApJ...585..823L}. 
The cold \ion{H}{I} traced by HINSA has been shown to be tightly associated with molecular components.
HINSA is an effective
method to detect cold \ion{H}{I} mixed with molecular hydrogen H$_2$.
Besides CO, other molecular tracers such as OH also have 
central velocity tightly  correlated with that of HINSA \citep{2003ApJ...585..823L},
although no clear correlation was found for the non-thermal velocity dispersion between OH and HINSA 
\citep{2021ApJS..252....1T}.

Since first detected in our Galaxy, the
HINSA science has been developed substantially with the observations of Arecibo 305 m Telescope 
\citep{2005ApJ...622..938G,2008ApJ...689..276K,
2010ApJ...724.1402K,2016A&A...593A..42T,2018ApJ...867...13Z}.  
In the optically selected nearby dark cores \citep{1999ApJS..123..233L}, 
the densities of H atoms ($n$(H)) appear slightly higher than the steady-state value from the balance between formation and destruction \citep{2003ApJ...585..823L}.
Guided by $^{13}$CO $J=1-0$, the HINSA detection rate 
(the fraction of $^{13}$CO emitting clouds where HINSA is detected) can be as high as $\sim$80 percent \citep{2008ApJ...689..276K}.  

The Five-hundred-meter Aperture Spherical radio Telescope \citep[FAST;][]{2011IJMPD..20..989N} located in the Southern-west of China
is the most sensitive radio telescope in the L band. 
In the era of FAST, HINSA science has a new opportunity for great advancement.
\citet{2020RAA....20...77T} conducted a  pilot \ion{H}{I} 21 cm spectra survey towards  17 
Planck Galactic cold clumps (PGCCs)  using the FAST in single point mode.
In that survey,  58\% PGCCs have detections of HINSA features, with the
detection rate slightly smaller than the rates   in Taurus/Perseus region \citep[77\%;][]{2003ApJ...585..823L}
and molecular cores \citep[80\%;][]{2010ApJ...724.1402K}.
This deviation may
arise from limited sample and different background \ion{H}{I} emission,
or reflect the different evolution statuses of these sources.

The PGCCs \citep{2011A&A...536A..23P,2016A&A...594A..28P} were 
released as one  of the  results of  Planck satellite \citep{2010A&A...520A...2T,2011A&A...536A...1P}, 
which provided an all sky survey in the submillimeter to millimeter range
with unprecedented sensitivity.
PGCCs are mostly cold quiescent samples but are not as dense as low mass cores, and they are at very early 
evolution stages
\citep{2012ApJ...756...76W,2012ApJS..202....4L,2015A&A...584A..93J,2019A&A...622A..32L,2020ApJS..247...29Z,
2021ApJ...920..103X}.
The detection of HINSA features in early cold molecular clouds like PGCCs \citep{2020RAA....20...77T} 
reveals the potential to improve our understanding of the atomic to molecular transition stages in this important sample. 

In the present work,  we have observed 
twelve PGCCs and five CCM abundant comparison objects, 
using the L-band 19-beam receiver of  the FAST.
The observed PGCCs are mainly chosen from PGCCs with HC$_3$N emission.
The comparison objects consist of one starless core L1521B and four star-forming sources. 
The 19 beam receiver installed on FAST \citep{2018IMMag..19..112L} can help us quickly map the target sources.
Multi-beam spectra and mapping observations are essential for the identification and analysis 
of HINSA, through comparing the spectral profiles and spatial morphologies of HINSA and molecular tracers.
Seven of the samples are observed in the snap-shot mode,
which is made up by  four sets of 19-beam tracking mode observations and consists of 76 pointings.
Another feature of this work is that the sources we observed  are carefully chosen as the 
carbon chain molecule (CCM) abundant clouds/clumps.
We will introduce  the basic information of the sample in Sect. \ref{sec:sample}.
The FAST \ion{H}{I} observations are described in Sect. \ref{sec:obs}.

PGCCs with CCM production regions are chosen because  they may tend to
contain abundant \HI, since hydrogen atom is one of the main components  of hydrocarbon molecules and N-bearing 
CCMs \citep[e.g.][]{2016ApJ...817..147T}. 
The initial searching for carbon chain molecule (CCM) production regions began in 1970s
\citep{1971ApJ...163L..35T}. 
Since then, CCMs were detected towards molecular cloud in virious environments.
In early and cold molecular cloud,  atoms and ions have not been totally depleted onto grains yet, and thus
play an important role on driving carbon chain chemistry \citep{1992ApJ...392..551S,2008ApJ...672..371S}.
In low mass star formation cores, molecular outflows and shocks
can dissociate molecules and fuel the environment with materials in moderate
temperature to drive the generations of CCMs \citep{2019A&A...627A.162W,2019MNRAS.488..495W}. 
However, it is not clear which gas component 
in PGCC is associated with the CCM production regions,  the warm one in accompany with turbulent shock and
stellar feedback or the cold one embedded in the central cold and dense region?
Searching for HINSA in PGCCs could help us understand this issue.
Another important reason for choosing a sample of CCM production sources  is that we may 
have a higher detection rate of HINSA towards these sources.


The HINSA features are extracted basically 
following the method of \citet{2008ApJ...689..276K}, denoted as ``method 1'' in this work. 
However, we will point out the caveats of method 1, and give an analytic
formula to quantify the requirement of high signal-to-noise (S/N) if method 1 is adopted.
An improved HINSA-extracting method is proposed  in this work, denoted as ``method 2''. 
The S/N requirement for method 2 is much lower than that for method 1.
The method 2 is sensitive to HINSA spectrum  with high velocity resolution (thus 
broad line width relative to  channel width) and low signal-to-noise  ratio (S/N)
or HINSA features with multiple velocity components.
The algorithms and the S/N thresholds of these two HINSA-extracting methods
are described in  Sect. \ref{sec:method}.

Applying the improved method to the observed \ion{H}{I} 21 cm spectra, HINSA features can be extracted in 14 among 
the 17 observed sources with a detection rate of 
82\%. The detection rate of HINSA in PGCCs is 11/12 (90\%), higher than that in the sample of \citet{2020RAA....20...77T}.   
The  fitted results of HINSA, HC$_3$N $J$=2-1 and CO spectra as well as the HINSA images are presented in 
Sect. \ref{sec:result}. 
Combining the information of HINSA, HC$_3$N and CO emission as well as the dust continuum,
the present work aims to improve the knowledge about 
the \ion{H}{I} abundances in PGCCs, the evolutionary statuses of PGCCs and the excitation mechanism
of HINSA in CO and CCM emission regions.
In Sect. \ref{sec:diss},
we discuss about the morphologies, abundances and excitations of \ion{H}{I}, CO and HC$_3$N in PGCCs.
We summarize the paper in Sect. \ref{sec:summary}.
 
\begin{table*}[htb]
\caption{Sources. \label{tab:sources}}
\centering
\begin{tabular}{cccccccccc}
\hline
source \tablefootmark{(1)} & RA & DEC& distance  & $D_{\rm dust}$ & $T_{\rm ECC}$ & $T_{\rm dust}$        &  $N^{\rm dust}$(H$_2$)  & Mod\tablefootmark{(2)} & $\theta_{ZA}$\tablefootmark{(3)}\\
       &  J2000        & J2000          &  kpc                 & \arcmin & K  &  K &  g cm$^{-2}$        &                   &    \degr                              \\
\hline
G159.2-20A1 & 03:33:24.06 & +31:06:59.31 &0.37(0.11)&6.3(0.5)&10.9(0.6)&14.9(0.6)& 40.5(1.9) & T&8.8\\
G160.51-17.07 & 03:46:51.18 & +32:42:28.95&0.40(0.12)&6.9(0.6)&9.7(0.1)&19.2(0.9)& 7.2(0.4) & T&10.3\\
G165.6-09A1 & 04:30:57.4 & +34:56:18.8 &0.46(0.15)&7.1(0.4)&17.2(1.5)&22(1.3)& 9.2(0.4) & S&16.5\\
G170.88-10.92 & 04:40:32.71 & +29:55:42.61&0.45(0.18)&7.7(0.4)&10.5(0.5)&18.6(0.7)& 7.6(0.4) & T&14.8\\
G172.8-14A1 & 04:33:06.18 & +25:58:41.1 &0.43(0.19)&8.5(0.6)&11.3(0.6)&15.5(0.7)& 12.2(0.4) & S&10.2\\
G173.3-16A1 & 04:29:25.19 & +24:32:45.3 &0.41(0.19)&5.3(0.3)&8.2(0.5)&13.3(0.4)& 31.8(1.9) &T&12.5\\
G174.06-15A1&  04:32:50.26 &+24:23:55.7 &0.41(0.20)&6.0(0.2)&8.1(0.4)&16.1(0.4)& 25.9(1.6) & S & 17.3\\
G174.08-13.2 & 04:41:34.77 & +26:01:46.3&0.44(0.21)&6.8(0.3)&8.8(0.6)&17(0.3)& 25.5(1.4) & S&7.5\\
G174.4-15A1 & 04:33:56.07 & +24:10:26.5 &0.42(0.20)&6.4(0.3)&9.6(0.3)&16.7(0.5)& 10.1(0.7) & S&12\\
G175.34-10.8&  04:53:25.84 &+26:35:21.7 &0.52(0.23)&7.9(0.2)&12.7(0.9)&17.4(0.7)& 4.1(0.2) & S &    16.1 \\
G178.98-06.7&  05:17:37.38 &+26:05:53.2 &0.59(0.33)&6.5(0.2)&10.1(0.5)&17.4(1.1)& 12.6(0.5) & T & 16.4 \\
G192.2-11A2 & 05:31:28.9 & +12:30:20.8 &0.42(0.07)&4.5(0.0)&12.3(1.0)&19.4(1)& 8.5(0.6) & S&14.3\\
L1489E\tablefootmark{(4)} & 04:04:47.50 & +26:19:12.0 &0.14&2.5&--&17.8(0.7)& 11.9(0.7) &T&9\\
L1521B\tablefootmark{(5)} & 04:24:12.70 & +26:36:53.0 &0.14&--&--&13.1(0.4)& 14.1(1.1) &T&9\\
05413-0104\tablefootmark{(6)}  &  05:43:51.50 &-01:02:52.0  &0.4&--&--&17.9(1)& 7.4(0.4) & T & 31.4 \\
HH25 MMS\tablefootmark{(7)}     &  05:46:07.50 &-00:13:35.7  &0.4&--&--&15.8(0.7)& 26.8(0.7) & T & 29.2 \\
CB34\tablefootmark{(8)}       &  05:47:02.30 &+21:00:10.2 &1.5&--&--&17.3(0.6)& 8.7(0.6) & T & 9.7\\
\hline
\end{tabular}\\
\flushleft
{\footnotesize
$^{(1)}${The 1st to 3rd columns list the source names and Equatorial coordinates. 
PGCCs all have names starting with the letter 'G' followed by their Galactic coordinates.
The 4th to 8th columns list the distance, source size ($D_{\rm dust}$), Planck dust temperture of a whole
clump ($T_{\rm ECC}$), dust temperature ($T_{\rm dust}$ ) extracted from the Planck all-sky map, and column density
of H$_2$ ($N^{\rm dust}$(H$_2$))  derived from the dust emission.
The values in the parentheses are 1-$\sigma$ uncertainties.
}
$^{(2)}${This column remark the scanning modes for \HI~observations by FAST, with T and S representing tracking mode and
snapshot mdoe, respectively.}
$^{(3)}${The zenith angle of the source when observed with FAST.}
$^{(4)}${The distance of L1489 region is quoted from \citet{1988ApJ...324..907M,2019A&A...627A.162W}, 
while the angular size is obtained based on Herschel 500 $\mu$m map \citep{2014ApJ...793....1Y,2019A&A...627A.162W}. }
$^{(5)}${The distance of L1521B is quoted from \citet{2004ApJ...617..399H}.}
$^{(6)}${IRAS 05413-0104 is the driven source of the known collimated  bipolar jet HH 212, with a distance of 400 pc \citep{2019AJ....158..107R}. } 
$^{(7)}${The distance of HH25 MMS is quoted from \citet{1998MNRAS.298..644G,2004A&A...426..503W}. }
$^{(8)}${The distance of  CB34 is quoted from  \citet{1994ApJS...92..145Y,2002A&A...383..502K}. }
}
\end{table*}
 
\section{Sample} \label{sec:sample}
This work focuses on the HINSA observations towards PGCCs and the HINSA extracting methods.
Besides twelve PGCCs, our sample consists of  five comparison objects chosen from literature. They are also CCM abundant
sources, including
one starless core L1521B as well as four star forming sources. 
The star forming sources are chosen from an outflow catalogue 
\citep[][and the references there in]{2004A&A...426..503W}.

Among the twelve PGCCs, nine has been observed in HC$_3$N $J$=2-1 using the 
Tian Ma radio telescope (TMRT)\footnote{\url{http://65m.shao.cas.cn/}},
as part of the survey of searching for Ku-band CCMs emission in PGCCs leaded by Yuefang Wu. The main line
of HC$_3$N $J$=2-1 is $F$=3-2  (18196.3104  MHz). 
Eight of the twelve PGCCs have valid  HC$_3$N $J$=2-1 detections.

All of the twelve PGCCs have single point observations in $J$=1-0 of CO and its isotopomers 
by \cite{2012ApJ...756...76W} using the Purple Mountain Observatory (PMO) 13.7 m telescope,
and 11 of the PGCCs have also been mapped in  these transitions 
by \cite{2012ApJ...756...76W}.
The analysis in this work focuses on the 
PGCCs.  We do not try to give any quantitative analysis of those comparison objects.
Instead, comparing between the PGCCs and the comparison objects 
can lead to more clear conclusion about the effects of the core statues and  star formation activities
on the morphologies of HINSA features.
The \ion{H}{I} spectra from sources of different types can also help to 
test the robustness  of the HINSA extracting algorithm.
 
The names and coordinates of sources in our sample are listed in the first three columns of
Table \ref{tab:sources}. The  distribution of the sources in the Galactic plane is shown in Fig. \ref{galac-crop}.
The background of Fig. \ref{galac-crop} shows the dust temperature  decomposed  from the
Planck continuum images \citep{2016A&A...594A...1P,2016A&A...596A.109P}\protect\footnote{\url{https://wiki.cosmos.esa.int/planckpla2015}}
by \citet{2016A&A...596A.109P} using 
the Generalized Needlet Internal Linear Combination \citep[GNILC;][]{2011MNRAS.418..467R} component-separation method.
The PGCCs tend to locate at the margins of the emission regions of the Planck 353 $\mu$m continuum (Fig. \ref{galac-crop}).

For the twelve PGCCs in our sample, their distances  were calculated using the 
Bayesian Distance Calculator\footnote{\url{http://bessel.vlbi-astrometry.org/node/378}}
\citep{2016ApJ...823...77R} based on their coordinates and velocities. 
The Bayesian calculator constrains the probability density function (PDF) 
of distance based on four types of information: kinematic distance (KD), the spiral arm model (SA), 
Galactic latitude (GL) and parallax source (PS). 
The velocity dispersion has been taken into account to give the uncertainty of kinematic distance.
Their angular size ($D_{\rm dust}$) and  dust temperatures ($T_{\rm ECC}$) can be quoted  
from the Planck early cold core (ECC) catalog  \citep{2011A&A...536A..23P}, which is the pioneer sample of PGCCs. 
For the other five sources, their distances are adopted from literature (see Table \ref{tab:sources}),
and their $D_{\rm dust}$ are given if the Hershcel continuum data are available.

The distance, $D_{\rm dust}$ and $T_{\rm ECC}$ were listed in the 4th to 6th columns of 
Table \ref{tab:sources}.

\section{Observation and archive data}\label{sec:obs}

\subsection{\HI~observations using the FAST}
The FAST\footnote{\url{https://fast.bao.ac.cn/}} is a ground-based radio telescope built in Guizhou province of Southwest
China \citep{2011IJMPD..20..989N}, which is  the most sensitive single dish telescope.
The tracking accuracy is about 0.2\arcmin~\citep{2019SCPMA..6259502J}.
For zenith angle $\theta_{ZA}<$26.4\degr, the illuminated aperture is 300 meters.
A 19 beam receiver in the L-band  is employed. The full-width at half-maximum (FWHM) 
beam size is about 3\arcmin~around 1420 GHz for $\theta_{ZA}<$26.4\degr,
which is about half of the angular spacing between two neighboring beams, 5.87\arcmin .
Our observations (2019a-020-S) are conducted on August 5th and 11th, 2019.
The on-source time was 10 minutes for each source. 
Our targets all have $\theta_{ZA}$ smaller than 20\degr~during the observations as listed in Table \ref{tab:sources}.
It costs approximately equal to or less than 10 minutes for changing the targeted sources. 
The spectral mode with 1024 k channels evenly spaced between 1 GHz and 
1.5 GHz were adopted for our observations, corresponding to a channel spacing about 0.1 km s$^{-1}$ for \HI~line with rest frequency 
about 1420 GHz.

\begin{figure*}[htbp]
\centering
\includegraphics[width=0.7\linewidth]{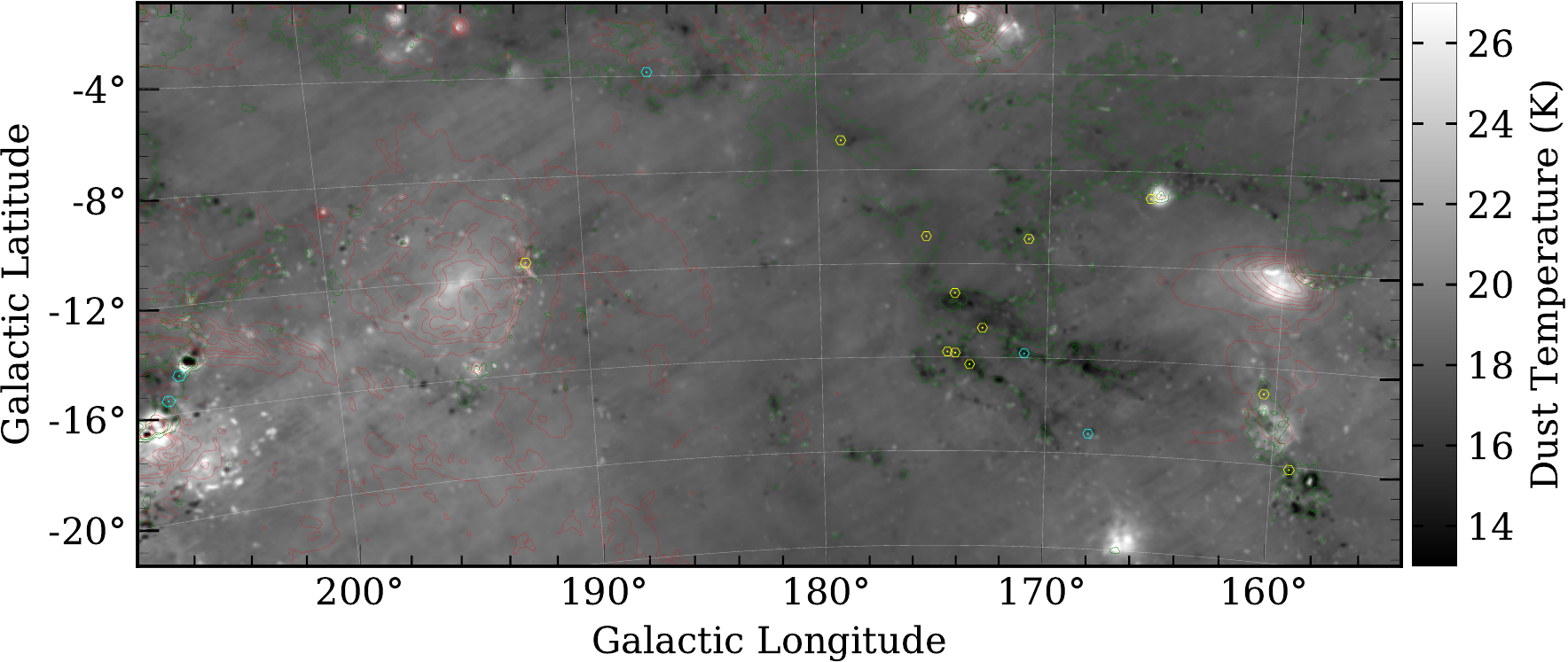}
\caption{
Galactic longitude-latitude positions of the 17 sources in our sample.
Yellow and cyan hexagons represent the 12 PGCCs and the other 5 sources, respectively. 
The size of the hexagon represents the covering area of the  19-beam receiver of the FAST.
The  dot within each hexagon  shows targeted locations of the central beam (Beam 1; Fig. \ref{fast_beams}).
The background image shows  Planck dust temperature distribution.
The green contours with levels (3,7,11) Myr/sr represent the Planck 850 $\mu m$ continuum.
The red contours represent the H$_\alpha$ emission \citep{2003ApJS..146..407F} in unit of R (10$^6/4\pi$  photons cm$^{-2}$
s$^{-1}$ sr$^{-1}$), with levels from 15 to 215 stepped by 40.  
\label{galac-crop} }
\end{figure*}

\begin{figure}[htb]
\centering
\includegraphics[width=0.75\linewidth]{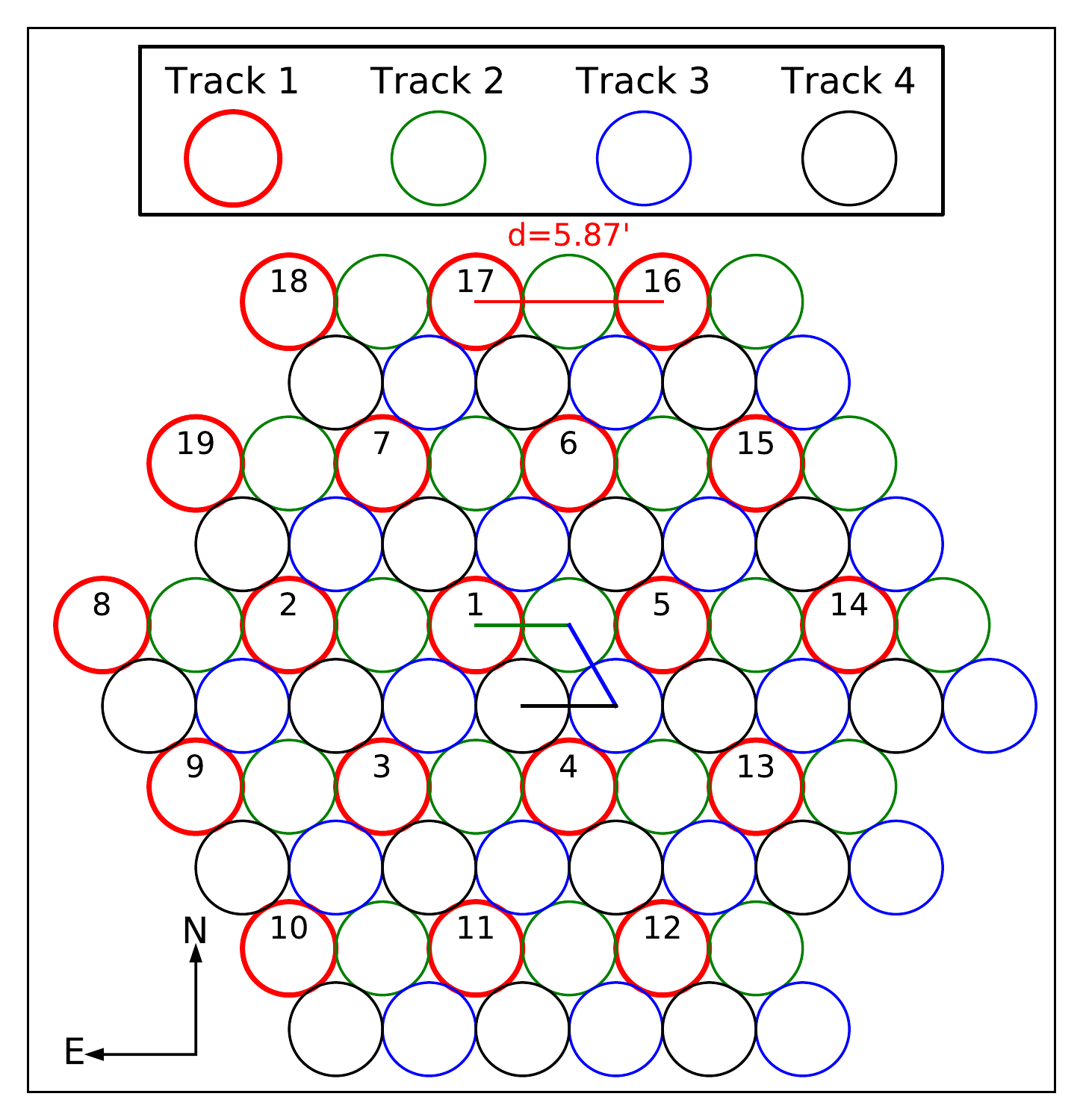}
\caption{The configuration of the 19 beam receiver of FAST \citep{2018IMMag..19..112L}. 
Eash snapshot mode observation consists of four single point mode observations. 
Equatorial coordinate system is adopted for the tracking mode, while galactic coordinate system is adopted for the snapshot mode. \label{fast_beams}}
\end{figure}

Seven of our sources were observed  using the tracking mode with the central beam (beam 1) always pointing on the center of the target,
while the other 10 sources were observed with the snapshot mode (see Table \ref{tab:sources}). 
The snapshot mode is made up by  four sets of tracking mode observations.
In snapshot mode, the targeted locations of the central beams between  two successive tracking mode observations
are slightly shifted (see Fig. \ref{fast_beams}).
The map obtained in snapshot mode is  spatially half-Nyquist sampled.

\begin{figure}[htbp]
\centering
\includegraphics[width=0.8\linewidth]{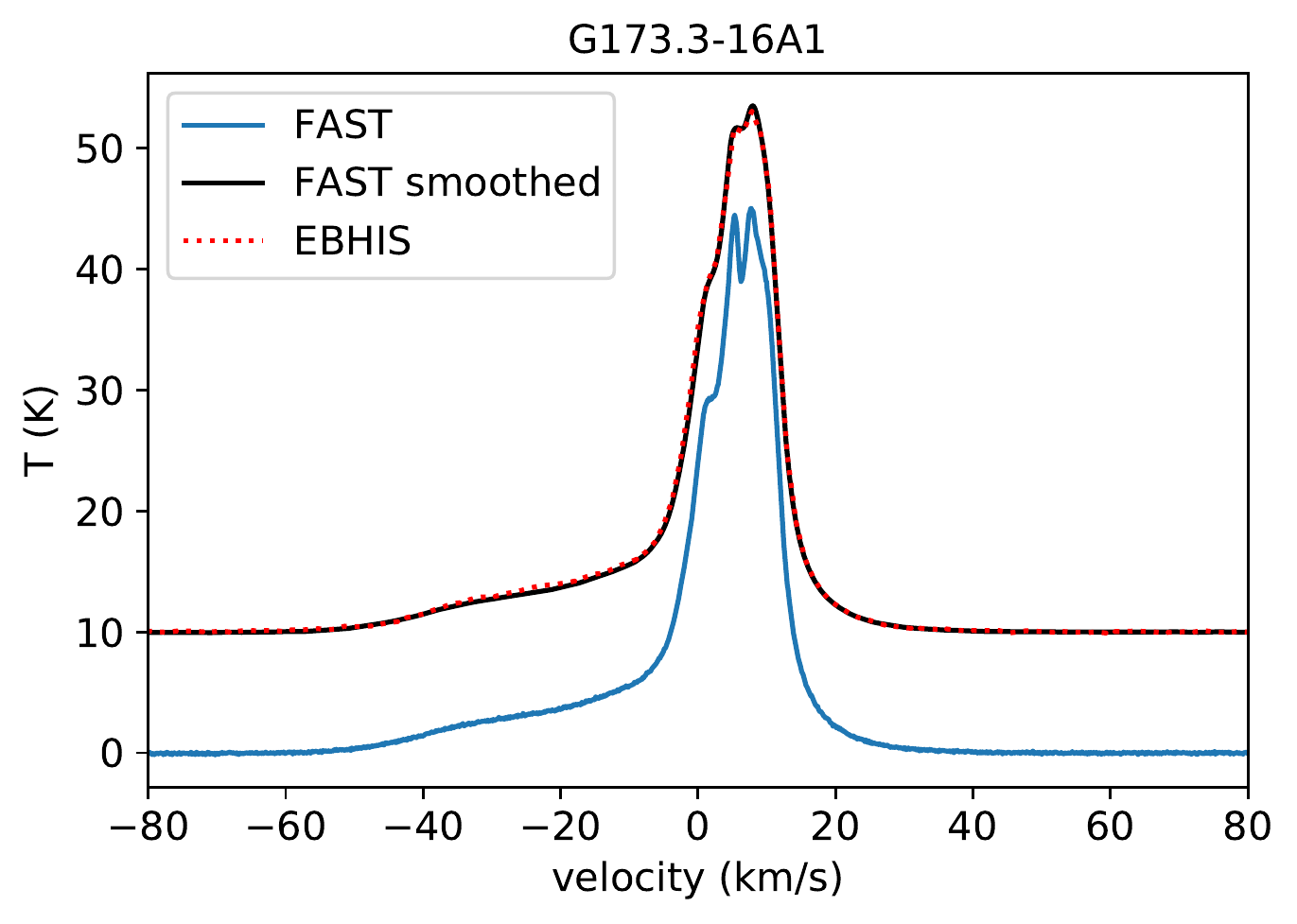}
\caption{Comparison between the \ion{H}{I} 21 cm spectra of G173.3-16A1 obtained by the FAST and the 
the Effelsberg-Bonn HI Survey \citep[EBHIS;][]{2016A&A...585A..41W}.
The blue solid line is the  \ion{H}{I} 21 cm spectrum obtained by the FAST, which is calibrated adopting
a beam efficiency of 0.85 for extended source. 
The red dashed line is the spectrum from EBHIS. The black solid line is the FAST \ion{H}{I} 21 cm spectrum smoothed 
to have a similar velocity resolution with the EBHIS spectrum.
The red dashed line and the black solid line have been moved up by 10 K to avoid overlapping.
\label{HI_calicompare_fig}  }
\end{figure}

The power spectra are recorded at every 0.1 second.
For calibration, the high-level noise (10 K) was injected lasting one second for every two seconds. 
The system temperature is about 20 K for $\theta_{ZA}<20\degr$. 
For point source, the beam efficiency is constant about $0.6$ for $\theta_{ZA}<26.4\degr$, and decreases 
to 0.5 as $\theta_{ZA}$ increasing to about 36\degr~\citep{2019SCPMA..6259502J}.
For extended source, the  beam efficiency is 0.85.
Fig. \ref{HI_calicompare_fig} shows the comparison between the \ion{H}{I} 21 cm spectra of G173.3-16A1 obtained by the FAST and 
the Effelsberg-Bonn HI Survey \citep[EBHIS;][]{2016A&A...585A..41W}.
They are consistent with each other if a beam efficiency of 0.85 is adopted for the FAST observation.

Binned with a channel spacing 0.1 km s$^{-1}$, the \ion{H}{I} 21 cm spectra we obtained have rms noise 
30-40 mK for sources observed in tracking mode,
and  60-80 mK for sources observed in the snapshot mode.

\begin{table*}
\centering
\caption{CO parameters. \label{tab:co_par}}
\begin{tabular}{ccccccccccc} 
\hline
source & $n$\tablefootmark{(1)} & $T_{12}$\tablefootmark{(2)}  & $V_{13}$& $\Delta V_{13}$& $V_{18}$ & $\Delta V_{18}$ & $\tau_{13}$& $\tau_{18}$ & D$_{\rm CO}$ & $N^{\rm CO}$(H$_2$) \\
       &    & K        &  km $^{-1}$  &  km $^{-1}$  & km $^{-1}$  &  km $^{-1}$& & &  \arcmin& 10$^{21}$ cm$^{-2}$\\
 \hline
G159.2-20A1 &1& 10.8(5) & 6.7(2) & 1.99(1) & 6.7(2) & 1.19(1) & 2.2(1) & 0.40(2) &10& 30.3(5) \\
 &2& 5.0(2) & 4.6(2) & 1.15(6) & 4.5(2) & 1.4(5) & 0.39(2) & 0.028(8) &7.9& 1.3(1) \\
G160.51-17.07 &1& 13.4(1) & 10.4(2) & 1.36(1) & 10.5(2) & 1.02(6) & 0.720(9) & 0.080(4) &7.2& 6.9(1) \\
G165.6-09A1 &1& 6.9(1) & -0.5(2) & 1.51(6) & -0.8(2) & 0.6(1) & 0.88(3) & 0.19(3) &13& 4.6(4) \\
G170.88-10.92 &1& 7.10(7) & 2.8(2) & 1.8(1) & -6.1(9) & 4(2) & 0.082(5) & 0.008(3) &3.2& 1.5(1) \\
 &2& 5.31(8) & 6.1(2) & 0.70(1) & 6.0(2) & 0.46(2) & 2.05(4) & 0.29(1) &12& 4.6(1) \\
G172.8-14A1 &1& 7.7(1) & 5.8(2) & 0.98(1) & 5.7(2) & 0.46(1) & 1.20(2) & 0.29(1) &12.5& 5.9(1) \\
 &2& 9.1(1) & 6.8(2) & 0.63(1) & 6.7(2) & 0.54(4) & 0.69(1) & 0.104(6) &10.5& 2.9(1) \\
G173.3-16A1 &1& 8.7(2) & 6.5(2) & 1.55(9) & 6.3(2) & 0.7(1) & 0.97(5) & 0.30(4) &9.4& 10.6(8) \\
G174.06-15A1 &1& 10.0(1) & 6.3(2) & 1.29(1) & 6.3(2) & 0.63(2) & 1.37(2) & 0.282(9) &11& 10.3(1) \\
G174.08-13.2 &1& 6.6(2) & 5.3(2) & 1.26(2) & 5.3(2) & 0.55(2) & 1.24(6) & 0.74(4) &--& 15.8(5) \\
G174.4-15A1 &1& 4.8(1) & 5.3(2) & 0.93(6) & 5.9(2) & 1.1(5) & 0.71(4) & 0.09(2) &9.3& 3.3(4) \\
 &2& 10.1(1) & 6.8(2) & 1.16(3) & 6.8(2) & 0.55(9) & 0.93(2) & 0.13(1) &7& 4.3(3) \\
G175.34-10.8 &1& 5.0(2) & 5.5(2) & 0.56(2) & 5.5(2) & 0.30(8) & 2.3(1) & 0.24(5) &8& 2.4(4) \\
G178.98-06.7 &1& 8.2(2) & 7.7(2) & 0.80(1) & 7.7(2) & 0.40(1) & 1.58(5) & 0.41(1) &11.7& 7.6(2) \\
G192.2-11A2 &1& 14.1(1) & 10.1(2) & 1.34(1) & 10.0(2) & 0.90(4) & 0.89(1) & 0.104(4) &5.3& 8.6(2) \\
\hline
\end{tabular}\\
\flushleft
{\footnotesize
$^{(1)}${The $n_{th}$ velocity components.}
$^{(2)}${The footnotes 12, 13 and 18 represents that the parameters are for $^{12}$CO, $^{13}$CO and C$^{18}$O.}
A number in parentheses indicates the 1-$\sigma$ uncertainty in the last digit.
}
\end{table*}

\subsection{HC$_3$N $J$=2-1 observations of nine PGCCs } \label{dec_tmrt}
Among the twelve PGCCs, nine have been observed in HC$_3$N $J$=2-1 using the 
Tian Ma radio telescope (TMRT) as part of the survey of  Ku-band CCMs emission 
in PGCCs lead by Yuefang Wu. 
The TMRT is a  65-m diameter fully steerable radio telescope located in Shanghai 
\citep{2016ApJ...824..136L}. The pointing accuracy is better than 10\arcsec. 
The main beam efficiency is 0.60 in the Ku band from 12 to 18 GHz \citep{2015AcASn..56...63W}. 
The front end is a cryogenically cooled receiver covering a frequency range
of 11.5--18.5 GHz. The digital backend system (DIBAS) \citep{2012AAS...21944610B}  is employed,  
which supports a variety of observing modes for molecular line observations. 
The mode 22 was adopted for our observation. In this mode, each of the three banks (Bank A, Bank B and Bank C)  has eight subbands
with a bandwidth of 23.4 MHz and
16384 channels. The center frequency of each subband is tunable to an accuracy of 10 kHz.  The calibration
uncertainty is 3 percent \citep{2015AcASn..56...63W}.

The Ku band observations cover the transition of \hctn~$J$=2-1 (18.196226 GHz).
The velocity resolution is  0.023 km s$^{-1}$ in the 18 GHz band.
Observations were conducted in July of 2017 and August of 2018.

\subsection{Archive data of CO and dense gas tracers} \label{sec_arcdata}
The spectral of the $J=1-0$ transitions of $^{12}$CO as well as its isotopomers $^{13}$CO and C$^{18}$O were
extracted from  \citet{2012ApJ...756...76W}.
Following up the releasing of PGCCs, \citet{2012ApJ...756...76W} observed the three CO lines
towards 673 PGCCs using the PMO 13.7 m telescope\footnote{\url{http://www.dlh.pmo.cas.cn/}}.
The morphologies and dynamic properties of PGCCs are
well understood thanks for the CO mapping observations \citep{2012ApJ...756...76W,2012ApJS..202....4L,
2013ApJ...775L...2L,2013ApJS..209...37M,2016ApJS..224...43Z,2020ApJS..247...29Z}.
All the twelve PGCCs have single point observations of these three CO lines. 
Among them, all except for G174.08-13.2 have been mapped in these 
lines. 
The CO spectra can help to identify HINSA through  providing 
us the initial guesses of the velocity and line width of the HINSA 
feature (see Sects. \ref{sec:method} and \ref{sec_result_extract_hinsa}). 

Observations of emission lines of dense gas tracers towards PGCCs are rare.
Among the sources in this work, two PGCCs G165.6-09A1 and G174.4-15A1 have been mapped in C$_2$H $N$=1-0 (87.316925 GHz)
and N$_2$H$^+$ $J$=1-0 (93.173397 GHz) using  the PMO 13.7 m telescope \citep{2019A&A...622A..32L}.
The G192.2-11A2 has been mapped in HCO$^+$ $J$=1-0 (89.188523 GHz) using the 
Instituto de
Radioastronom\'{i}a Milim\'{e}trica (IRAM) 30 m 
telescope\footnote{\url{https://www.iram-institute.org/EN/astronomers.php?}} (Project ID: Delta03-13).
The observation of HCO$^+$ $J$=1-0 in G192.2-11A2 was conducted on May 3 of 2014,
with a velocity resolution of 0.16 km s$^{-1}$.



\section{Methods of Extracting HINSA} \label{sec:method}
The \ion{H}{I} 21 cm spectrum towards a molecular cloud is a mixture of  foreground and background emission 
and the absorption from cold \HI~component in that molecular  cloud. It is not easy to
extract the absorption components from an \ion{H}{I} 21 cm spectrum, since the spectrum without
absorption features ($T_ \ion{H}{I}$) is unknown. A reference point having the same foreground and background \ion{H}{I} emission 
with the target source but showing no absorption features is difficult to found.

Combining Eqs. (1) and (5) from \citet{2003ApJ...585..823L}, the observed continuum-subtracted \ion{H}{I} spectrum brightness temperature along the HINSA sight line is 
\begin{equation}
\begin{aligned}
&T_{\rm r} = [T_ce^{-\tau_b}+T_b(1-e^{-\tau_b})]e^{-\tau}e^{-\tau_f} \\
&\quad\quad\quad\quad\quad\quad\quad +T_{\rm ex}(1-e^{-\tau})e^{-\tau_f}+T_f(1-e^{-\tau_f}) - T_c
\end{aligned}
\end{equation}
Here, $\tau$ is the optical depth of the cold \ion{H}{I} component,
$\tau_b=p(\tau_f+\tau_b)$, $\tau_f$ and $\tau_b$ denote the  foreground and background \ion{H}{I} optical depth, and $T_{\rm c}$ is
the continuum brightness temperature. 
$T_{\rm ex}$, $T_b$ and $T_f$ are the  excitation temperatures
of the cold \ion{H}{I} component, the background warm \ion{H}{I} component and the foreground
warm \ion{H}{I} component, respectively.
$T_\ion{H}{I}$ can then be expressed as $T_r|_{\tau=0}$.

If $T_f = T_b$ and both $\tau_f << 1$ and $\tau_b << 1$, the absorbed  component of \ion{H}{I} spectrum can be expressed as in  
Eq. (8) of \citet{2003ApJ...585..823L}
\begin{equation} \label{eq_tab_2}
T_{ab}=T_\ion{H}{I}-T_r = [pT_\ion{H}{I}+(T_c-T_{ex})(1-\tau_f)](1-e^{-\tau})
\end{equation}
Eq. (\ref{eq_tab_2}) leads to the expression of $T_\ion{H}{I}$ as 
\begin{equation}
T_{\ion{H}{I}} =\frac{T_{\rm r}+(T_c-T_{\rm ex})(1-\tau_f)(1-e^{-\tau})}{1-p(1-e^{-\tau})} 
\label{eq:T_HI_define} 
\end{equation}
To avoid over-fitting, parameters $p$, $T_{\rm ex}$, $\tau_f$  and $T_{\rm c}$ 
should be known parameters or fixed to be  reasonable values.

The excitation temperature ($T_{\rm ex}$) of HINSA is assumed to be the value  
of molecular gas $\sim$ 10 K, which is much smaller than $T_r$. 
$T_c$ should be smaller than $T_{\rm ex}$ and $T_r$,
if it is adopted as the typical value of the $L$ band intensity of the Galactic background emission $\sim 3.5$ K
\citep{2003ApJ...585..823L}.  
We extract the L-band background continuum intensities for our sources from the Bonn-Stockert survey
\citep{1982A&AS...48..219R,1986A&AS...63..205R}, 
which gives a mean value of 3.9 K with a standard deviation of 0.2 K.
Neglecting $T_c$ is  reasonable and may lead to an overestimation of the optical depth of HINSA by 
10 percent.

Each of our sources has a short distance $d$ compared with the Galactics scale.  Because of the short distance and not high Galactic latitude $b$ (Table \ref{tab:sources}), the height from the Galactic plane
$d\sin(b)$  is small compared with the scale height of  
the Galactic \ion{H}{I} layer \citep[$\sim$350 pc;][]{1984ApJ...283...90L,2003PASJ...55..191N}.
Most of the \ion{H}{I} emitters along the line of sight should locate behind the absorption source. 
We believe that p should be close to 1 for most of our targets, and we adopt p=1 hereafter.
We note that, if $p < 1$, this assumption may lead to underestimations of the HINSA optical depth and column density.
In optical thin limit, the obtained optical depth of HINSA would be approximately $p\tau_{\rm real}$, where $\tau_{\rm real}$
is the real optical depth.

If p=1, $\tau_f$=0, and $T_{\rm ex}$  and $T_c$ are ignored, Eq. (\ref{eq:T_HI_define}) can be simplified as 
\begin{align}
 T_{\ion{H}{I}}  &=  T_{\rm r} e^{\tau} + (T_{c}-T_{\rm ex})(e^\tau-1) \label{eq:T_HI_define_middle}\\
                 &\sim T_{\rm r} e^{\tau} \label{eq:T_HI_define_simple}
 \end{align} 
Despite of its simple form, Eq. (\ref{eq:T_HI_define_simple}) does not deviate much from Eq. (\ref{eq:T_HI_define})
for HINSA with small optical depth.

The key step of extracting HINSA is to make reasonable estimations of the optical depths of cold \HI~($\tau$) in Eq. (\ref{eq:T_HI_define})
or Eq. (\ref{eq:T_HI_define_simple}), and to obtain the corresponding $T_{\ion{H}{I}}$ with the highest  probability to represent the real 
unabsorbed spectrum.
The HINSA extracting methods (method 1 and method 2) described below  are
optimized for   HINSA with $T_{\rm ex} << T_r$
and $T_{c} << T_r$. 
Eq. (\ref{eq:T_HI_define_simple}) 
helps to estimate the signal-to-noise (S/N) requirement of the method 1 and
throw light on the motivation of the method 2.
Great care is required if these methods are implemented to extract absorption features
of HISA clouds with warmer gas, 
significant foreground HI emission, 
or significant background continuum emission.

\subsection{Method 1} \label{sec:m1}
HINSA usually shows significant feature in the second derivative representation of $T_{\rm r}$ \citep{2008ApJ...689..276K}. 
It is assumed the $\tau$ can be expressed as (multi-)Gaussian function,
\begin{equation}
\large{ \tau(v) = \sum_{i=0}^{m-1} \tau_i e^{-\frac{\left(v-v_i\right)^2}{2\left(\sigma_i\right)^2}}  }
\end{equation}
where $\tau_i$, $\sigma_i$ and $v_i$ represent the optical depth, velocity dispersion and central velocity of the $i_{th}$ component,
respectively.
Free parameters ($\tau_i$, $\sigma_i$, $v_i$) are  needed to be fitted during  extracting  HINSA features.  
The initial  trial values of ($\tau_i$, $\sigma_i$, $v_i$) and the fixed parameter $T_{\rm ex}$ 
can be estimated from emission lines of other species,
especially those supposed to have space distributions similar to  the cold \ion{H}{I} in molecular cloud.

The Gaussian fitting tries to obtain a smooth curve which can be expressed as a straight line or other analytic function.
However, the $T_{\ion{H}{I}}$ is unknown and can not be assumed as a linear function of velocity or frequency.
Thus, the Gaussian fitting can not be directly applied to extract HINSA. 
Instead, the fitting process can be conducted under a relaxed judgment of the fitted curve ($T_{\ion{H}{I}}$), that is $T_{\ion{H}{I}}$
should looks smoothest.

$T_ \ion{H}{I}$ can be obtained through minimizing the sum of the square of second derivative of $T_{\rm r}$,
denoted as $\mathcal{R}$ in this work.
This method (method 1) is first highlighted by \citep{2008ApJ...689..276K} to extract HINSA. 
Here, $\mathcal{R}$ is expressed as
\begin{equation}
\begin{aligned}
\mathcal{R}(T_ \ion{H}{I} ) &=\left\Vert \frac{\partial^2T_ \ion{H}{I} }{\partial v^2}  \right\Vert_2= \int_{v_l}^{v_u}\left( \frac{\partial^2T_ \ion{H}{I} }{\partial v^2}  \right)^2dv\\
             & \cong \sum_{i=1}^N (T_ \ion{H}{I} ^{''})_i^2 dv
\end{aligned}
\label{R_define}
\end{equation}
where 
$v_u$--$v_l$ is the frequency or velocity range that we are interested in.
The $T_\ion{H}{I}''$ in Eq. (\ref{R_define}) represents the numerical version of the second derivative of $T_\ion{H}{I}$,
\begin{equation}
T_{\ion{H}{I}}^{''} = \frac{T_{\ion{H}{I},i+1}+T_{\ion{H}{I},i-1}-2T_{\ion{H}{I},i}  }{\Delta_{\rm ch}^2} \label{eqpp}
\end{equation} 
where $\Delta_{\rm ch}$ is the channel spacing of the spectrum.

The basic idea of this method  is to find a set of parameters ($\tau_i$, $\sigma_i$, $v_i$)
which will reasonably fill the dips of $T_{\rm r}$ through Eq. (\ref{eq:T_HI_define}). 
The $T_ \ion{H}{I}$ with minial $\mathcal{R}$ value
is the  smoothest $T_{\ion{H}{I}}$ which can represent the unabsorbed spectrum.

                

However, this method (also denoted as method 1) has theoretical defects and is noise-sensitive. 
We will give an analytic
formula to quantify the S/N threshold of method 1 in Sect. \ref{sec:m1}.
An improved method (method 2) will be proposed and described in Sect. \ref{sec:m3}.

\subsubsection{Short-coming of method 1} 
$T_\ion{H}{I}$ usually shows complex profile. 
To estimate the S/N of the HINSA detection in the second derivative representation ($\mbox{S/N}(T_{\ion{H}{I}}^{''})$),
$T_\ion{H}{I}$ is approximated by a Gaussian function with 
a line width of $\Delta_{w}$ and a peak intensity of $T_{\rm peak}$. 
Here, 
 $T_{\rm peak}$ and  $\Delta_{w}=N_{\rm ch}\Delta_{\rm ch}$ are adopted as the typical height and 
typical width  
of $T_\ion{H}{I}$, respectively. 
S/N$(T_{\ion{H}{I}}^{''})$ can then be estimated from Eq. (\ref{eqpp}) and
the second derivative of a Gaussian function  \citep[e.g. Eq. (2) of][]{2008ApJ...689..276K},
and expressed as
\begin{equation}
\mbox{S/N}(T_{\ion{H}{I}}^{''}) \sim \frac{\mbox{S/N}(T_{\ion{H}{I}}) 8 \ln(2)}{N_{\rm ch}^2\sqrt{6}} 
\sim \frac{2}{{N_{\rm ch}^2}} {\mbox{S/N}(T_{\ion{H}{I}}) }
\end{equation}
The $1/N_{\rm ch}^2$ term requires a very high S/N of the spectrum.
Another problem is that, when $T_{\rm r}$ is multiplied by a factor of $g$ during the fitting process,
the noise will also be amplified by the same factor. 
The value of this amplification factor $g$ can be estimated from Eq. (\ref{eq:T_HI_define_simple}) as 
\begin{equation}
g=\frac{\partial T_ \ion{H}{I} }{\partial T_{\rm r}}\sim e^\tau \label{eq:g1}
\end{equation}
In theory, the original $T_{\ion{H}{I}}$ will not be fully recovered through this method, since the $\mathcal{R}$ value 
contributed by the noise will also be amplified.
The deviation between the original $T_{\ion{H}{I}}$ and the fitted $T_{\ion{H}{I}}$ is systematic, and it can be reduced
but never completely eliminated  by improving S/N. 
 
\subsubsection{S/N threshold of method 1}
Here, we further explain why the method 1 can only partly extract the HINSA feature especially when S/N is low,
and  try to quantitatively assess the performances of method 1 when applied to spectra with different S/Ns.
Considering a linear spectrum which has an absorption feature with  a Gaussian-like optical depth 
\begin{equation}\tau=\tau_0\times exp(-v^2/2\Delta^2) \end{equation}  
The $\mathcal{R}$ value corresponding to a test value of $\tau_0$  (denoted as $\tau_{0,t}$) can be expressed as 
\begin{equation}
\mathcal{R}(T_ \ion{H}{I} ) = \left( \frac{\tau_0-\tau_{0,t}}{\tau_0} \right)^2\mathcal{R}(T_{\rm ab}) + e^{2\tau_{0,t}} \mathcal{R}(T_{\rm noise})
\end{equation}
The fitted value of $\tau_0$ (denoted as $\tau_{0,f}$) can be obtained through solving
equation
\begin{equation}
\left. \frac{ \partial \mathcal{R} }{\partial \tau_{0,t}} \right|_{\tau_{0,f}} \sim -2\left(1-\frac{\tau_{0,f}}{\tau_0}\right)\frac{1}{\tau_0}\mathcal{R}(T_{\rm ab}) + 2\mathcal{R}(T_{\rm noise}) =0 \label{eq_tau_0f}
\end{equation}
Here, we assume that $\tau_{0,f}$ is small and thus $e^{2\tau_{0,f}}\sim 1$.
A recovering factor ($rf$) is defined as 
\begin{equation}
rf = \frac{\tau_{0,f}}{\tau_0},
\end{equation}
which represents the proportion of the HINSA extracted through method 1. 
Eq. (\ref{eq_tau_0f}) leads to
\begin{equation}  rf \sim 1-\tau_0 \frac{\mathcal{R}(T_{\rm noise})}{\mathcal{R}(T_{\rm ab})} \sim 1-\frac{\tau_0N_{\rm ch}^4}{4} 
\left(\frac{\sigma(T_{\rm noise})}{T_{\rm ab}}\right)^2\end{equation}
$\sigma(T_{\rm noise})$ is the root mean square of the noise.
The signal-to-noise threshold for method 1  (S/N$^c$) is defined as the value of $T_{\rm ab}/\sigma(T_{\rm noise})$ when $rf=3/4$,
which leads to 
\begin{equation}
\mbox{S/N}^c \sim \sqrt{\tau_0} N_{\rm ch}^2  \label{snr_c_eq}
\end{equation}  
The recovering factor can then be expressed as
\begin{equation} \label{eqrf_}
rf = 1-\frac{1}{4}\frac{\mbox{S/N}^c}{\mbox{S/N}(T_{\rm ab})}
\end{equation}  
If S/N($T_{\rm ab}$) is smaller than S/N$^c$, $rf$ should be small and Eq. (\ref{eqrf_})
is only statistically valid.
Thus, it is not permitted to obtain the optical depth of HINSA  through dividing the fitted value by $rf$.
In such case the method 1 can only partly extract the HINSA feature or totally fails.

Taking   the channel spacing of our spectra $\Delta_{\rm ch}$ as 0.1 km/s,
the  absorption line width as 1 km s$^{-1}$ (corresponding to $N_{\rm ch}$=10), and $\tau=0.1$, Eq. (\ref{snr_c_eq}) leads to 
${\mathrm{S/N}}^c>30$. Keep in mind that ${\mathrm{S/N}}^c$ is defined on $T_{\rm ab}$ scale,
and it will be multiplied by a factor of 10 if defined on $T_{\rm r}$ scale.
Such high S/N can not always be achieved in observations.

\subsection{Method 2} \label{sec:m3}
The main difficulty of method 1 is that it recovers $T_{\ion{H}{I}}$ through multiplying $T_r$ by a factor of $\sim e^\tau$,
which will lead to the multiplication of noise simultaneously. 
To inhibit the increasing  of noise, we replace the multiplication 
on $T_r$ in Eq. (\ref{eq:T_HI_define_simple}) (or Eq. (\ref{eq:T_HI_define_middle})) by an addition operation.
After a simple manipulation, Eq. (\ref{eq:T_HI_define_simple}) can be rewritten as
\begin{equation}
T^f_ \ion{H}{I} =T_{\rm r}+ (1-e^{-\tau})T^f_ \ion{H}{I} \sim  T_{\rm r}+ (1-e^{-\tau})T^f_{\ion{H}{I}, smooth}
\label{eq:T_HI_define_anotherform} 
\end{equation} 
where $T^f_\ion{H}{I}$ means the fitted $T_ \ion{H}{I}$.
This equation highlights the possibility to keep the noise component in $T_\ion{H}{I}$ constant and 
equal to that in $T_{\rm r}$, if the  $T_\ion{H}{I}$ term in the right side is replaced by its smoothed 
version $T_{\ion{H}{I}, smooth}$.
It is reasonable since we expect that  $T_\ion{H}{I}$ is smooth and can be 
approached by low-order polynomial.

The fit process can be described as
\begin{itemize}
\item[1.] For each given test value of free parameters, calculate $T_\ion{H}{I} $ through 
Eq. (\ref{eq:T_HI_define}).
\item[2.] 
          Obtain $T_{\rm HI,smooth}$, a smooth version of $T_{\ion{H}{I}}$, e.g., by $n_{\rm th}$-order polynomial fitting.
\item[3.] Update $T_\ion{H}{I} $ using Eq. (\ref{eq:T_HI_define_anotherform}).
\item[4.] Calculate the $\mathcal{R}$ value.
\item[5.] Repeat above steps to find the minimal $\mathcal{R}$ value and its corresponding $T_{\ion{H}{I}}$.
\end{itemize}
\subsubsection{Advantage of method 2}
Since most of our HINSA features are weak, it is safe to assume that $\tau_0 << 1$,
which would produce a near-Gaussian absorption dip.
Method 2 is to recover $T_{\ion{H}{I}}$ through adding back a smoothed near-Gaussian component
to $T_r$. It splits the multiplying term $e^\tau T_{HI}$ in Eq. (\ref{eq:T_HI_define_simple}) 
into two terms, one keeps the noise and details, the other is smoothed.
What we try to do is to balance the resolution and S/N.
In method 2, the noise will not increase much, and 
the velocity resolution is kept since the smoothing is not conducted on $T_{\rm r}$,
but on $T_{\ion{H}{I}}$ which is expected to be smooth.
In this work, we use polynomial fitting to obtain a smooth version of $T_{\ion{H}{I}}^f$, 
but other methods such as convolution could also be used.

Once $T^f_{\ion{H}{I}}$ is obtained, the absorbed component can be calculated through 
\begin{equation}T_{\rm ab}=T^f_ \ion{H}{I} -T_{\rm r} \sim T^f_ \ion{H}{I}(e^\tau-1) \end{equation}
where $T^f_ \ion{H}{I}$ represents the fitted $T_ \ion{H}{I}$ and $T_{\rm ab}$ 
should be a positive quantity. If the HINSA is optically thin, the noise intensity of $T_{\rm ab}$  given by method 1
would be small with a value of only about $\tau T_{noise}$.
If method 2 is adopted,  the calaulated $T_{\rm ab}$ shoud looks further smoother since the amplification of noise in $T^f_{\ion{H}{I}}$ has been restrained.

\begin{figure*}
\centering
\includegraphics[width=0.24\linewidth]{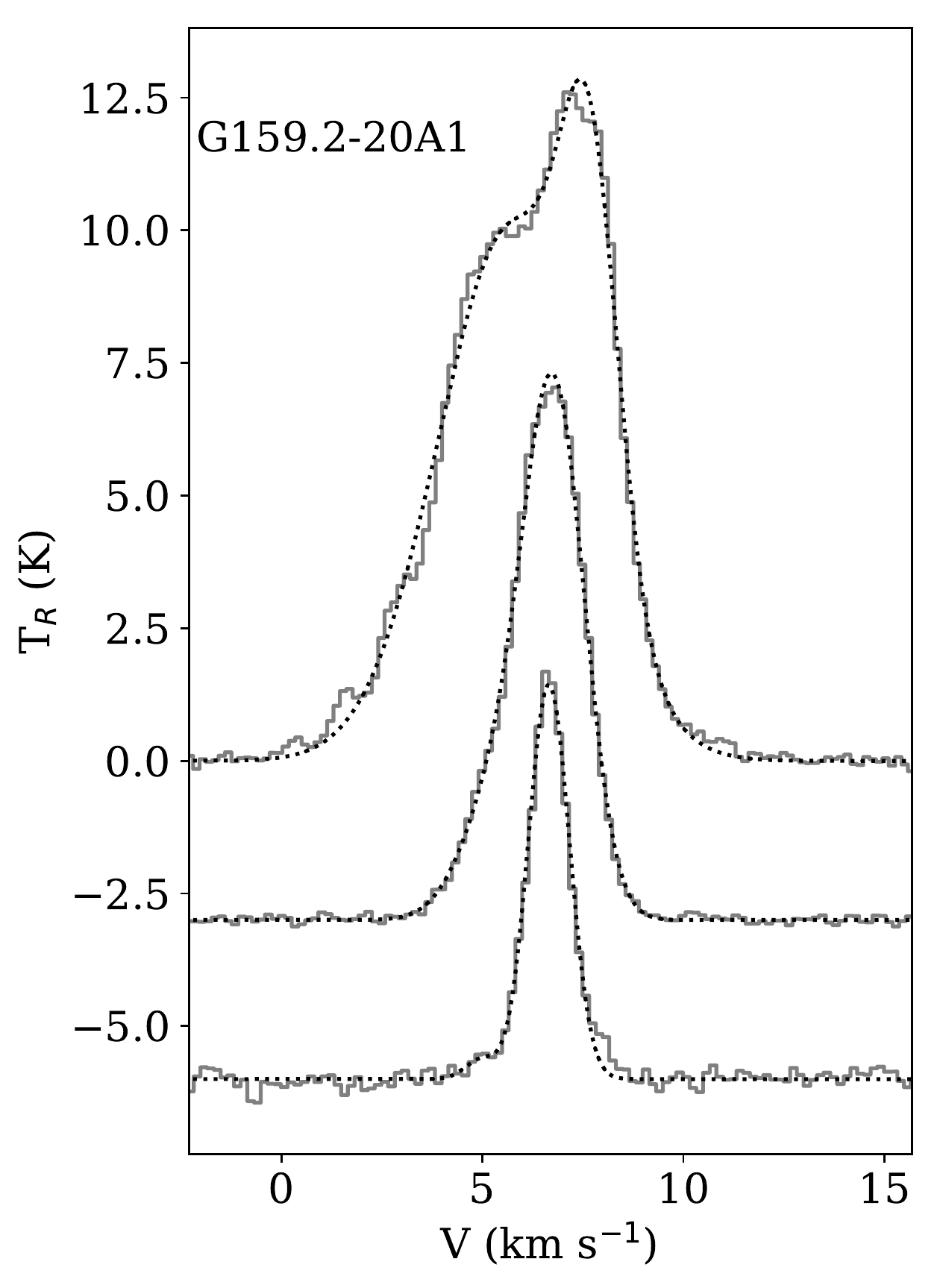}
\includegraphics[width=0.24\linewidth]{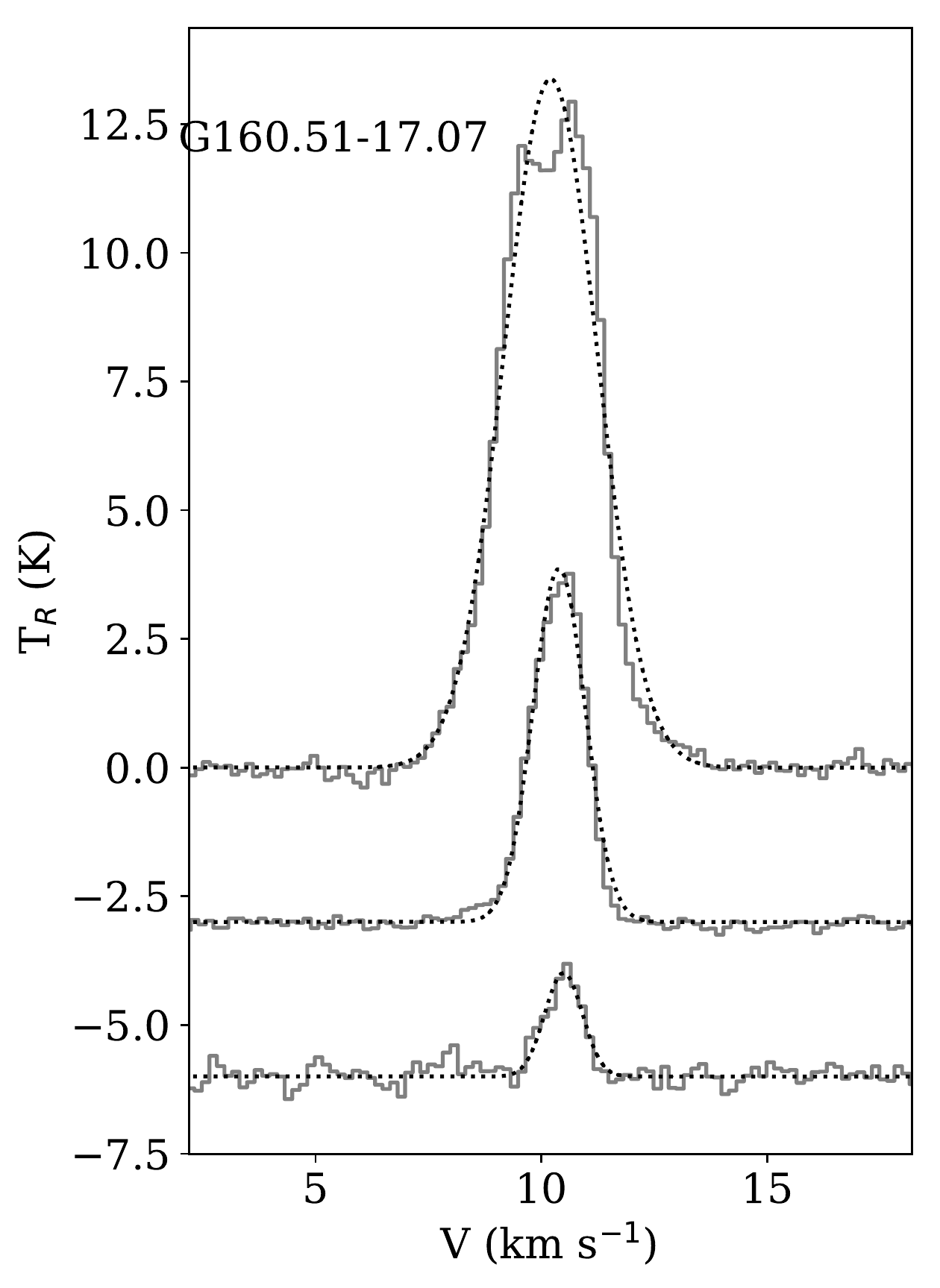}
\includegraphics[width=0.24\linewidth]{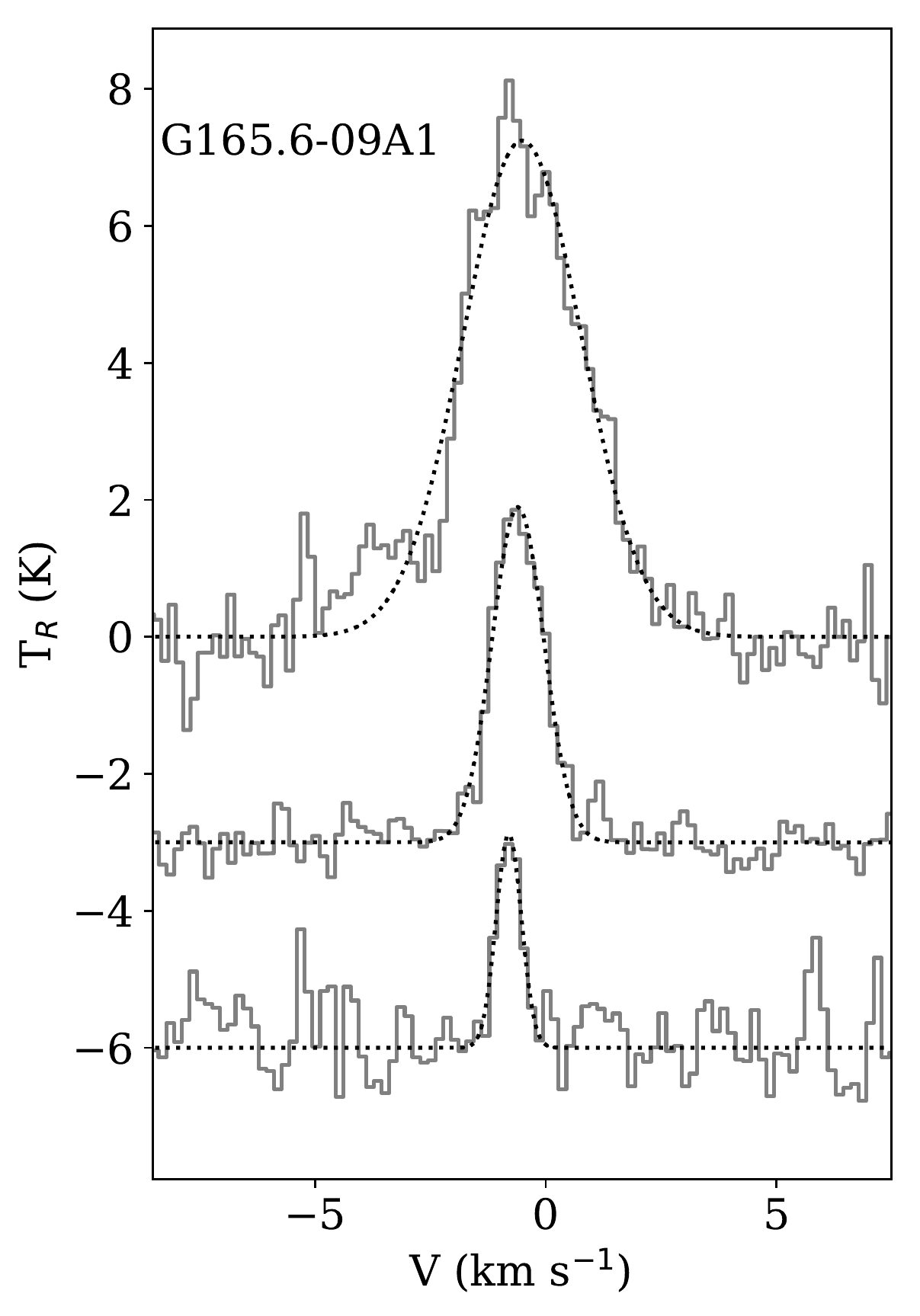}
\includegraphics[width=0.24\linewidth]{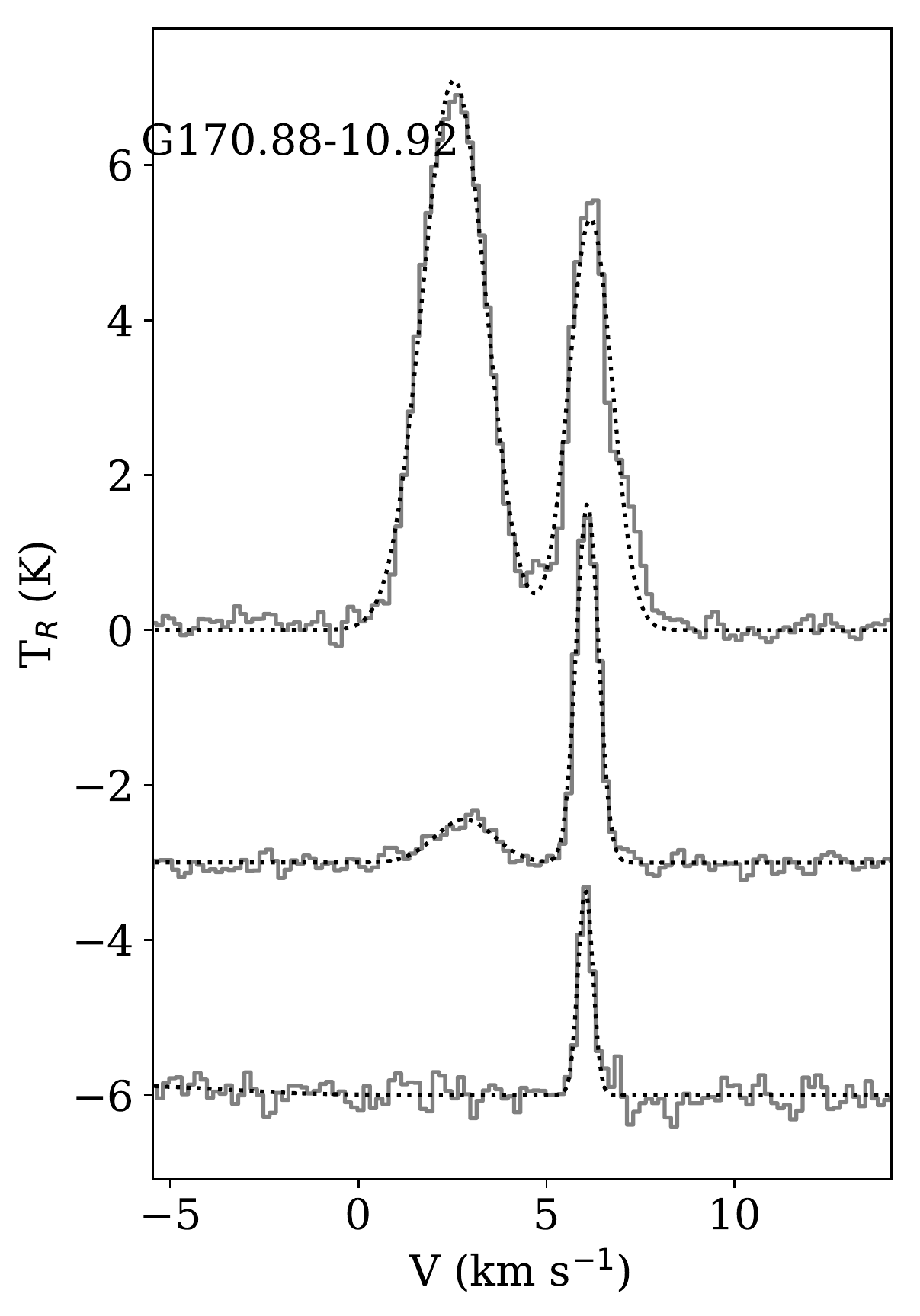}
\includegraphics[width=0.24\linewidth]{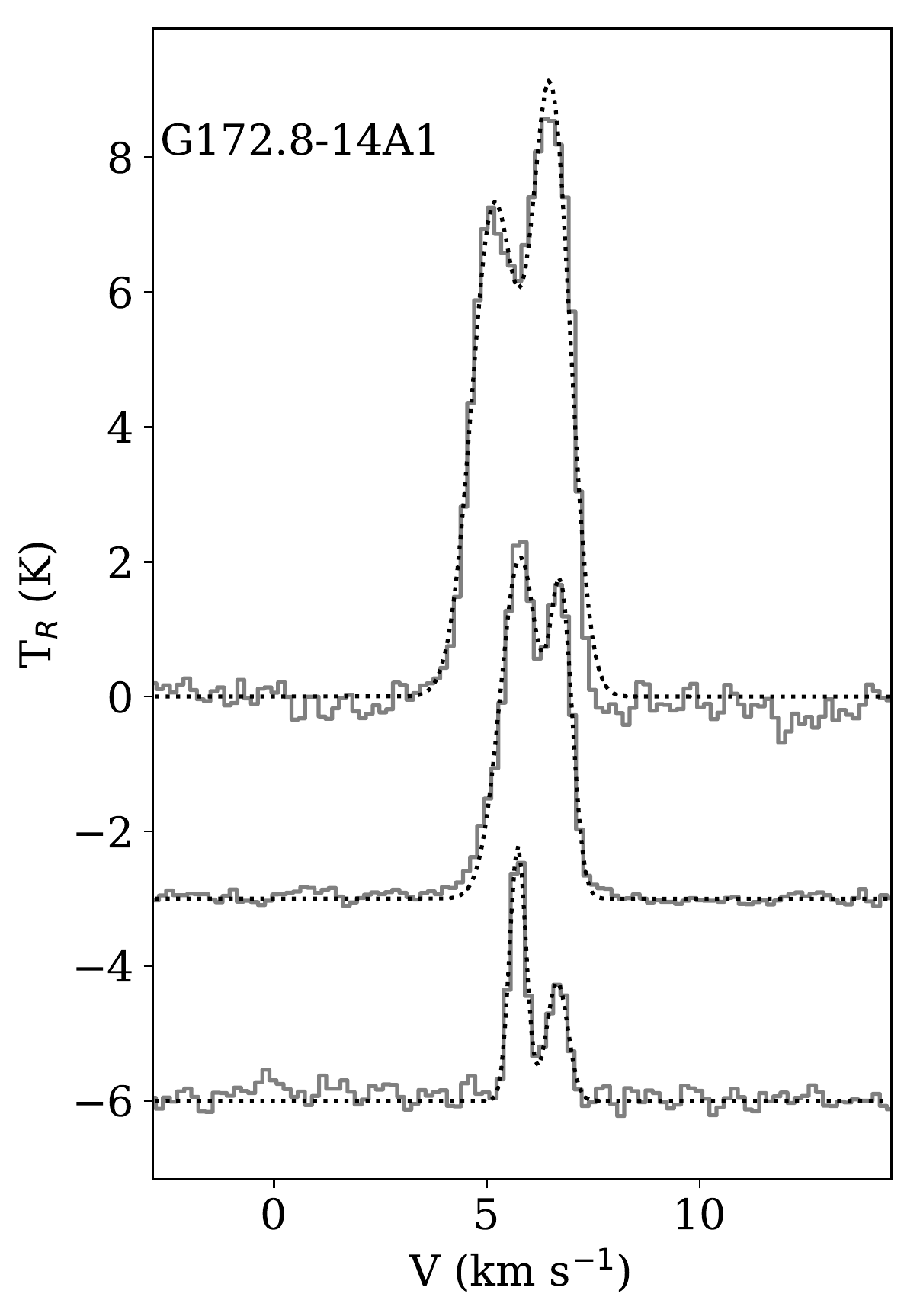}
\includegraphics[width=0.24\linewidth]{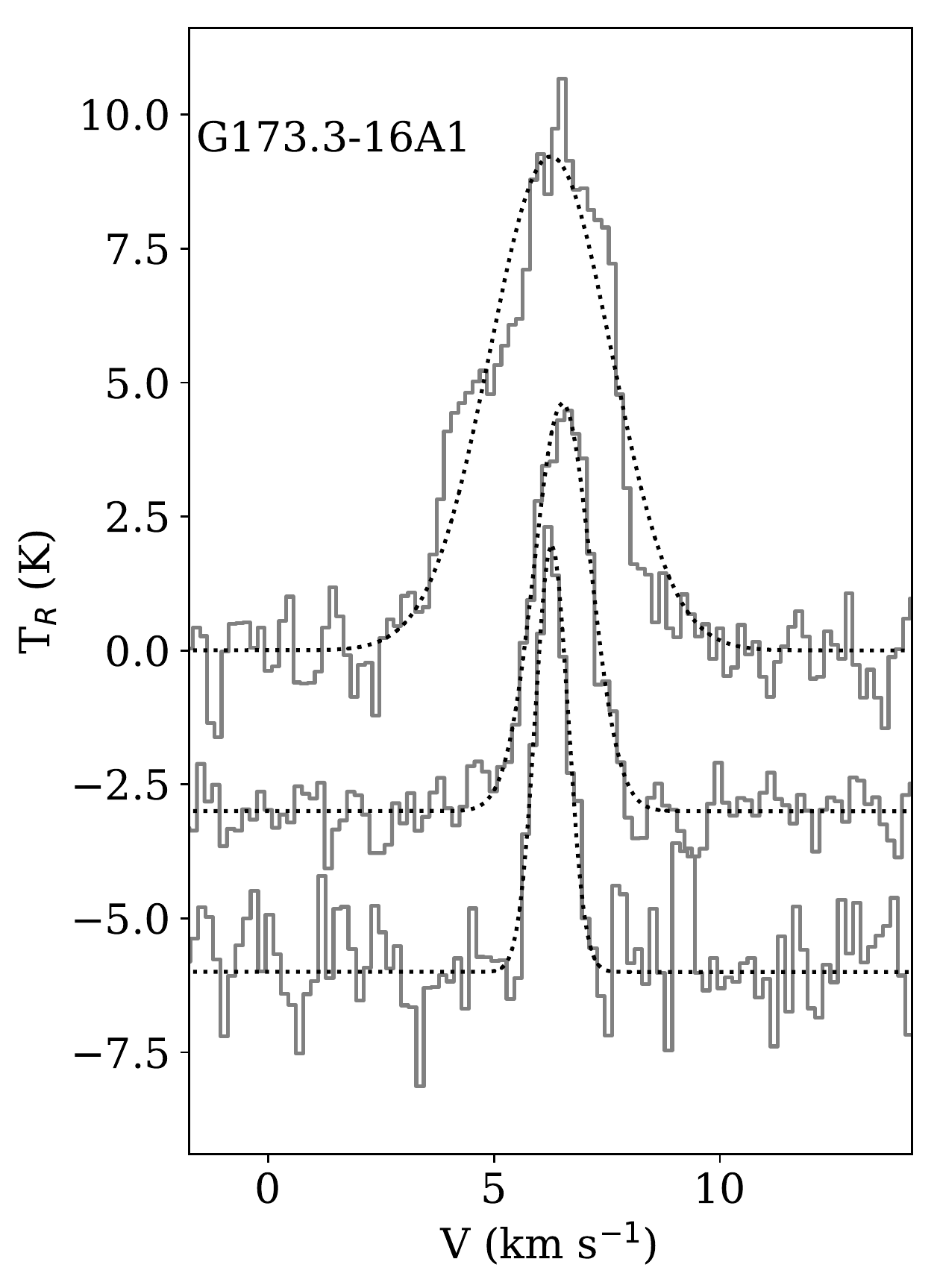}
\includegraphics[width=0.24\linewidth]{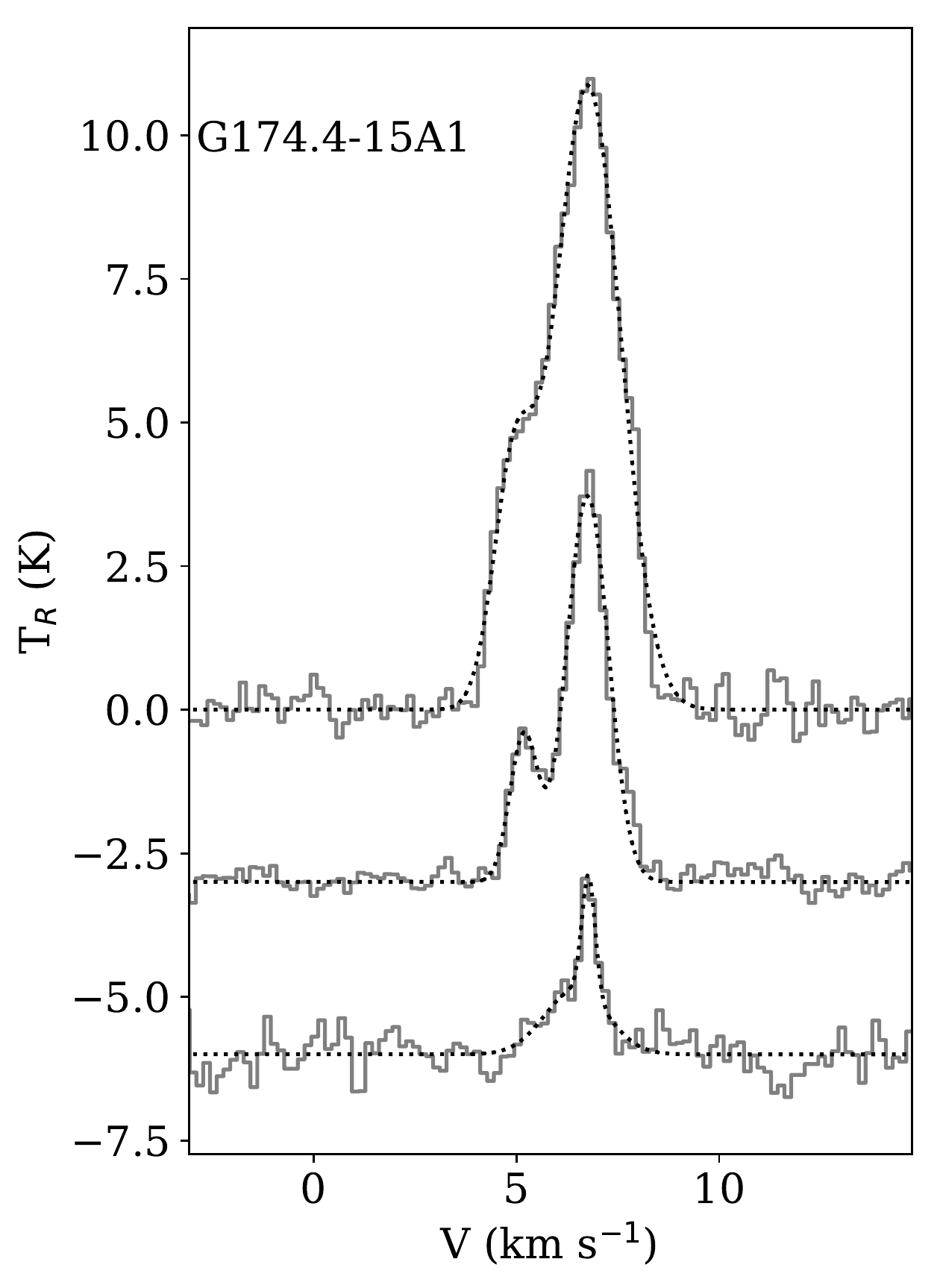}
\includegraphics[width=0.24\linewidth]{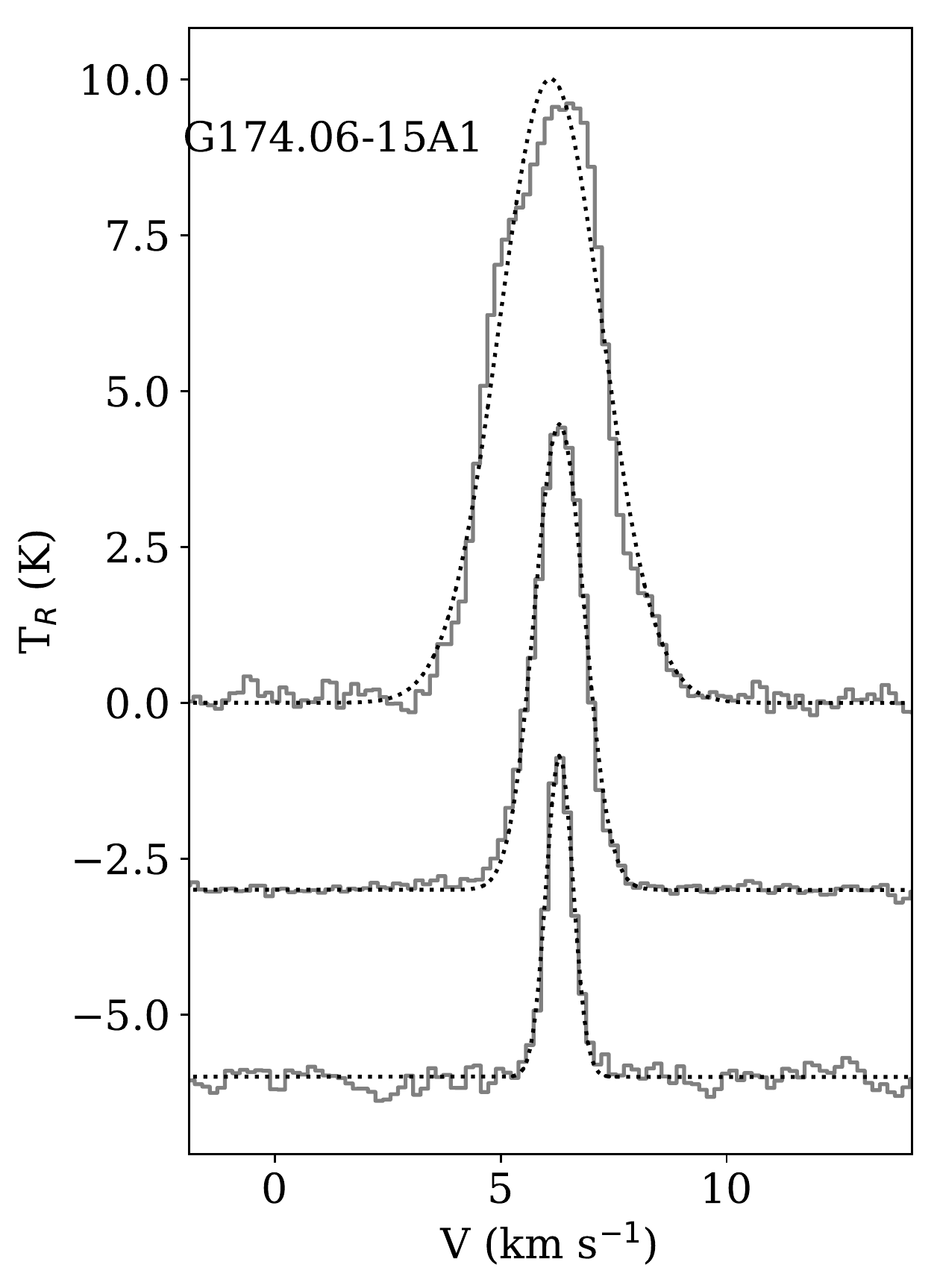}
\includegraphics[width=0.24\linewidth]{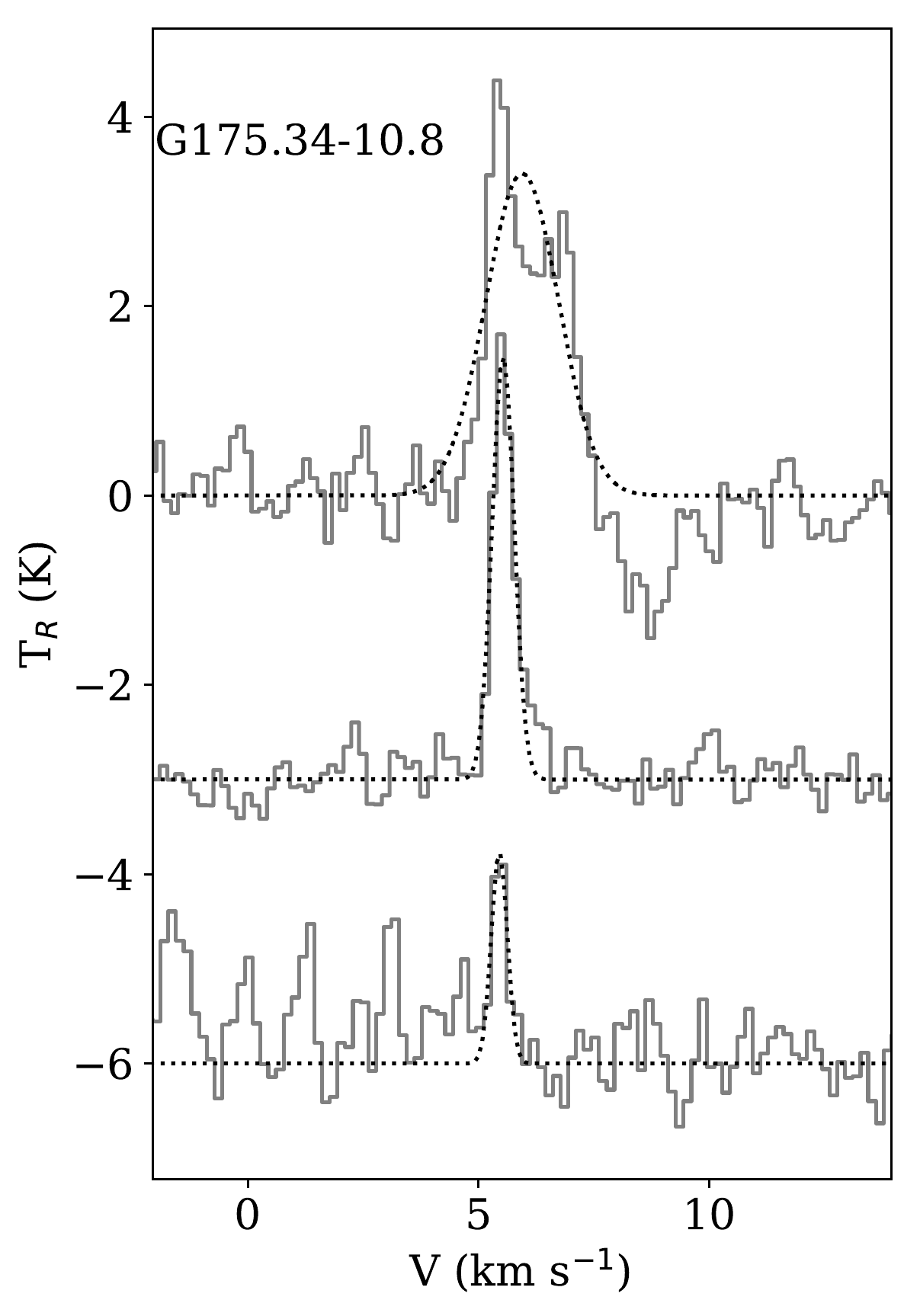}
\includegraphics[width=0.24\linewidth]{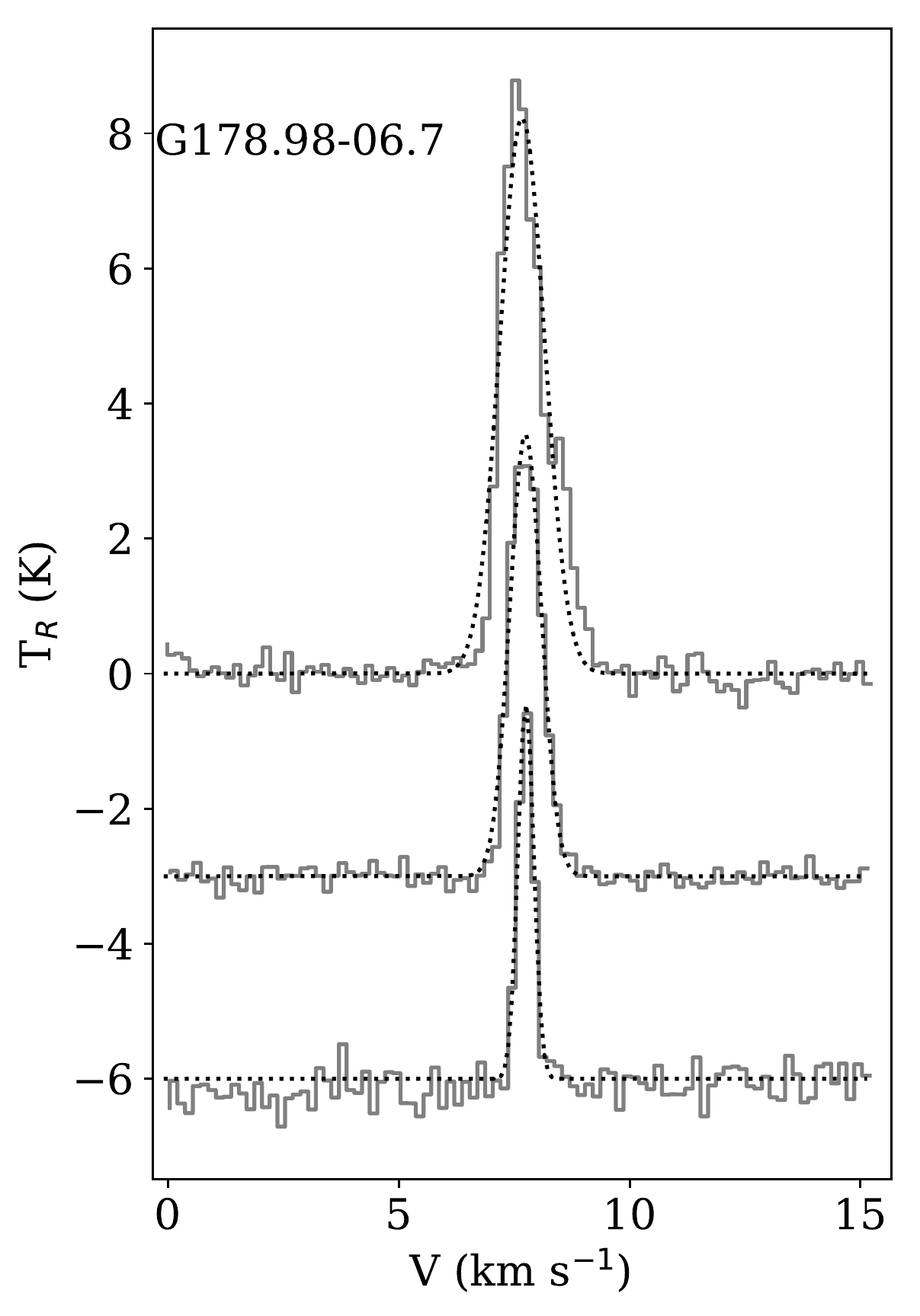}
\includegraphics[width=0.24\linewidth]{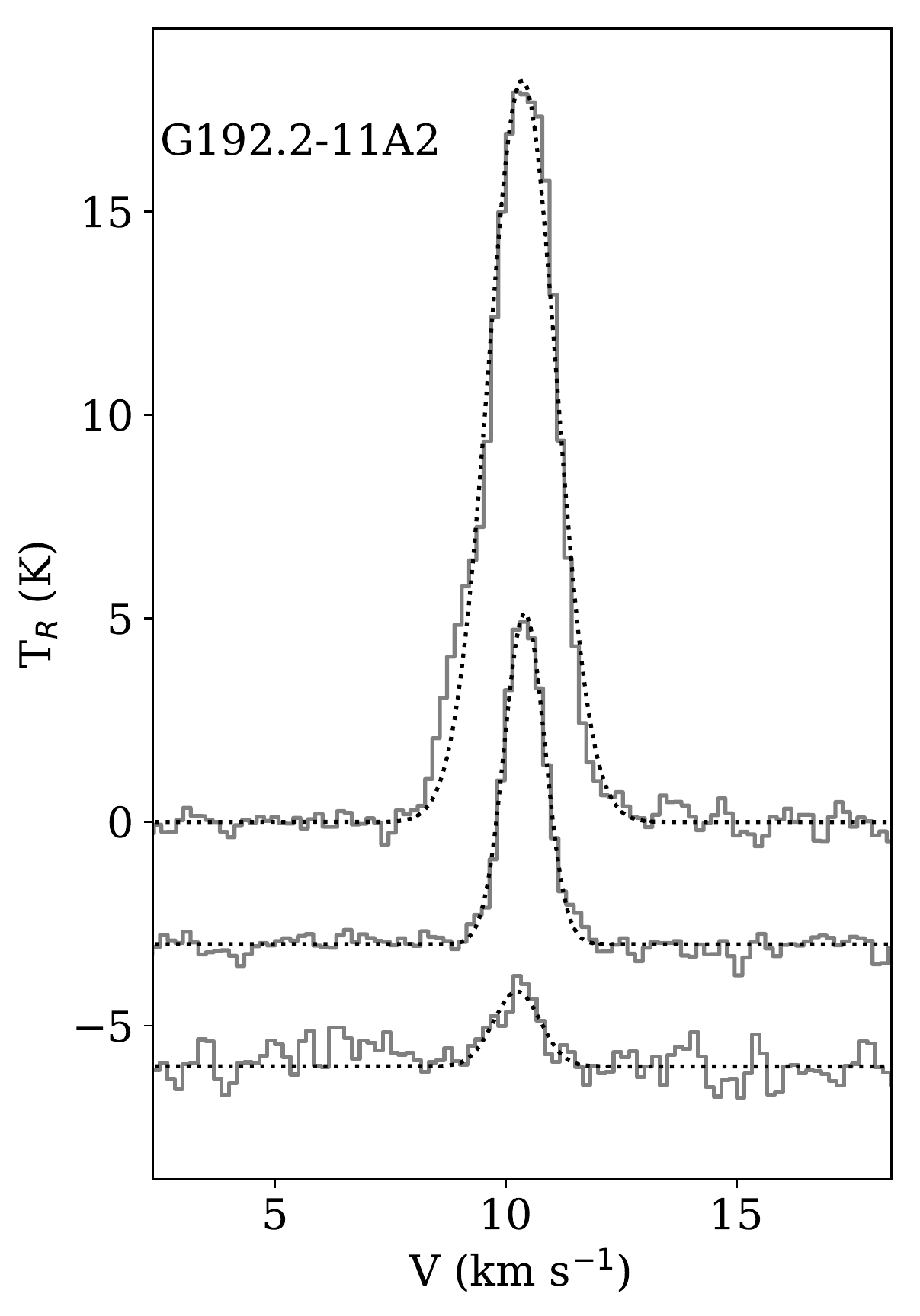}

\caption{The spectra of J=1-0 of $^{12}$CO, C$^{13}$O and C$^{18}$O \citep{2012ApJ...756...76W} are shown from upper to lower in each panel. \label{emission_line_spec}}
\end{figure*}

\begin{figure*}[htbp]
\centering
\includegraphics[width=0.48\linewidth]{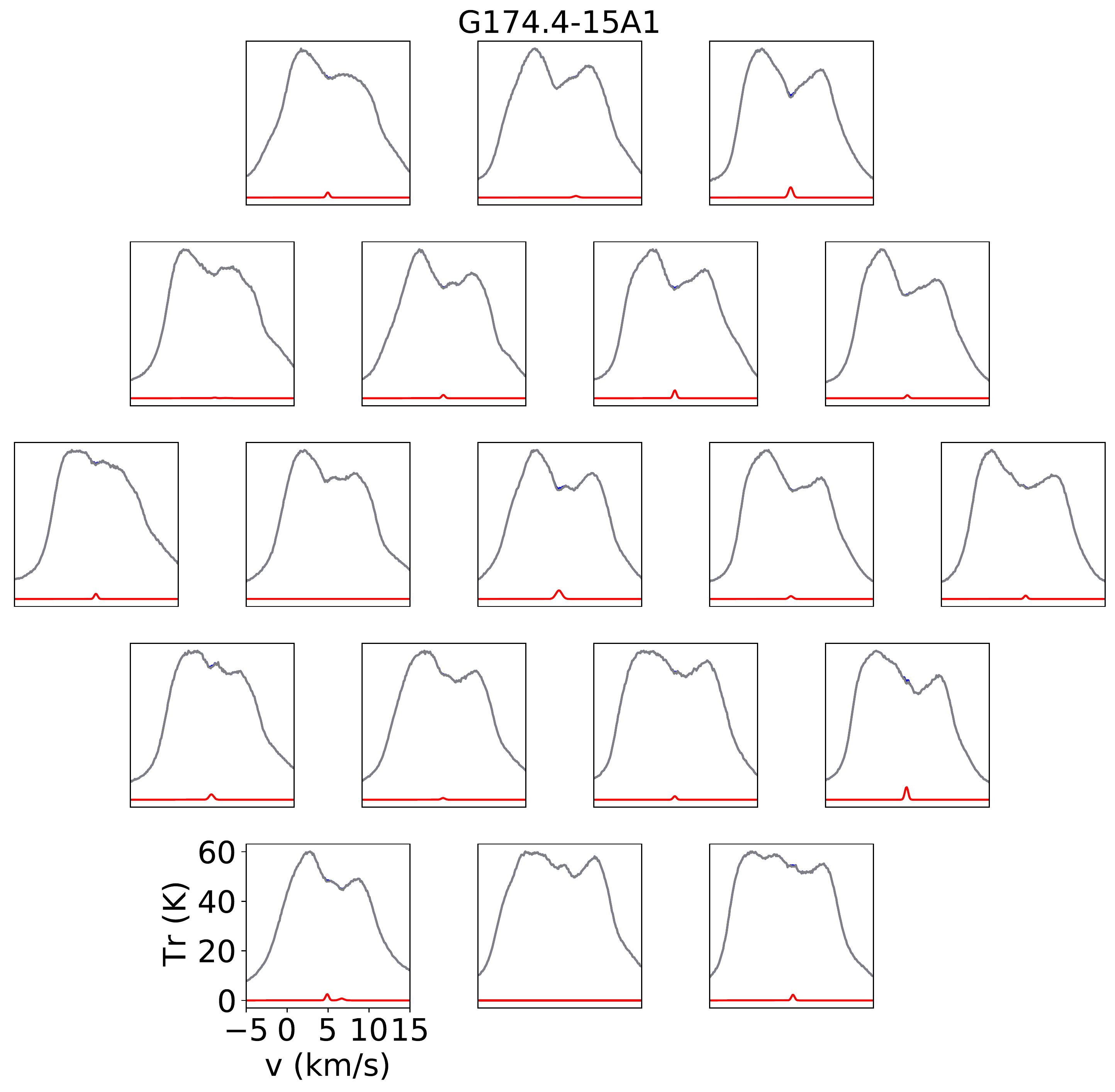}
\includegraphics[width=0.48\linewidth]{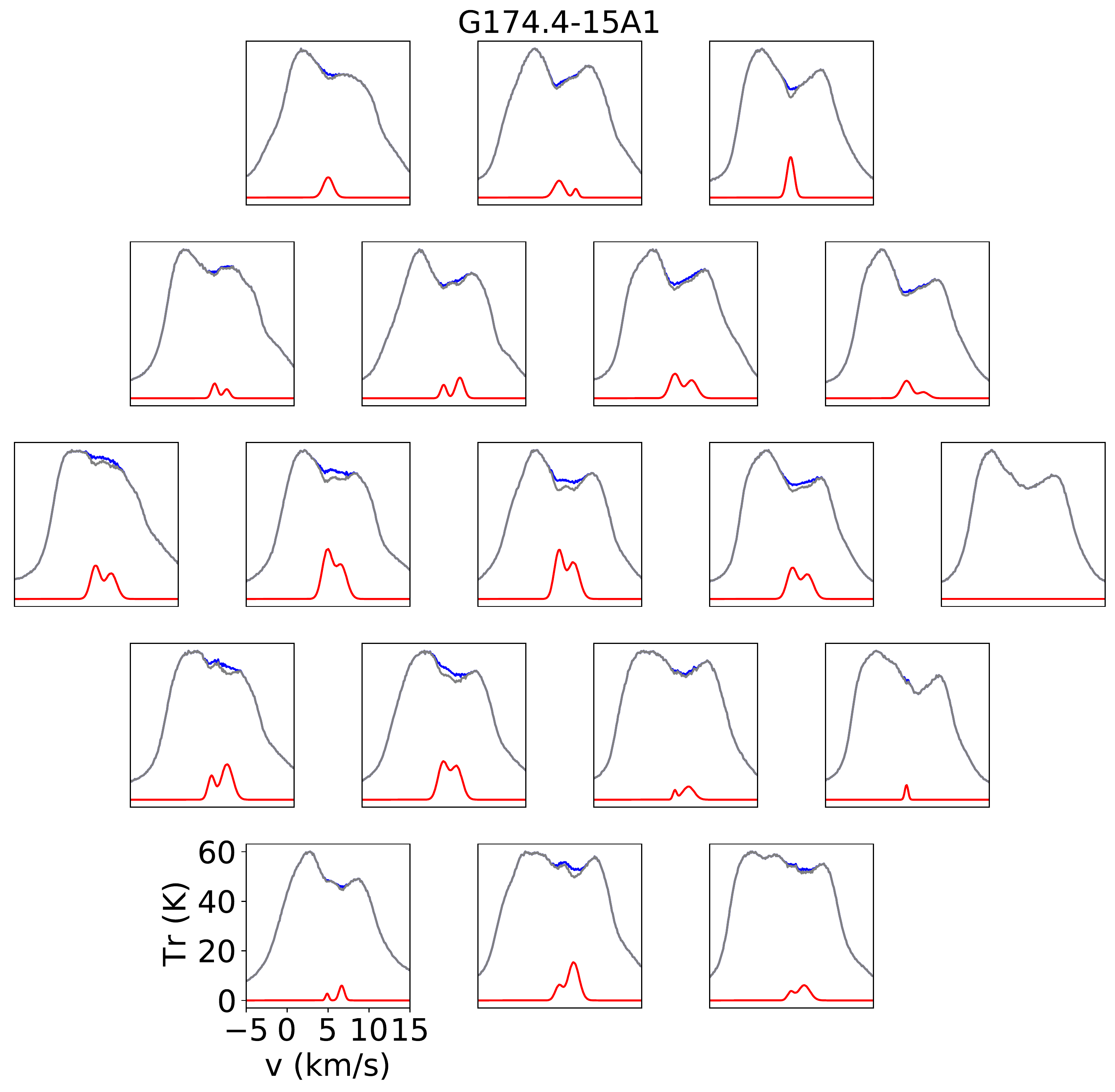}
\caption{HINSA features of G174.4-15A1 extracted by method 1 (left)  and method 2 (right).
The method 1 fails to fit the HINSA component because of the low S/N and the blending of two absorption components with similar 
velocities. 
In each panel, gray line is the observed spectrum $T_{\rm r}$, blue line is the recovered $T_{\ion{H}{I}}$,
and red line is $T_{\rm ab}$. $T_{\rm ab}$ shown here has been multiplied by a factor of 5.
 \label{example_HI_spe}}
\end{figure*}

\subsubsection{S/N threshold of method 2}
We try to quantitatively access the performance of method 2, and compare it with method 1.
The method 2 works when 
\begin{equation}
\mathcal{R}(T_{\rm ab}+T_{\rm noise}) > \mathcal{R}(T_{\rm noise})
\end{equation}
which leads to
\begin{equation}
\mathcal{R}(T_{\rm ab}) > 5\delta\left( \int 2 T_{\rm ab}^{''}T_{\rm noise}^{''} dv \right)
\sim 5 \delta\left( \int  2 T_{\rm ab}^{''''}T_{\rm noise} dv \right) \label{thresh_m2}
\end{equation}
Here $\delta(\vec{X})$ represents the standard deviation of $\vec{X}$.
It is important to note that $T_{\rm ab}$ is a fitted curve which does not contain noise.
If Eq. (\ref{thresh_m2}) is satisfied, we say that the absorption feature against the noise has a 
significance  higher than five-sigma,
when extracted using method 2.
Eq. (\ref{thresh_m2}) can be satisfied when S/N($T_{\rm ab}$) is larger than the  
signal-to-noise threshold for method 2 (S/N$^{c;n}$).
To derived S/N$^{c;n}$ from Eq. (\ref{thresh_m2}), we will use 
\begin{equation}
\lim_{v\to \infty} T_{\rm ab}'' = \lim_{v\to \infty} T_{\rm ab}  = 0
\end{equation}
and the approximation
\begin{equation}
\frac{(T_{\rm ab}'')^2}{T_{\rm ab}''''} \sim T_{\rm ab}.
\end{equation}
Eq. (\ref{thresh_m2}) can then be approximated as
\begin{equation}
(T_{\rm ab}'')^2 > 10 \frac{(T_{\rm ab}'')^2}{T_{\rm ab}} \delta\left(\int T_{noise} dv\right)
\sim 10 \frac{(T_{\rm ab}'')^2}{T_{\rm ab}} \frac{\delta{T_{noise}}}{\sqrt{N_{\rm ch}}}
\end{equation}
which leads to
\begin{equation}
\mbox{S/N}^{c;n} = \frac{T_{\rm ab}}{\delta(T_{noise})}  \sim \frac{10}{\sqrt{N_{\rm ch}}}
\label{ulti_label}
\end{equation}

From Eq. (\ref{ulti_label}), it can be said that the method 2 is analogous to the Gaussian fitting, 
since both of them have sensitivities proportional to the square root of line  widths.
In contrast to method 1, method 2 favours to extract features with relatively large line widths resultant from
multiple absorption components or  high velocity resolution.
The S/Ns of the spectra required by method 2 are much smaller than those by method 1.

\subsection{Comparison between  method 1 and method 2}
Method 2 is an updated version of method 1. 
It overcomes the shortcoming of method 1 through introducing a smoothing intermediate step, 
and thus has less strict requirement on S/N. If the $T_{\ion{H}{I}}$ spectral
structure is too complex, method 2 may lose resolution.

If the absorption dip deviates from Gaussian shape, 
the fitted unabsorbed 
spectrum ($T_{\ion{H}{I}}^f$) may be larger or smaller than the real $T_{\ion{H}{I}}$.
For method 1, the fitted $T_{\ion{H}{I}}^f$ will then tend to be smaller than $T_{\ion{H}{I}}$, because 
solution of $T_{\ion{H}{I}}^f$ larger than $T_{\ion{H}{I}}$ will be more severely penalized by the amplification of noise.
For method 2,
the fitted $T_{\ion{H}{I}}^f$ has no system bias of being larger or smaller than the real value. 
Thus, method 2 is  less predicted 
and method 1 seems to be more robust when the absorption dip deviates much from Gaussian shape.
However, the results of the two methods in such cases are both somewhat unreliable. 

In practice, method 1 is applicable only when S/N $>$ S/N$^c$ (Sect. \ref{sec_result_extract_hinsa}).
This confirms the validity of the threshold of signal-to-noise ratio (S/N$^c$) defined in Eq. (\ref{snr_c_eq}) to describe the
requirement of method 1, and highlights the usefulness of method 2. 
For molecular clouds such as PGCCs, S/N$\sim$S/N$^c$ requires a FAST on-source integration  time of $\sim$10 minutes,
where S/N is the signal-to-noise ratio of the absorption feature in the observed \ion{H}{I} spectrum.
For FAST drift-mode observation \citep{2018IMMag..19..112L}, the effective integration  time is $\sim$ 40 seconds
\citep{2019SCPMA..6259506Z}.
Method 2 provides a possibility to extract HINSA features from spectra with such short integration time.

\begin{figure*}[htbp]
\centering
\includegraphics[width=0.48\linewidth]{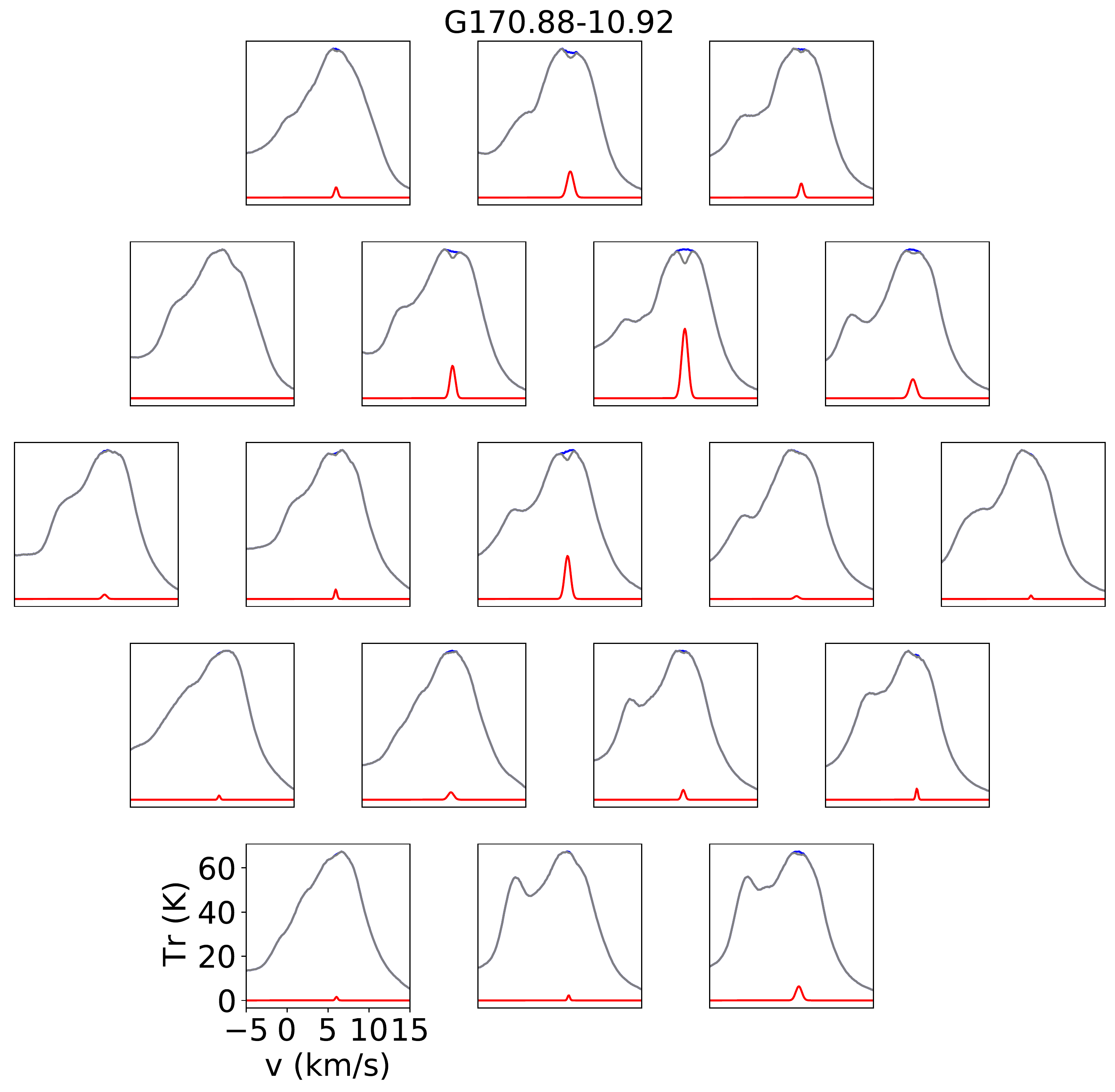}
\includegraphics[width=0.48\linewidth]{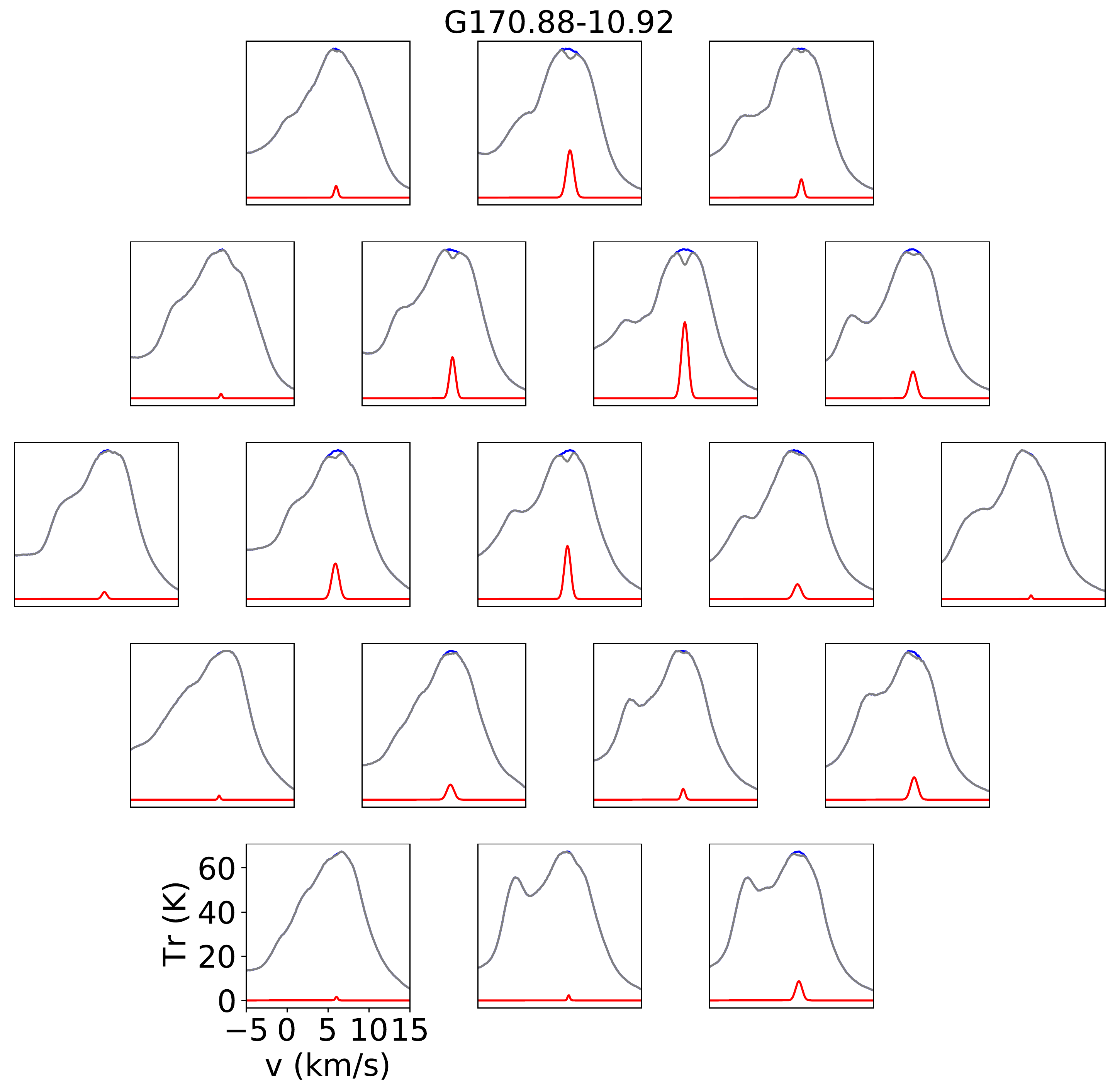}
\caption{Same as Fig. \ref{example_HI_spe} but for G170.88-10.92. 
 \label{example_HI_spe1}}
\end{figure*}

\section{Result} \label{sec:result}
\subsection{CO parameters} \label{sect_co}
Among the 12 PGCCs in our sample, all except for G174.08-13.2 have been mapped by \citet{2012ApJ...756...76W}
in $J$=1-0  of $^{12}$CO and its isotopomers (Sect. \ref{sec_arcdata}).
To match the  \ion{H}{I} 21 cm spectra,  the  $J$=1-0 spectra of $^{12}$CO and its isotopomers
of the 11 mapped PGCCs were smoothed to have an angular resolution of 3 arcmin.
Fig. \ref{emission_line_spec} shows the CO spectra extracted from the locations of the central beam of the \ion{H}{I} observations.  
Those spectra are fitted to obtain line parameters, which can help us to extract HINSA features from the \ion{H}{I} 21 cm spectra.
For G174.08-13.2, the CO parameters derived from its single point observations by \citet{2012ApJ...756...76W} are adopted.

Assuming $^{12}$CO $J=1-0$ is optically thick ($\tau>>1$), and the beam filling factor is unit, 
the excitation temperature can be expressed as \citep{2007pcim.book.....K} 
\begin{equation} \label{cotexeq}
T_{\rm ex} = \frac{h\nu}{k}/\ln\left\{1 + \left[  \frac{k}{h\nu} \left(T_{\rm CO}^{\rm peak}+J(T_{\rm bg})\right)  
\right]^{-1}    \right\}
\end{equation}
where $T_{\rm CO}^{\rm peak}$ is the peak brightness temperature of $^{12}$CO $J$=1-0,
the background temperature ($T_{\rm bg}$) can be taken as 2.73 K,
\begin{equation} J(T;\nu)   = \frac{h\nu}{k}\frac{1}{ exp(h\nu/kT)  -1}\end{equation}

Among the 11 sources observed by \citet{2012ApJ...756...76W}, 9 have single peaked spectra of $^{13}$CO $J$=1-0,
and we fit them with single Gaussian profile.
The other three sources, G170.88-10.92, G172.8-14A1 and G174.4-15A1, have double peaked spectra of $^{13}$CO $J$=1-0,
and fittings of double Gaussian profiles are applied. Gaussian fitting is also applied to the spectra of 
$^{12}$CO $J$=1-0 and C$^{18}$O $J$=1-0, with the number of velocity components same as that of  $^{13}$CO $J$=1-0.

The column densities of $^{13}$CO and C$^{18}$O  are calculated following the method described in \citet{2012ApJ...756...76W}.
Adopting the typical abundance ratios
 $X$[H$_2$]/$X$[$^{13}$CO] = $89\times 10^4$  \citep{1980ApJ...237....9M,2013A&A...554A.103P} and
 $X$[H$_2$]/$X$[C$^{18}$O] = $7\times 10^6$ \citep{1982ApJ...262..590F}, 
the H$_2$ column densities $N^{\mathrm{^{13}CO}}$(H$_2$)
and $N^{\mathrm{C^{18}O}}$(H$_2$) can be calculated.

The peak brightness temperature of $^{12}$CO $J$=1-0, Gaussian parameters of  $^{13}$CO $J$=1-0 and $^{12}$CO $J$=1-0,
and the gas column densities derived form the CO spectra are listed in Table \ref{tab:co_par}.

\subsection{Extracting HINSA} \label{sec_result_extract_hinsa}
Both method 1 and method 2 are applied to extract HINSA features in our spectra.
The free parameters to be fitted are $\tau_i$, $v_i$, $\sigma_i$ with $i=0,1,...,m-1$ (Sect. \ref{sec:method}), 
and $m$ is the number of velocity components
determined from CO emission or quoted from literature. 
For all sources, $\tau_i$ is initially set as 0.1.
$T_{\rm ex}$ is fixed as 10 K and it will only introduce a small value of uncertainty 
(Sect. \ref{sec_errors}).

For PGCCs, initially,  $v_i$ is adopted as 
the  central velocity of $^{13}$CO $J$=1-0 (Table \ref{tab:co_par}),  and
$\sigma_i$ is adopted as 
\begin{equation}
\sigma_{i,init} = \sqrt{\frac{ \Delta V({^{13}\rm CO})^2}{8\ln(2)}+\frac{kT}{{m_{\rm H} }} - \frac{kT}{m_{^{13}\rm CO} } }
\end{equation}
Here, the $\Delta V$ represents the FWHM of the spectral line, and the correction for the
different molecular weight between $^{13}$CO and \ion{H}{I} has been conducted.

For comparision objects, the initial guesses of central velocities are adopted as the system velocities quoted from literature. 
The system velocities are 6.6 km s$^{-1}$ for L1489 \citep{2019A&A...627A.162W}, 
6.4 km s$^{-1}$ for L1521B \citep{2004ApJ...617..399H}, 9.0 km s$^{-1}$ 
for IRAS 05413-0104 \citep{2021A&A...648A..83Z}, 10.0 km s$^{-1}$ for HH25 MMS \citep{1997IAUS..182P.120G,2004A&A...426..503W}, 
and 0.5 km s$^{-1}$ for CB34 \citep{2003MNRAS.344.1257C}.
The initial values of line widths are set as 1 km s$^{-1}$.

We use the L-BFGS-B algorithm  implemented by the Python package Scipy\footnote{\url{https://www.scipy.org/}} 
to minimize $\mathcal{R}$ and obtain corresponding fitted parameters.

\begin{figure*}[thb]
\centering
\includegraphics[width=0.9\linewidth]{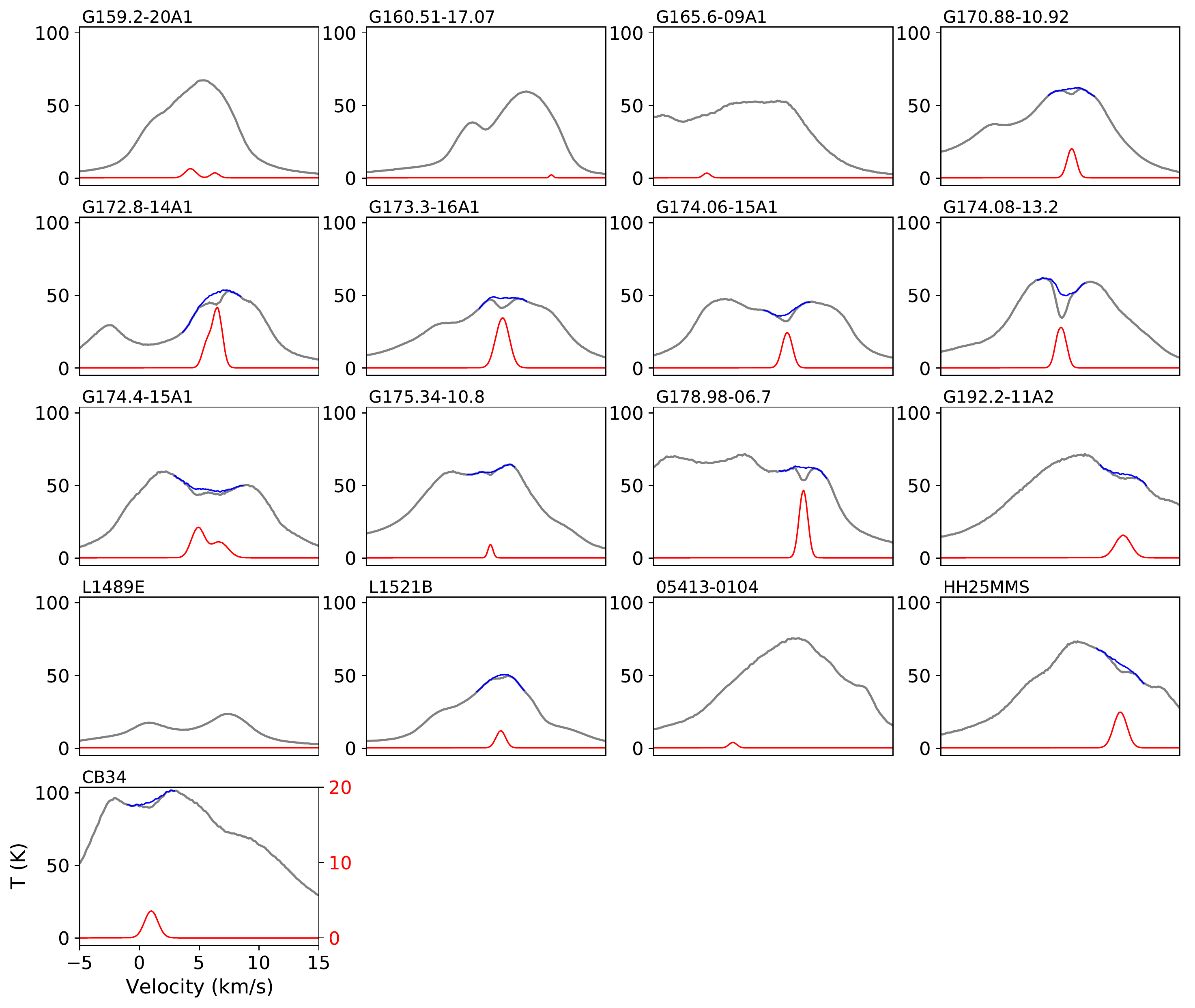}
\caption{\label{HI_beamI} The \ion{H}{I} 21 cm spectra at the center beam. 
In each panel, gray line is the observed spectrum $T_{\rm r}$, blue line is the recovered $T_{\ion{H}{I}}$ using method 2,
and red line is $T_{\rm ab}$.  
Values of $T_{\rm ab}$ on the y-axis range from 0-20 K for all sources except G174.08-13.2, which uses 0-60 K.
The orange and blue squares represent the sources that can be and can not be successfully fitted using method 1.
}
\end{figure*}

\subsubsection{Performances of method 2 towards individual objects}
The fitting results of G174.4-15A1  are shown in Figs. \ref{example_HI_spe} 
as examples to 
present the different performances of the two methods. 
For G174.4-15A1, there are two gas components with velocities traced by CO emission close to each other (Table \ref{tab:co_par}).
Method 1 can partly extract HINSA corresponding to the bluer velocity component ($V_{\rm LSR}\sim$ 5 km s$^{-1}$), 
but fails  to extract HINSA corresponding to the redder velocity component ($V_{\rm LSR}\sim$ 6.8 km s$^{-1}$). 
This result is expected by Eq. (\ref{snr_c_eq}).  For the bluer component, S/N($T_{\rm ab}$) is similar to S/N$^{c}$. 
However, for the redder component,  S/N($T_{\rm ab}$) is smaller than S/N$^{c}$ (see Table \ref{tab:hp}).
If method 2 is adopted, both of the HINSA features  can be extracted.
This example comfirms that method 2 can extract absorption features from \ion{H}{I} spectra with low S/N and blended 
velocity components.   

To test the robustness of method 2, we also compare the fitting result of the two methods towards the spectra of G170.88-10.92. 
The \ion{H}{I} spectrum of G170.88-10.92 
shows a single dip with high S/N. 
The fitted results of the two methods are very close to each other (see Fig. \ref{example_HI_spe1}).
For the center beam, the optical depths fitted by the two methods have a deviation
smaller than 10 percent. The performances of the two methods are similar to each other when the S/N is high.
The fitting result of method 2 should be reliable. 
This is confirmed by the similarity between the distributions of CO emission lines and \ion{H}{I} 
absorption features for both G174.4-15A1 and G170.88-10.92 
(see Sect. \ref{sec_hinsa_map}). 

\begin{figure}
\centering
\includegraphics[width=0.85\linewidth]{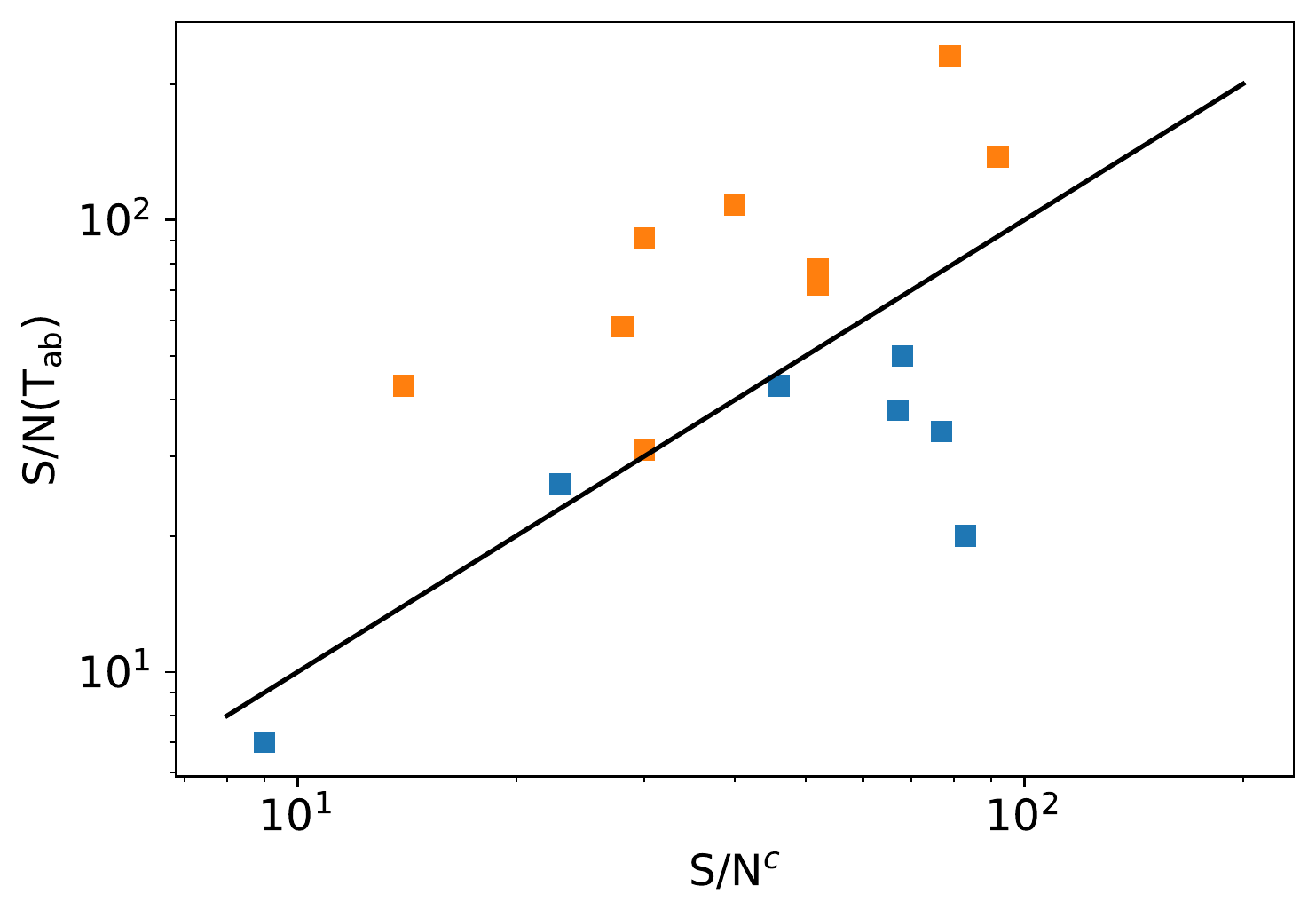}
\caption{The distribution of S/N$^c$ and S/N($T_{\rm ab}$) for the \ion{H}{I} spectra of the central beams
(Table \ref{tab:hp}). Here,  S/N($T_{\rm ab}$) is the ratio between the intensity of absorption dip
revealed by method 2 and the spectra noise, and  S/N$^c$ is the signal-to-noise threshold required
by method 1 (Sect. \ref{sec_perf_m2}).
The orange and blue squares represent the sources that can and cannot be successfully fitted using method 1, respectively.
The black line represents x=y. As predicted,
the black line is the separation line between the two cases. \label{fig_snr_comp} }
\end{figure}

\begin{table*}
\centering
\caption{HINSA parameters fitted using method 2.\label{tab:hp}}
\begin{tabular}{lcccccccc}
\hline
        source\tablefootmark{(1)}  & $n$  &      $\tau_0$\tablefootmark{(2)}  & $V_{\rm LSR}$  & $\Delta$ $V$ &     $T_{\rm ab}$ &  S/N($T_{\rm ab}$) & S/N$^{c}$ & N(HINSA)\\
               &    &                &    km s$^{-1}$& km s$^{-1}$ & K  &  &  &10$^{18}$ cm$^{-2}$ \\ 
\hline
   G159.2-20A1 &  1 &  0.01(1) &   6.48(8) &  0.8(1) &   0.55(5) &   -- &           -- &  0.10(8) \\
               &  2 &  0.03(1) &   4.24(5) &   1.1(1) &   1.30(5) &   26 &          23 &   0.5(2) \\
 G160.51-17.07 &  1 &  0.01(1) &  10.45(8) &   0.4(1) &   0.40(4) &   -- &           -- &  0.08(5) \\
   G165.6-09A1 &  1 &  0.02(1) &   -0.6(1) &  0.7(1) &    0.6(1) &    7 &           9 &   0.4(2) \\
 G170.88-10.92$^*$ &  1 &  0.09(1) &   5.95(6) &  0.89(7) &   3.88(4) &   91 &          30 &   1.4(1) \\
   G172.8-14A1$^*$ &  1 &  0.10(1) &   5.60(5) &   0.89(1) &    3.1(1) &   31 &          30 &   1.7(3) \\
               &  2 &  0.24(1) &   6.50(5) &  0.93(7) &   7.87(8) &   78 &          52 &   4.2(3) \\
   G173.3-16A1$^*$ &  1 &  0.26(1) &   6.36(5) &  1.26(6) &  7.88(5) &  138 &          92 &   4.4(1) \\
  G174.06-15A1$^*$ &  1 &  0.20(1) &   6.15(6) &  0.94(7) &   4.52(5) &   82 &          52 &   2.6(1) \\
  G174.08-13.2$^*$ &  1 &  0.60(1) &   5.09(5) &  0.92(5) &  16.1(7) &  230 &          79 &   8.0(1) \\
   G174.4-15A1 &  1 &  0.17(1) &   5.10(5) &  1.39(5) &    5.0(1) &   38 &          67 &   3.4(2) \\
               &  2 &  0.12(1) &   6.80(5) &  1.41(5) &    3.8(1) &   20 &          83 &   2.8(2) \\
  G175.34-10.8$^*$ &  1 &  0.07(1) &   5.35(6) &   0.70(1) &   2.79(6) &   43 &          14 &   1.0(1) \\
  G178.98-06.7$^*$ &  1 &  0.22(1) &   7.52(5) &  0.82(6) &   8.93(8) &  108 &          40 &   3.1(1) \\
   G192.2-11A2 &  1 &  0.07(1) &   10.2(1) &   1.4(1) &    2.63(9) &   34 &          77 &   1.9(3) \\
        L1489E &  1 &   (0.01) &     -- &   -- &    -- &   -- &          -- &       -- \\
        L1521B$^*$ &  1 &  0.07(1) &   6.21(6) &  0.95(7) &   2.27(4) &   58 &          28 &   1.4(2) \\
    05413-0104 &  1 &  (0.01) &    -- &   -- &   -- &   -- &          -- &   -- \\
       HH25 MMS &  1 &  0.12(1) &  10.05(7) &   1.3(1) &    4.73(9) &   50 &          68&   4.1(4) \\
         CB34 &  1 &  0.05(0.01) &   0.81(4) &   1.3(1) &   3.50(8) &   43 &          46 &   -- \\
\hline
\end{tabular}\\
\flushleft
{\footnotesize
$^{(1)}$ {The superscript ``*'' means HINSA feature can be extracted using method 1 from the
\ion{H}{I} 21 cm spectrum of the corresponding source. }
$^{(2)}$ { If $\tau_0$ is 0.01, the corresponding fitting result is unreliable 
(see Eq. (\ref{thetau0_eq}) in 
Sect. \ref{sec_lowerlimitforfitting}).
A number in parentheses indicates the 1-$\sigma$ uncertainty in the last digit.}
}
\end{table*}

\subsubsection{Performances of method 2 towards central beams} \label{sec_perf_m2}

The method 2 was adopted to obtain the HINSA parameters for all the \ion{H}{I} 21 cm spectra.
For each source, the observed spectrum of the central beam (beam 1) as well as the extracted HINSA are shown 
in Fig. \ref{HI_beamI}. 
Among the 12 PGCCs, 10 have valid HINSA features extracted from the central beam spectra using method 2.
The exceptions are G160.51-17.07 and G165.6-09A1. 
It corresponds to a HINSA detection rate of 83\%.

The fitted parameters of HINSA, including the optical depth,
central velocity, line width of HINSA  and the height of the absorption dip ($T_{\rm ab}$) 
are listed in  Table \ref{tab:hp}. 
S/N($T_{\rm ab}$), defined as the ratio between the peak intensity of $T_{\rm ab}$ fitted through method 2 and the noise intensity 
$T_{noise}$, is also listed in Table \ref{tab:hp}.

For central beam spectrum that can be successfully fitted using method 2,   
the S/N threshold of method 1 (S/N$^c$) is calculated 
through Eq. (\ref{snr_c_eq}).
To obtain S/N$^c$, the values of $\tau_0$ and $\Delta V$ fitted through
method 2  are used, and $N_{\rm ch}$ is adopted as $\Delta V/\Delta_{\rm ch}$.
Most of our spectra have S/N($T_{\rm ab}$)  comparable to or even larger than S/N$^c$. 
The sources whose HINSA features can be partly 
extracted using method  1 are labeled in Table \ref{tab:hp}.
From Fig. \ref{fig_snr_comp},  it is clear that spectra with successful fittings by method 1 all
have S/N($T_{\rm ab}$)$>$ S/N$^c$.
This implies that the S/N threshold of method 1 given by Eq. (\ref{snr_c_eq}) is reasonable,
and it is essential to apply method 2 to extract HINSA features from the \ion{H}{I} spectra
of this present work.

If method 1 is applied, 
HINSA features can only be extracted in seven PGCCs (Table \ref{tab:hp}), corresponding to an extracting rate of 58\%.
This value is similar to the value of \citet{2020RAA....20...77T} but lower than
the HINSA detection rate by method 2.


\begin{figure*}[!thb]
\centering
\includegraphics[width=0.4\linewidth]{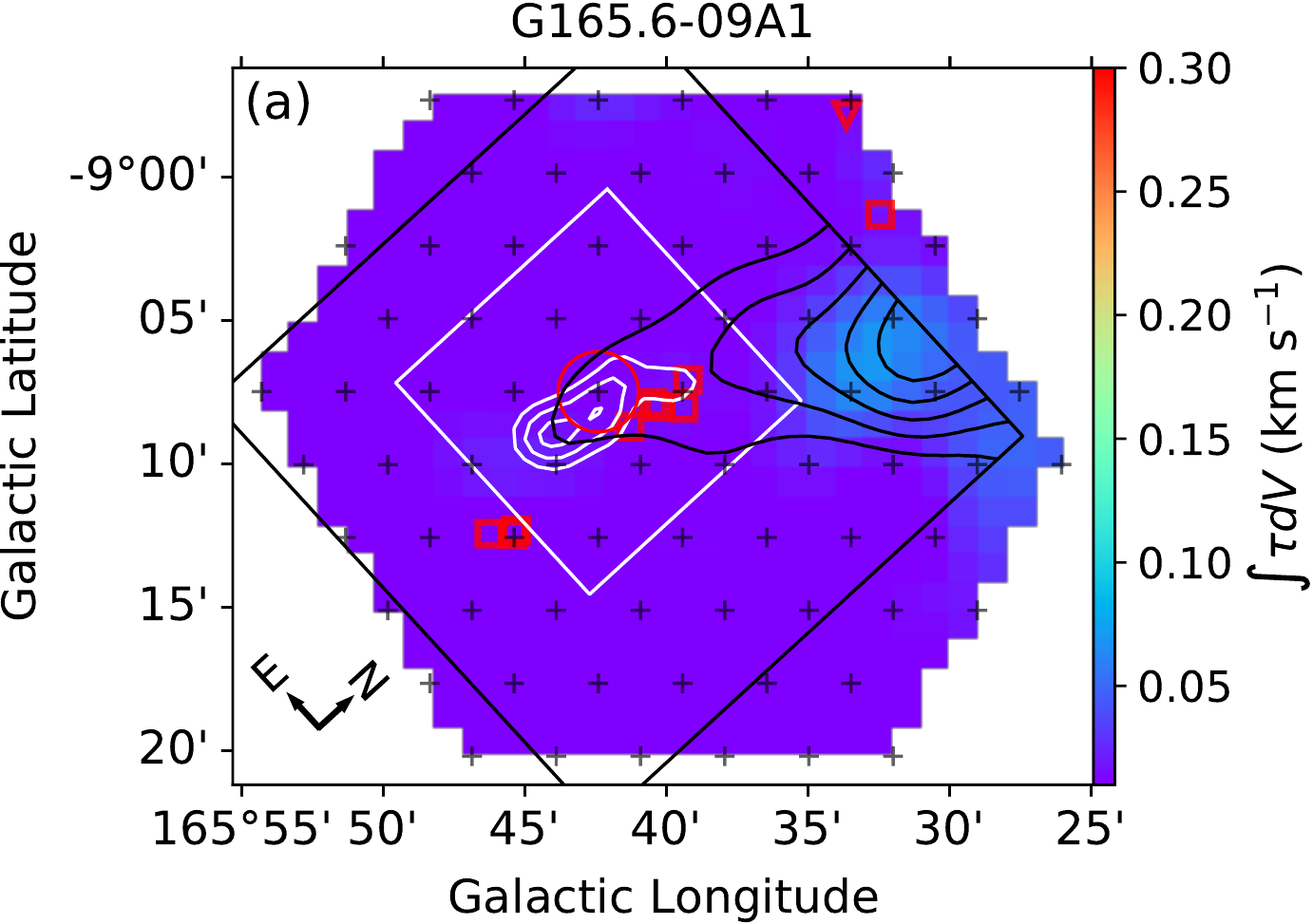}
\includegraphics[width=0.4\linewidth]{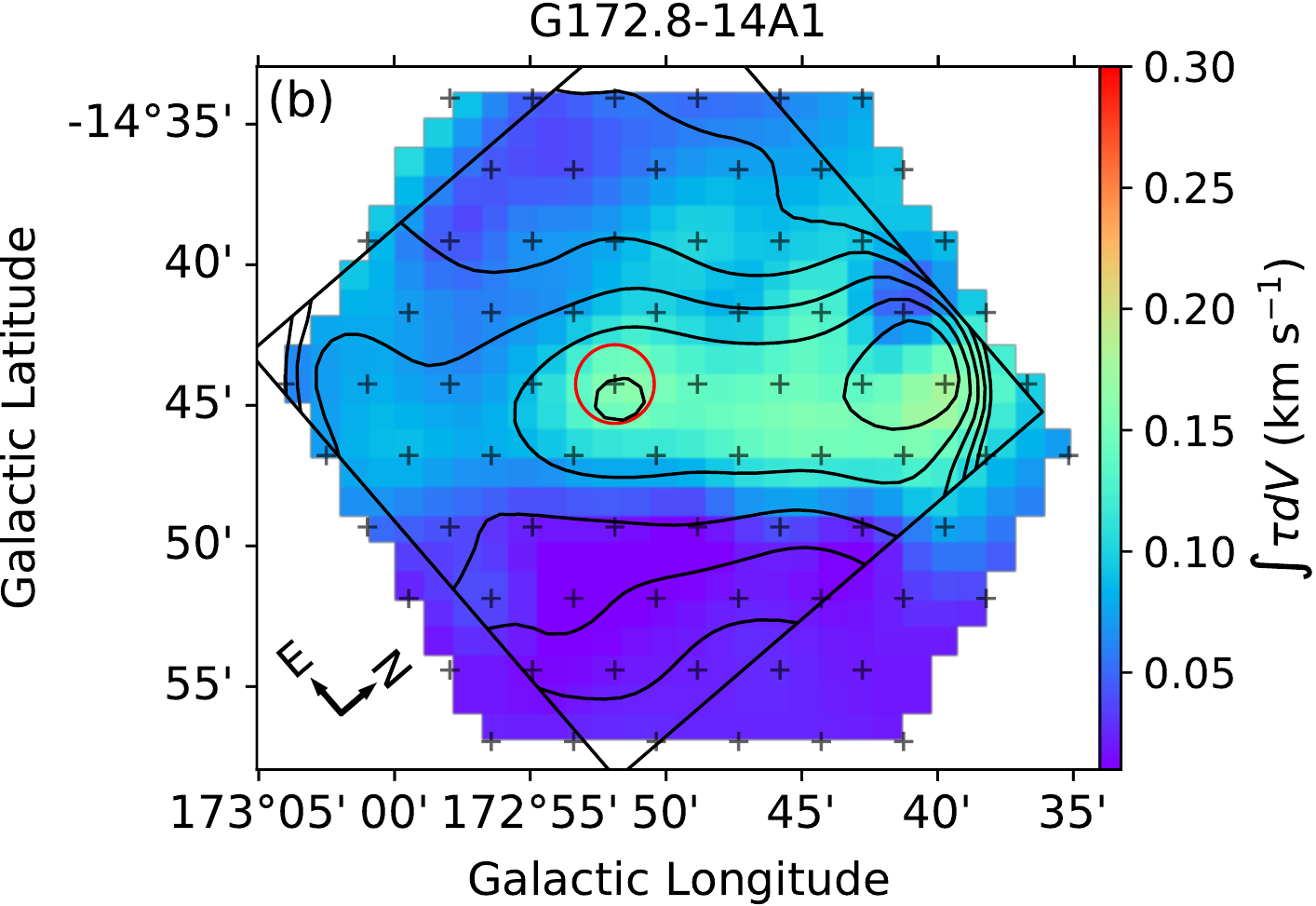}\\
\includegraphics[width=0.4\linewidth]{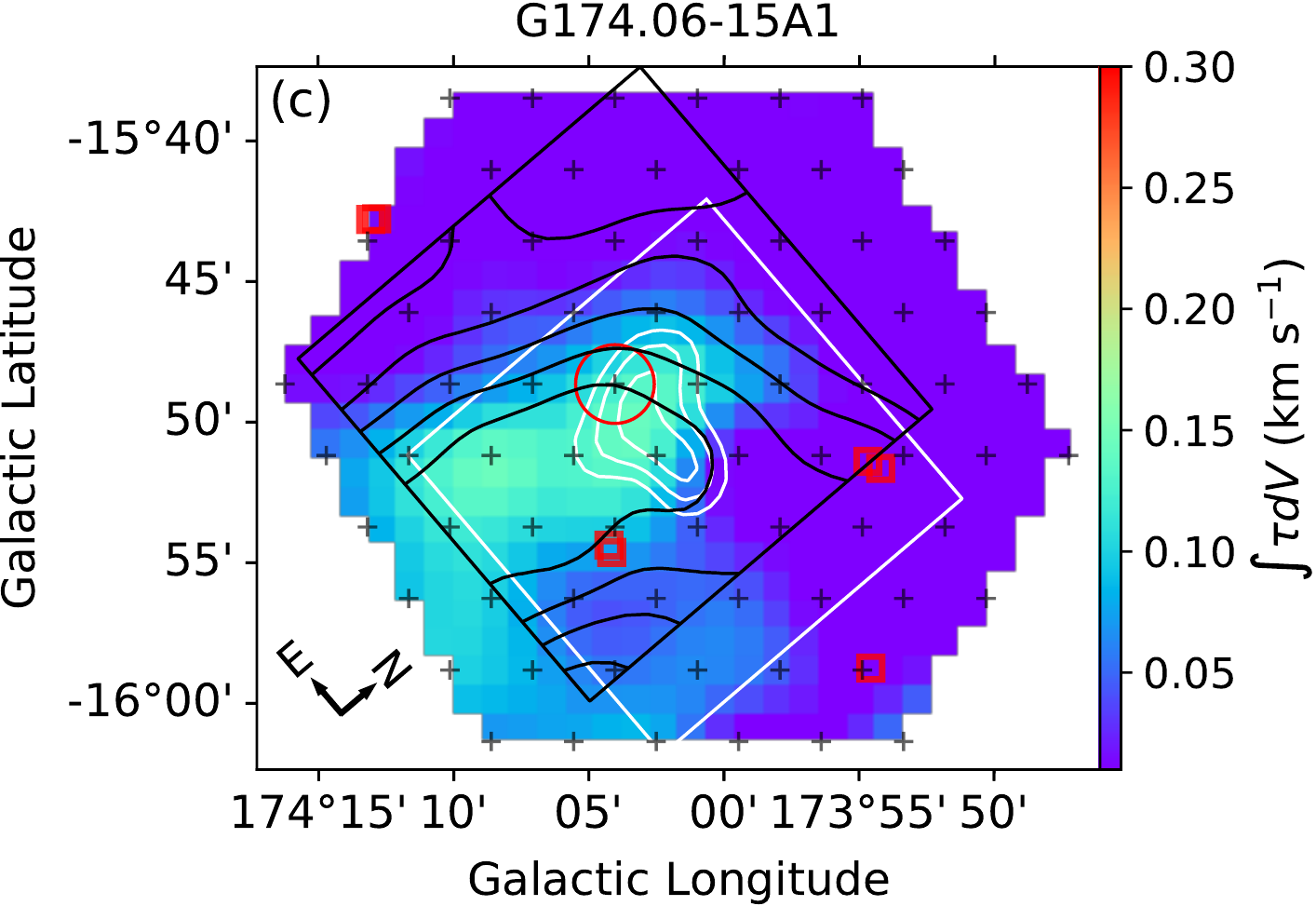}
\includegraphics[width=0.4\linewidth]{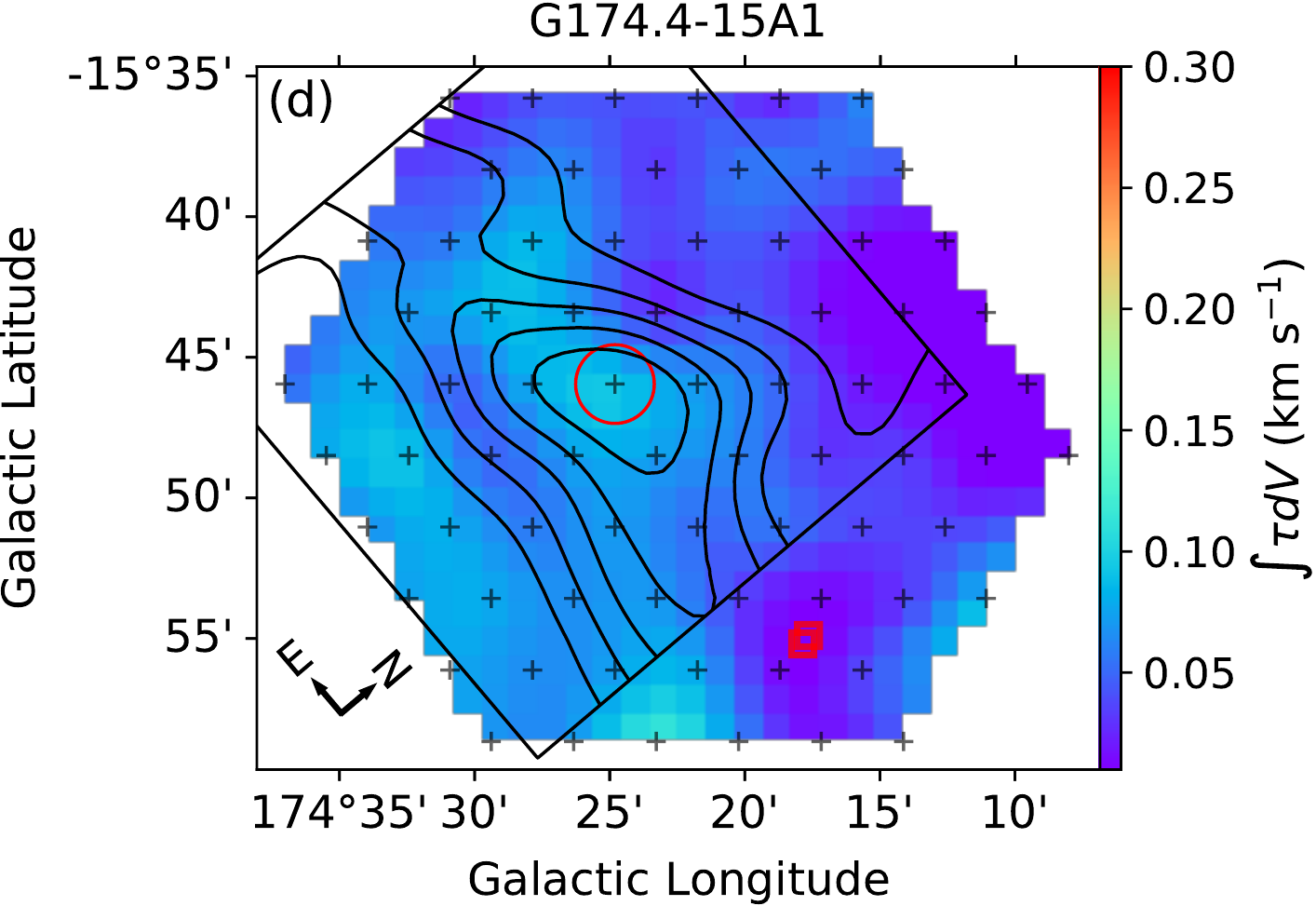}\\
\includegraphics[width=0.4\linewidth]{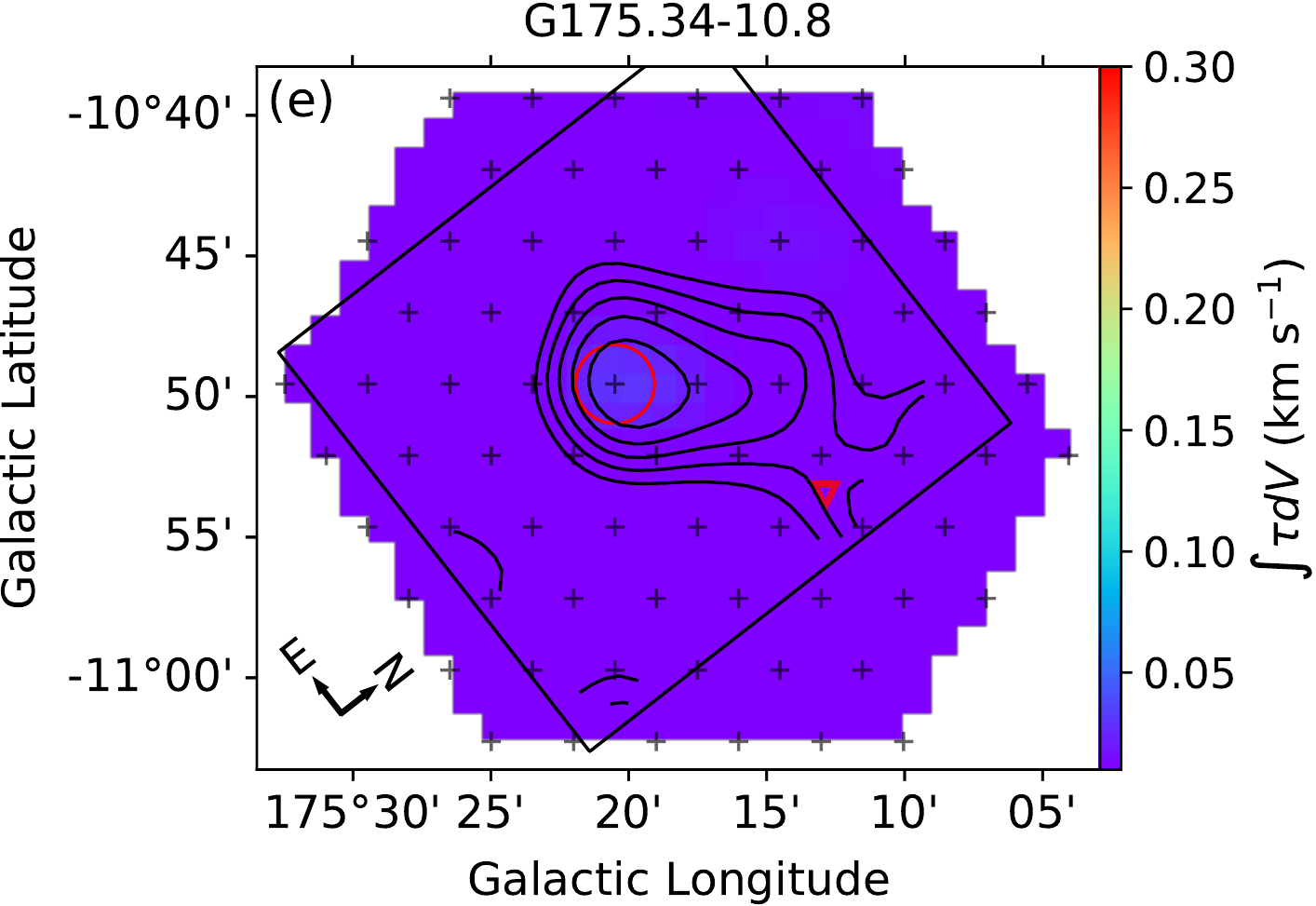}
\includegraphics[width=0.4\linewidth]{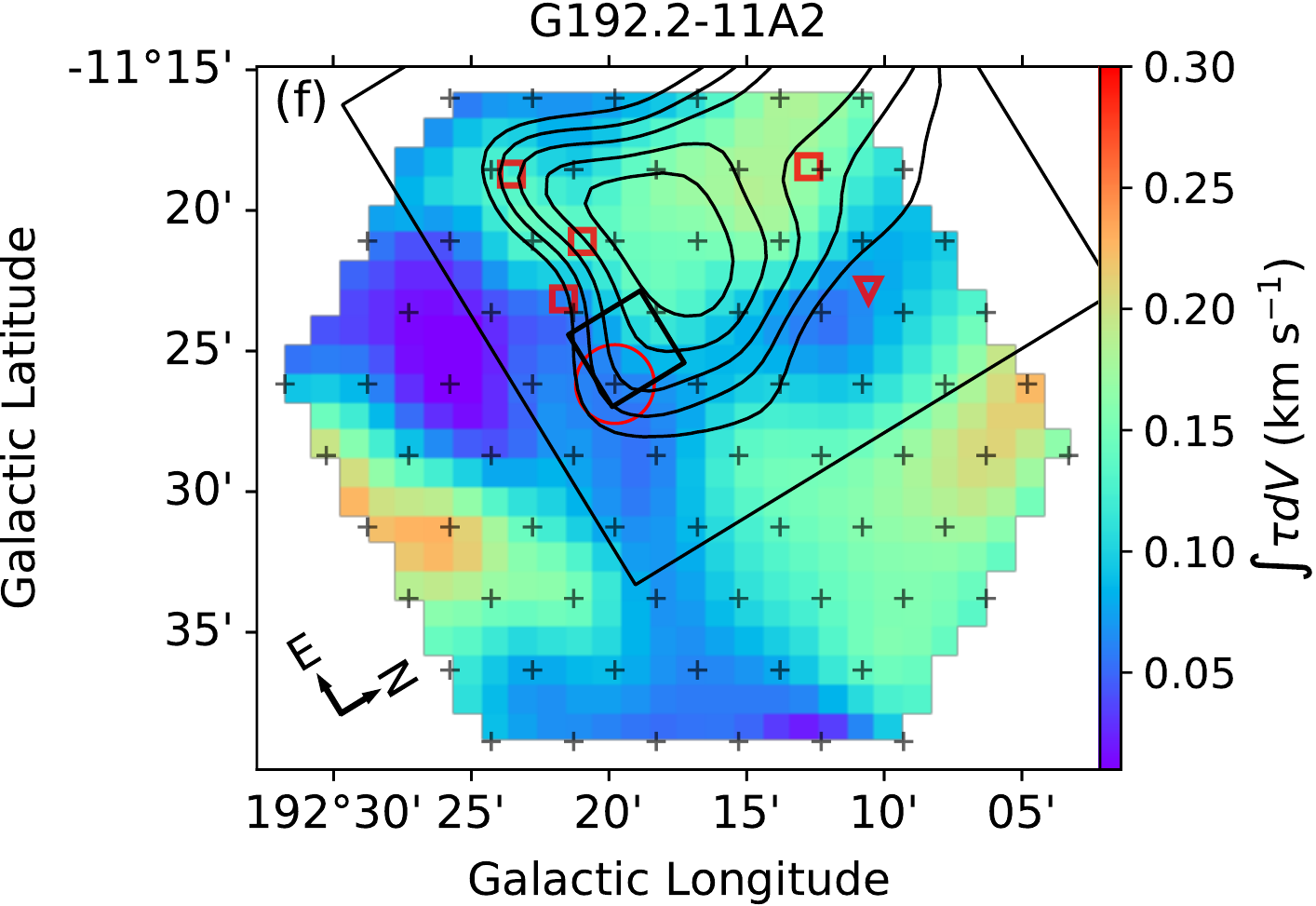}\\
\includegraphics[width=0.4\linewidth]{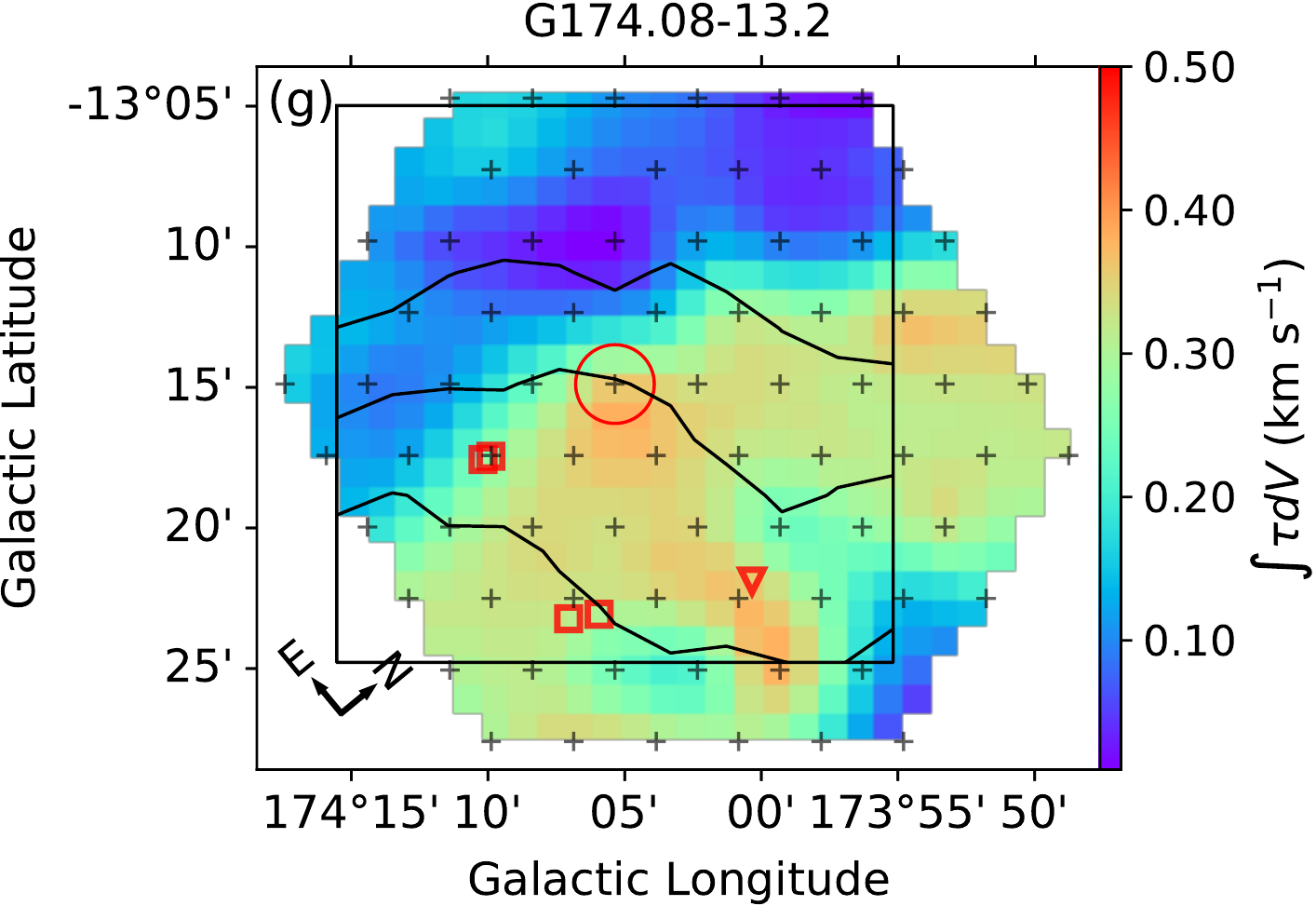}
\caption{Backgrounds show the strengths of HINSA measured in the snapshot mode, extracted using method 2. 
Black contours represent $^{13}$CO J=1-0 emissions,
which have been smoothed to the angular resolution of \ion{H}{I} observations. The big black boxs represent the areas covered by CO data.
The Blue crosses represent the sampled points of the FAST and the red circle marks the center beam.  
Small red boxes and triangles represent Class I/II and Class III YSOs. 
In panel (a) and (c), the white contours  represent the emission 
of C$_2$H $N$=1-0 within the white rectangles \citep{2019A&A...622A..32L}.
In panel (f), the small black rentangle represents the mapped region of the IRAM HCO$^+$ $J$=1-0 and C$^{18}$O $J$=2-1 observations shown
in Fig. \ref{iram_map}.
\label{HI_CO_fig} } 
\end{figure*}

\begin{figure*}
\centering
\includegraphics[width=0.9\linewidth]{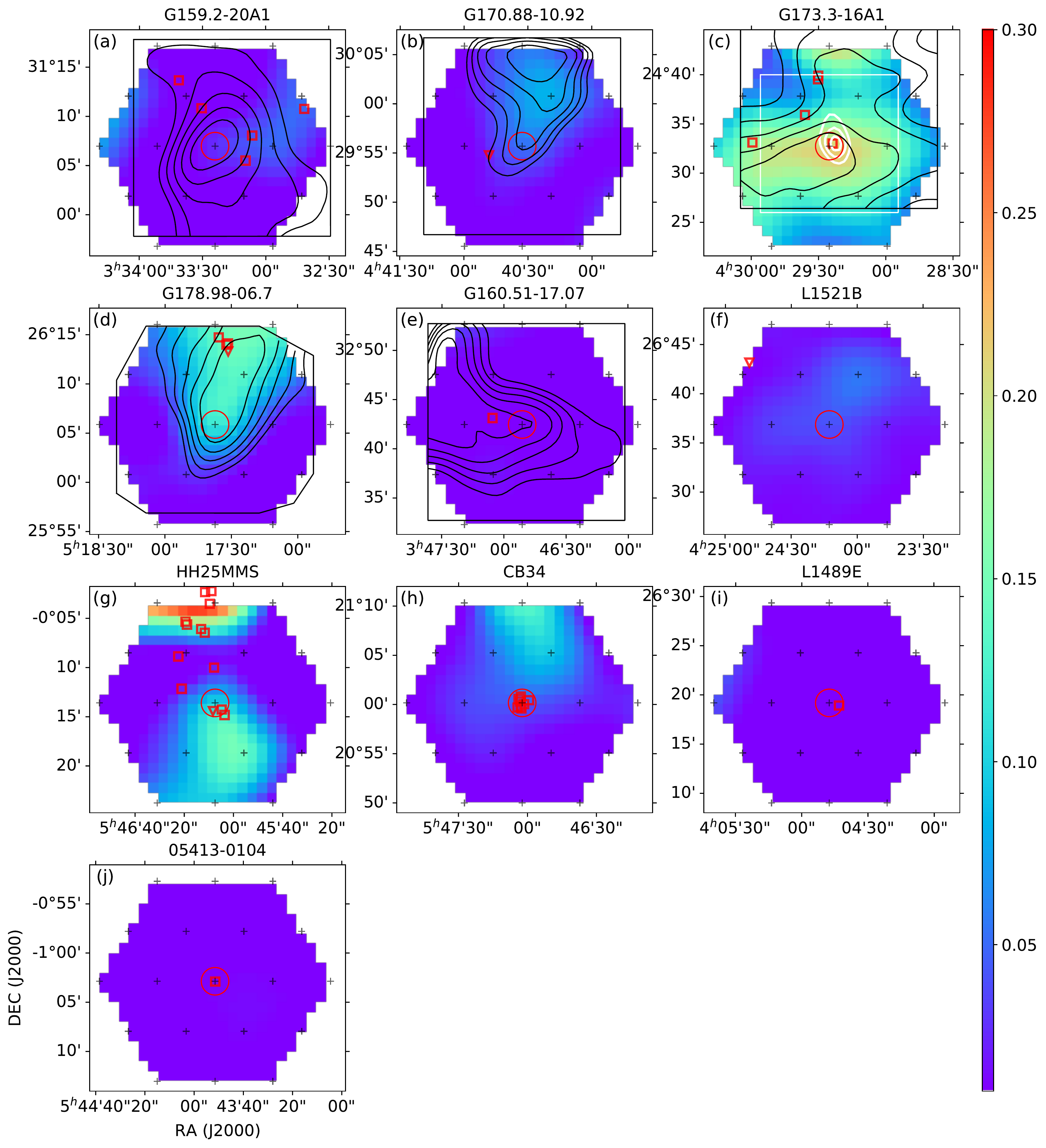}
\caption{Backgrounds show the strengths of HINSA measured in the tracking mode.
Symbols have the same meanings with those in Fig. \ref{HI_CO_fig}. \label{HI_CO_fig_1}}
\end{figure*}

\begin{figure*}
\centering
\includegraphics[width=0.8\linewidth]{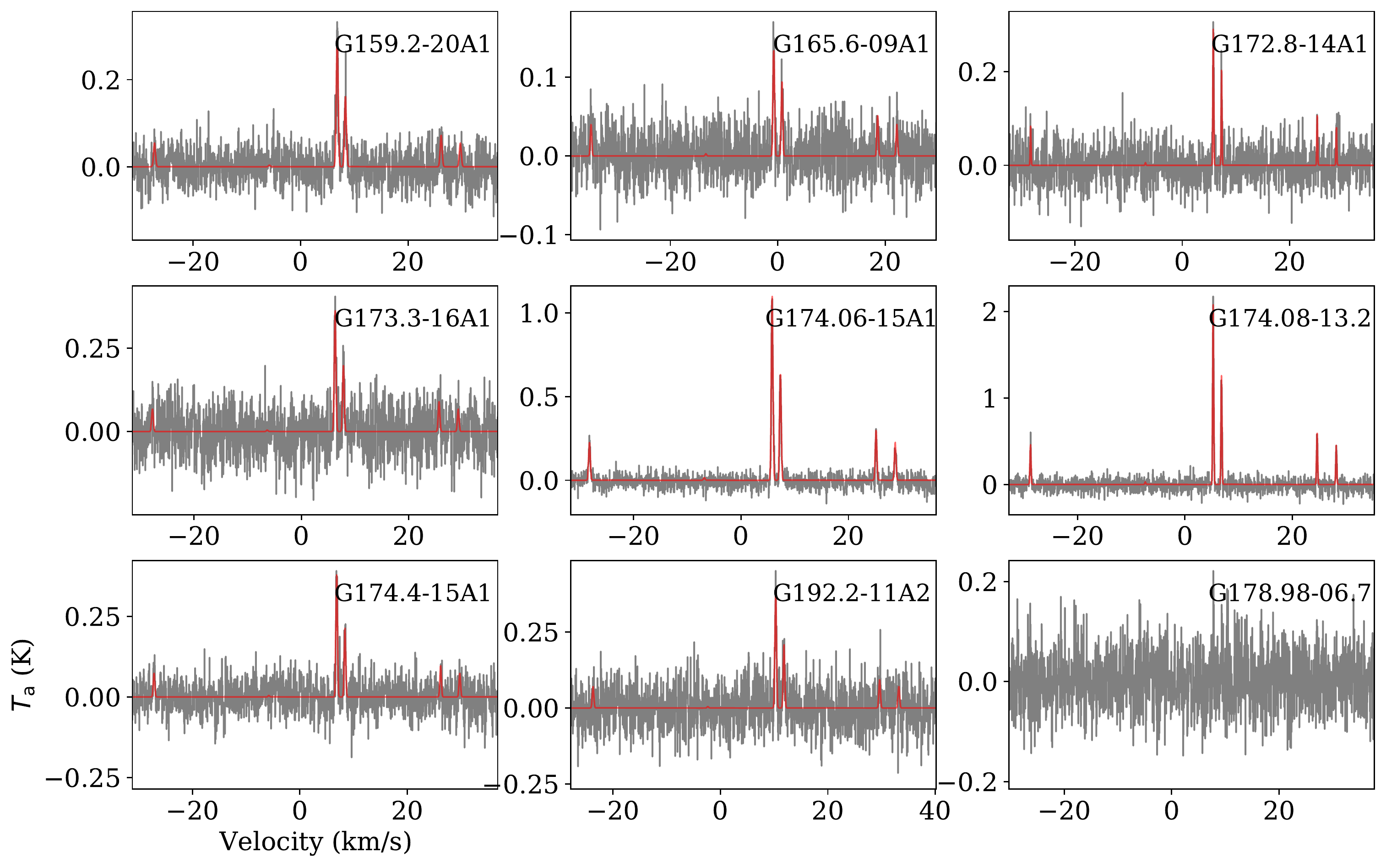}
\caption{HC$_3$N spectra binned  with a velocity resolution of 0.05 km s$^{-1}$.
The red lines shows the results of HFS fitting. \label{hc3n_spec_fig}}
\end{figure*}

\subsubsection{HINSA maps} \label{sec_hinsa_map}

The maps of the integrated intensity of the optical depth of HINSA features extracted by method 2  are overlaid by the
contours of CO emission.
The sources observed in snapshot mode and tracking mode are shown in Figs. \ref{HI_CO_fig} and  \ref{HI_CO_fig_1}, respectively.
Comparing the emission regions of HINSA and molecular lines can help us to confirm the detections of HINSA.
It will also help to study the correlation between the distribution of cold \ion{H}{I} and other species.

We find that HINSA features can be extracted in the north-west margin of the map of G165.6-09A1. 
This makes
G160.51-17.07 the only observed PGCC that does not show any HINSA feature.
The detection rate of HINSA features in PGCCs could be as high as 11/12 ($\sim$90\%) using method 2 if \ion{H}{I} 
spectral of all beams are taken into consideration.

Among 5 comparasion objects, only L1251B, HH25 MMS and CB34 have HINSA features, with a detection rate lower than the value in PGCCs.
The morphologies and environments of HINSA in PGCCs and comparasion objects are further discussed in Sect. \ref{sec_dis_me}.

\subsubsection{Uncertainties of HINSA parameters} \label{sec_errors}
The $T_{\rm ex}$ of CO is close to 10 K (Sect. \ref{sec_co_dust_emission}). The $T_{ECC}$ of all sources except 
G165.6-09A1 have values about 10$\pm$2 K. Thus, it is reasonable to adopt the excitation temperature of HINSA as 10 K
(Sect. \ref{sec_result_extract_hinsa}).
For $T_r\sim 50$ K and the uncertainty of excitation
temperature $\sigma(T_{\rm ex}) = 2$ K, the relative error of the fitted optical depth can be estimated as
\begin{equation}
\frac{\sigma(\tau)}{\tau} \sim \frac{\sigma(T_{\rm ex})}{T_r-T_{\rm ex}} \sim 0.05,
\end{equation}
which is smaller than the values transferred from the data errors (Table \ref{tab:hp}).

It is important to note that the \ion{H}{I} column density traced by HINSA is the value for
cold \ion{H}{I} mixed within a molecular cloud. If there are no background emission with $T_{\ion{H}{I}}>T_{\rm ex}$
there will be no HINSA, and this may underestimate the column density of cold \ion{H}{I} and the detection rate of HINSA. 

Although method 2 significantly reduces the requirement for S/N, the HINSA features may still be difficult to extract
if $T_{\ion{H}{I}}$ is too complex. If the value of the fitted optical depth is too low, it would be 
not easy to judge if the extracted absorption feature is a real  
HINSA signal. The lower limit of $\tau$ to be safely extracted  will be discussed in Sect. \ref{subsec_lsafe}.

\subsubsection{Lower limit of $\tau$ to be safely extracted} \label{sec_lowerlimitforfitting} \label{subsec_lsafe}
Assuming both $T_{\ion{H}{I}}$ and  $T_{\rm ab}$ have Gaussian shapes with their central velocities aligned, and
their line widths are 10 km s$^{-1}$ and 1 km s$^{-1}$, respectively, 
$T_{\rm r}$ will have a dip-like structure only when 
\begin{equation}
\left|T_{\ion{H}{I}}^{''}\right|_{V_{\rm LSR}} < \left|T_{\rm ab}^{''}\right|_{V_{\rm LSR}}
\end{equation}
which leads to
\begin{equation} \label{thetau0_eq}
\tau_0 = \left.\frac{T_{\rm ab}}{T_{\ion{H}{I}}}\right|_{V_{\rm LSR}} > \frac{\Delta V^2(T_{\rm ab})}{\Delta V^2(T_{\ion{H}{I}})} = 0.01
\end{equation}
Only HINSA with $\tau$ larger than 0.01 can be safely extracted.
If the fitted optical depth $\tau_{0,f}$ has value smaller than 0.01, the fitting results should be considered as unreliable. 
The fitted absorption feature may be a
misleading signal introduced by the irregularity of the background \ion{H}{I} emission.
If the fitted optical depth is smaller than or equal to 0.01, 
it is listed in Table \ref{tab:hp} as 0.01, which means that the fitting result is not reliable.

\subsubsection{Column dnesity of \ion{H}{I} traced by HINSA}
The column density of \ion{H}{I} can be calculated through \citep{2003ApJ...585..823L}
\begin{equation}
N(\ion{H}{I})=1.9\times 10^{18}\tau_0 \frac{T_{\rm ex}}{\rm K} \frac{\Delta V}{\rm km\ s^{-1}}\ {\rm cm}^{-2}
\end{equation}
The total gas column densities ($N$(H$_2$)) are adopted as the values derived from CO spectra (Sect. \ref{sect_co})
to calculate the abundances of cold \ion{H}{I}.

The values of column densities and abundances of \ion{H}{I} are also listed in Table \ref{tab:hp}.
We give discussions about the \ion{H}{I} abundances of PGCCs in Sect. \ref{dis_abhi}.

\subsection{HC$_3$N}

\begin{table}
\small
\caption{HC$_3$N $J$=2-1 parameters. \label{hc3n_par_table}}
\begin{tabular}{lccccc}
\hline
source & $T_{\rm peak}$ & $V_{\rm LSR}$ & $\Delta V$ & $\tau$\tablefootmark{(1)} & N\\
       & K     & km s$^{-1}$ &  km s$^{-1}$ & &  10$^{12}$ cm$^{-2}$ \\
\hline
G159.2-20A1 & 0.29 & 6.82(1) & 0.42(3) & 0.1(--) & 5.94(3) \\
G165.6-09A1 & 0.13 & -0.74(1) & 0.30(3) & 1.5(6) & 3.6(2) \\
G172.8-14A1 & 0.29 & 5.780(6) & 0.17(1) & 1.3(5) & 4.1(2) \\
G173.3-16A1 & 0.36 & 6.29(1) & 0.35(3) & 0.1(--) & 6.20(8) \\
G174.06-15A1 & 1.10 & 5.810(2) & 0.344(6) & 0.4(1) & 20.77(3) \\
G174.08-13.2 & 2.09 & 5.270(1) & 0.215(4) & 0.5(1) & 26.86(4) \\
G174.4-15A1 & 0.38 & 6.870(8) & 0.326(2) & 0.2(3) & 6.3(1) \\
G192.2-11A2 & 0.38 & 10.30(1) & 0.33(3) & 0.1(--) & 6.12(7) \\
\hline
\end{tabular}\\
\tablefoottext{1}{The optical depth of the main line of HC$_3$N $J$=2-1. The value of 0.1 means that
the line is optically thin with $\tau \le 0.1$.
A number in parentheses indicates the 1-$\sigma$ uncertainty in the last digit.} 
\end{table}

Nine PGCCs have been observed in  HC$_3$N $J$=2-1 using the TMRT 65-m telescope (Sect. \ref{dec_tmrt}).
To compare the intensities and line widths of HINSA and HC$_3$N in PGCCs, the spectra of HC$_3$N $J$=2-1 are 
shown in Fig. \ref{hc3n_spec_fig}.  
The 1-$\sigma$ noise levels are about 45 mK when the spectra are binned to have a channel width of $\sim$0.05 km s$^{-1}$. 
All of those PGCCs except for G178.98-06.7 have valid detection of HC$_3$N $J$=2-1. 
Although the CO spectra of several PGCCs such as G159.2-20A1, G172.8-14A1 and G174.4-15A1 have double velocity components
with seperations $\sim$1 km s$^{-1}$, 
all of the detected HC$_3$N $J$=2-1 spectra have single velocity component.

The HC$_3$N $J$=2-1 spectra are fitted using the Hyperfine Structure (HFS) fiting procedure provide by 
Gildas/CLASS\footnote{\url{https://www.iram.fr/IRAMFR/GILDAS/}}.
If the fitted optical depth of the spectrum tends to be smaller than 0.1, 
the fitting procedure will give a fitting result with optical depth adopted as 0.1. 
The fitted parameters include the peak intensity ($T_{\rm peak}$), centeral velocity ($V_{\rm LSR}$), 
line width ($\Delta V$) and the optical depth ($\tau$)
of the main line of HC$_3$N $J$=2-1. These parameters are listed in Table \ref{hc3n_par_table}. 

G174.06-15A1 and G174.08-13.2 have peak brightness temperature  of HC$_3$N $J$=2-1 larger than 1 K.
Their optical depths are 0.4$\pm$0.1  and 0.5$\pm$0.1 respectively. 
Assuming that the beam filling factor $f$ is unit and the background temperature $T_{\rm bg}=2.73$ K,
the  excitaion temperatures of HC$_3$N $J$=2-1  would be derived as 6.4 K and  7.7 K for G174.06-15A1 and G174.08-13.2, respectively.
For G165.6-09A1 and G172.8-14A1, the fitted optical depths are larger than 1. The reason may be  that the 
S/Ns are not high enough for these two sources.
The   HC$_3$N $J$=2-1  of G159.2-20A1, G173.3-16A1 and G174.08-13.2 are optically thin with optical depths smaller than 0.1.

Under local thermodynamic equilibrium (LTE) assumption, 
the equation to calculate the molecular column density  is \citep {1991ApJ...374..540G,2015PASP..127..266M}
	 	\begin{eqnarray}
	 		N= \frac{3k}{8\pi^3v}\frac{Q}{S_{ij}\mu^2}
              \frac{J(T_{\rm ex})exp(\frac{E_{\rm up}}{kT_{\rm ex}})}{J(T_{\rm ex})-J(T_{\rm bg})} \frac{\tau}{1-e^{-\tau}}
	 	         \int T_{\rm r} d\upsilon
        \end{eqnarray}
where $B$, $\mu$, and $Q$ are the rotational constant,
the permanent dipole moment and the partition function, respectively.
These transition parameters can be  obtained from  ``Splatalogue'' \footnote{\url{https://splatalogue.online/}}.
The beaming factors may be much smaller than 1 for most sources except for G174.06-15A1 and G174.08-13.2.
The excitation temperatures are adopted as 5 K to calculate the beam-averaged column densities of HC$_3$N. 
The column densities of HC$_3$N are also listed in Table \ref{hc3n_par_table}. 
For G165.6-09A1 and G172.8-14A1, if the excitation temperature given by HFS fittings is adopted, the 
value of the calculated column densities will only change by less than 5 percent. 
The average abundance of HC$_3$N is 2.2$\times$10$^{-9}$ with a standard error
of 0.8$\times$10$^{-9}$. 


\section{Discussions} \label{sec:diss}

\subsection{CO and dust emission}  \label{sec_co_dust_emission}

\begin{figure*}
\centering
\includegraphics[width=0.4\linewidth]{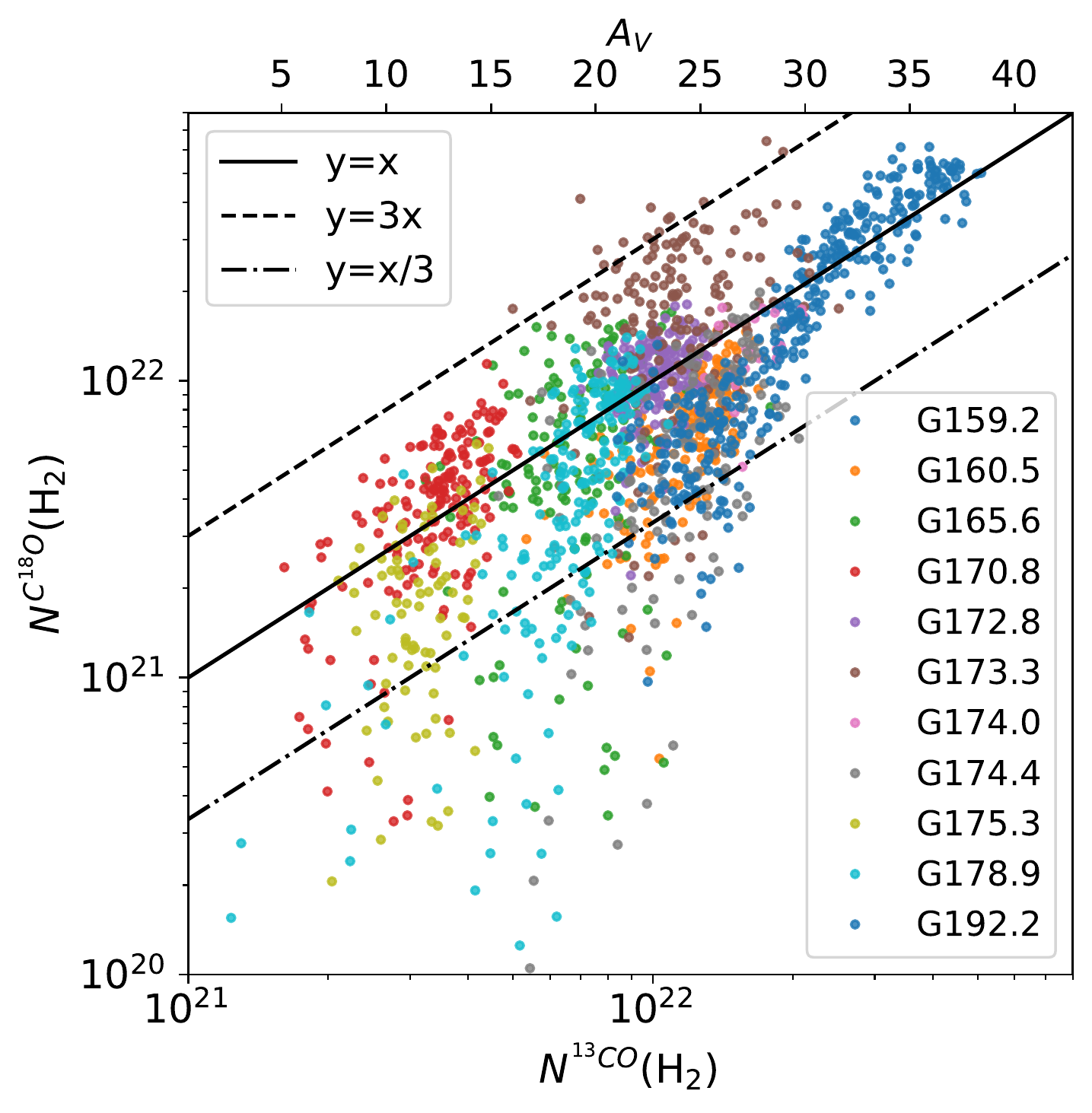}
\includegraphics[width=0.4\linewidth]{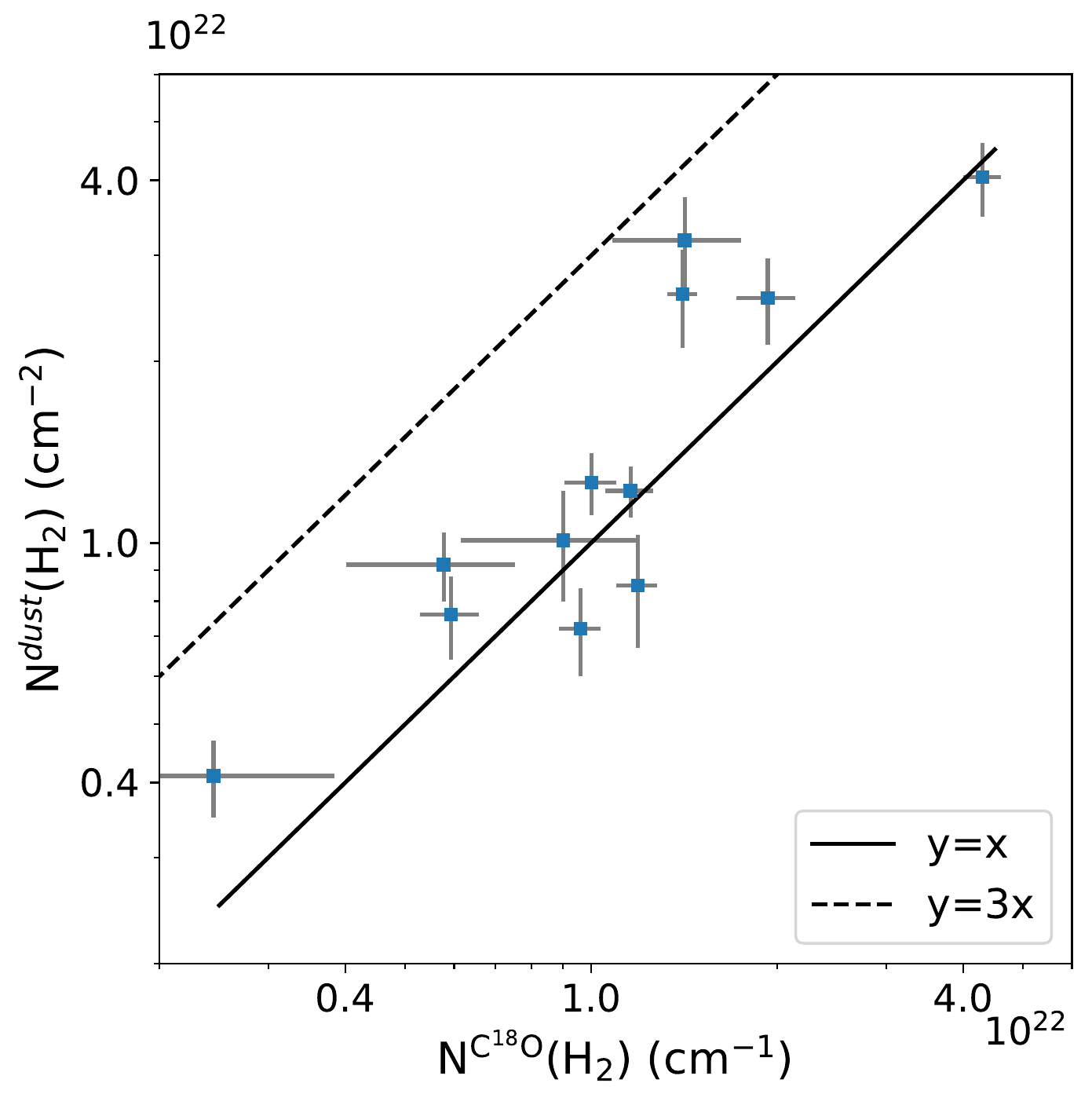}
\caption{
Upper: Correlation between the column densities of H$_2$ derived from  $^{13}$CO ($N^{\rm ^{13}CO}({\rm H_2})$)
and C$^{18}$O ($N^{\rm C^{18}O}({\rm H_2})$) at the positions
of the FAST beams for 11 PGCCs (G174.08-13.2 is not included because of the lacking of CO mapping observations).
Data of different sources are shows in different colors. 
Lower: Correlation between the H$_2$ column densities at the map centers derived from dust emissions ($N^{\rm dust}({\rm H_2})$)
and C$^{18}$O  ($N^{\rm C^{18}O}({\rm H_2})$).
\label{CO_dust_N_fig} }
\end{figure*}

The upper panel of Fig. \ref{CO_dust_N_fig} shows the correlation between the H$_2$ gas column densities
derived from $^{13}$CO  ($N^{\rm ^{13}CO}$(H$_2$)) and C$^{18}$O 
($N^{C^{18}O}$(H$_2$)).
$N^{\rm ^{13}CO}$(H$_2$) and $N^{\rm C^{18}O}$(H$_2$) are consistent with each other within a factor of 3.
For sources with low column densities ($N^{^{13}CO}$(H$_2$)$<10^{22}$), the gas column densities derived from C$^{18}$O
tend to be smaller than those derived from $^{13}$CO.
This may result from the low S/Ns for C$^{18}$O $J$=1-0 when the gas column densities are small.
Another explain is that C$^{18}$O is easier to be photo-dissociated by the external radiation field compared with $^{13}$CO,
especially for sources with low column density and extinction like PGCCs \citep{2014A&A...564A..68S,2014A&A...566A.120S,2011MNRAS.412.1686S}.

\citet{2016A&A...594A...1P,2016A&A...596A.109P} decomposed the all-sky maps of foreground dust column density and temperature.   
For sources in this present work, we extracted the values of dust temperatures ($T_{dust}$) from the temperature map  and 
the gas column densities ($N^{dust}$(H$_2$))
from the dust column density map, according to their coordinates. 
Here, a gas-to-dust mass ratio of 100 is assumed to calculate the gas column density ($N^{dust}$(H$_2$))  from the
dust column density.
$T_{dust}$ and $N^{dust}$(H$_2$) are listed in the 7th and 8th column of 
Table \ref{tab:sources}, respectively.
The lower panel of Fig. \ref{CO_dust_N_fig} shows the correlation between $N^{dust}$(H$_2$) and
$N^{C^{18}O}$(H$_2$). The gas column densities derived from dust emissions and CO spectra are consistent with each other, 
and their ratios vary within a factor of 3.

From $^{13}$CO intensity  maps 
which have been smoothed to have angular resolutions of 3 arcmin,
we measured the angular diameters of the $^{13}$CO emission regions ($D_{\rm CO}$).
$D_{\rm CO}$ is comparable or even larger than $D_{dust}$, the diameter of the Planck dust emission region (see Sect. \ref{sec:obs}). 
This means that the CO emissions, if smoothed to have similar angular resolutions  to those of Planck continuum, 
will look as extended as the Planck dust emission regions. 
CO emission is coupled well with the dust.  
The CO cores/subclumps extracted from CO intensity maps  may be 
parts of the continuous CO emission structures with column densities locally enhanced.
This is in contrast to the case of C$_2$H. In G165.6-09A1, G174.06-15A1 and  G173.3-16A1,
the emission regions of C$_2$H (also have been smoothed to have angular resolutions of 3 arcmin) 
are much more compact than CO emission regions (Figs. \ref{HI_CO_fig} and \ref{HI_CO_fig_1}). 

\begin{figure}[t]
\centering
\includegraphics[width=0.8\linewidth]{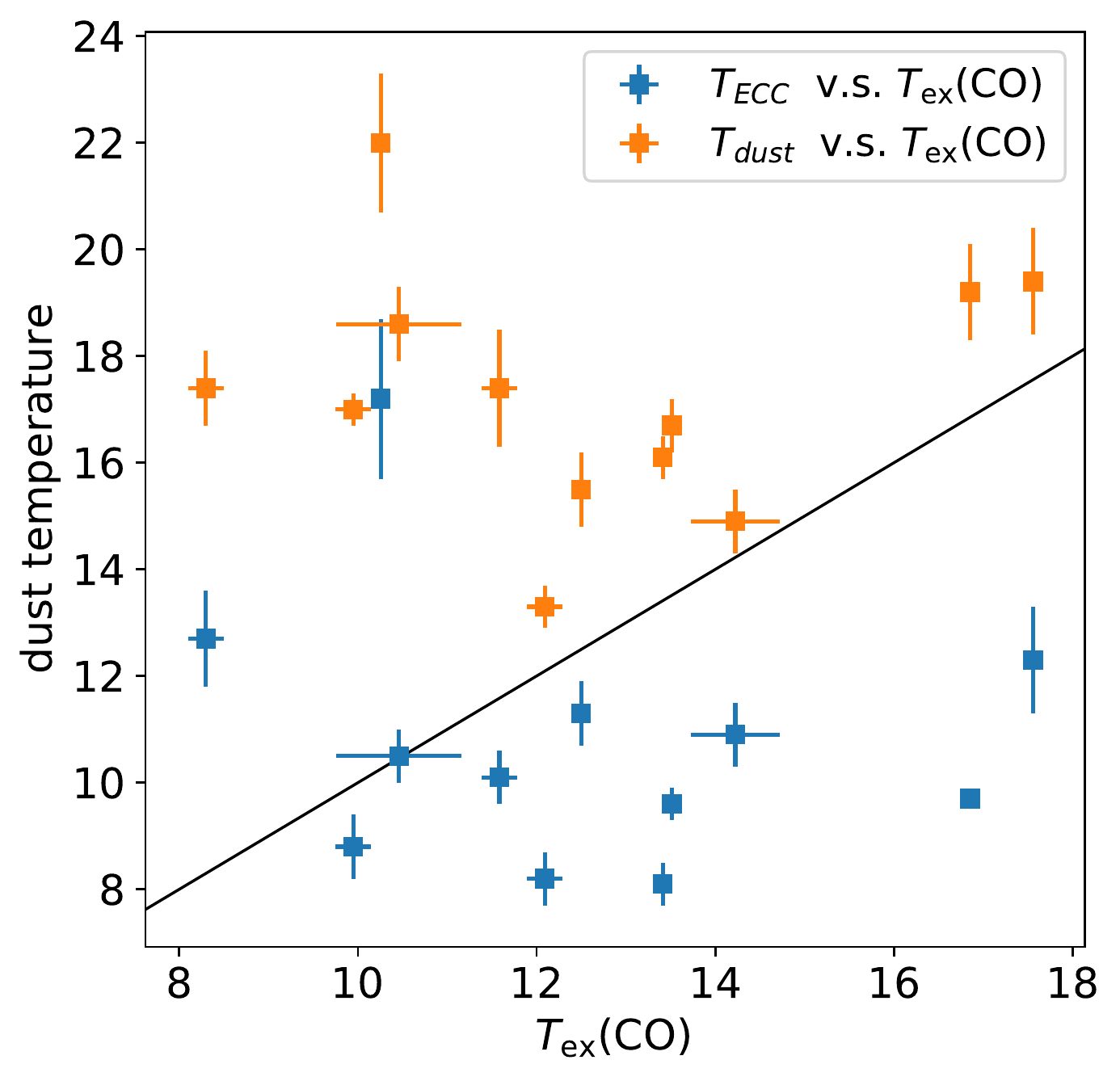}
\caption{Comparision bewteen CO $J$=1-0 excitation temperatures and Planck dust temperatures.   
The black line represnets $y=x$.
\label{temp_comp_fig}
}
\end{figure}

Fig. \ref{temp_comp_fig} shows the correlation between CO $J$=1-0 excitation temperatures ($T_{\rm ex}$(CO)) and
dust temperatures $T_{ECC}$ and $T_{dust}$ (see Sect. \ref{sec:obs} and Table \ref{tab:sources}). 
$T_{\rm ex}$(CO)  is estimated from the peak brightness temperatures of $^{12}$CO $J$=1-0 ($T_{12}$) through
Eq. (\ref{cotexeq}),
assuming the $^{12}$CO $J$=1-0 lines are optically thick and the beam filling factors are equal to unit.
$T_{\rm ex}$(CO) tends to be larger than the dust temperatures of 
the PGCCs derived from the SED fittings of the  
continuum fluxes ($T_{ECC}$), but smaller than the values of $T_{dust}$ extracted from the decomposed dust maps
\citep{2016A&A...594A...1P,2016A&A...596A.109P}. 
The mean values of $T_{\rm ex}$(CO), $T_{ECC}$ and $T_{dust}$ (in unit of K) are 12.6(0.8), 10.8(0.7) and 17.3(0.7), respectively.
The numbers in brackets are the values of corresponding standard errors.
$T_{ECC}$ should be the temperatures of the coldest  dust components, since  the large scale emissions have been filtered out before 
the applying of SED fitting. On the contrary,  contribution of the extended
warmer dust components makes the  value of $T_{dust}$ to be higher. Part of CO emission may come from more diffuse region besides the
regions associated with the coldest dust components.
It can also explain the extended morphology of CO emission.
This reflects the nature of PGCC, the coldest and quiescent molecular cloud with a mixture of 
relatively diffuse and dense components. This is also consistent to the high detection rate of HINSA especially in HC$_3$N harboring PGCCs
(Sect. \ref{sec_dv}). 

\subsection{Morphologies and  environments of HINSA } \label{sec_dis_me}

\begin{figure}[t]
\centering
\includegraphics[width=0.8\linewidth]{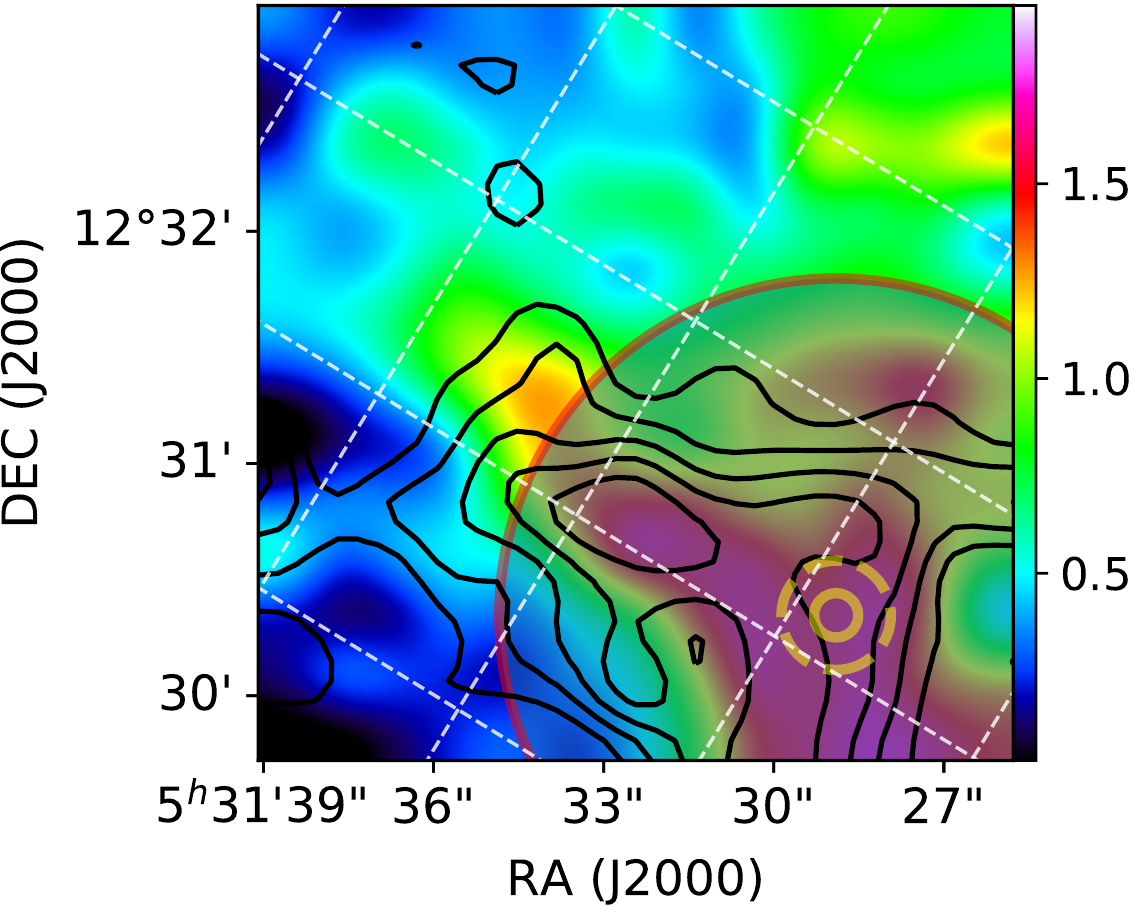}
\caption{The cutout region of G192.2-11A2, represented by the black box in Fig. \ref{HI_CO_fig}(f). 
The background and black contours present the emissions of C$^{18}$O J=2-1
and HCO$^+$ J=1-0 measured with the IRAM 30 m telescope (Xu et al. in preparation), respectively. 
The contours have levels from 0.5 to 0.9 stepped by 0.1 in unit of K km s$^{-1}$. 
The red circle shows the location  of the central beam of the \ion{H}{I} 
observation. The yellow dashed circle and solid circle represent the beam size of  HCO$^+$ J=1-0  and C$^{18}$O J=2-1, respectively.
The white dashed  lines are the grid lines of the Galactic coordinate. \label{iram_map} }
\end{figure}

In the scales of source sizes,
the morphologies of HINSA emission are similar to those of $^{13}$CO for most PGCCs detected with HINSA features 
(Figs. \ref{HI_CO_fig} and \ref{HI_CO_fig_1}).
For G165.6-09A1 and G174.4-15A1, the $^{13}$CO $J$=1-0 emission and the HINSA features are weak at the map centers.   
This is not caused by pointing bias. 
In panel (a) and panel (c) of Fig. \ref{HI_CO_fig}, the blue contours  represent the emission 
of C$_2$H $N$=1-0 \citep{2019A&A...622A..32L}.
At the map center of G165.6-09A1, emissions of C$_2$H $N$=1-0 were detected.
Fig. \ref{iram_map} shows the HCO$^+$ $J$=1-0 and C$^{18}$O $J$=2-1 emission towards the map center of G174.4-15A1. 
The HCO$^+$ $J$=1-0 and C$^{18}$O $J$=2-1 emission tends to be enhanced 
in the south-west corner (Fig. \ref{iram_map}), in constract of the CO $J$=1-0 emission extending to north (panel (f) 
of Fig. \ref{HI_CO_fig}). These results show that HINSA features are  more tightly correlated with the emission of CO  than
that of dense gas tracers.

However, there are also several exceptions to the similarity between the emission regions of CO and HINSA.
For example, in G159.2-20A1 (see the first panel of Fig. \ref{HI_CO_fig_1}),  this correlation is weak. 
For a molecular cloud, its large-scale enviroment and the star formation activities within it may
have influence on the existence and distribution of HINSA.

HINSA features are not detected ($\tau_0\lesssim 0.1$) at the central beam towards G165.6-09A1 and G160.51-17.07, and
marginally detected ($\tau_0\sim 0.3$) towards G159.2-20A1. 
There are no central-beam HINSA detected in  G165.6-09A1 and G159.2-20A1, while G160.51-17.07
has no detected HINSA anywhere.
The three sources (G165.6-09A1, G160.51-17.07, G159.2-20A1) all have $T_{\rm dust}$ larger than 19 K,
the largest compared with other PGCCs (Table \ref{tab:sources}).
All PGCCs are located at the margin of large-scale dust emission 
regions traced by Planck 353 GHz continuum (Fig. \ref{galac-crop}).
G160.51-17.07, the only PGCC showing no HINSA feature, is located within an H$_{\alpha}$ emission region
represented by red contours in Fig. \ref{galac-crop}.
These results all demonstrate the importance of the environments to the detection of HINSA
in PGCCs and the local deviation between HINSA and CO emission regions. 

Star formation activities also have important influence on the detections of HINSA.
Among 12 PGCCs, 11 have detections of HINSA and the detection rate is 90\%. 
HINSA features are also detected in starless cores L1521B. 
In four star formation regions, L1489E, IRAS 05413-0104, CB34 and HH25 MMs,
HINSA features are only detected in CB34 and HH25 MMS. The HINSA features detected in HH25 MMS are not 
very reliable. HH25 MMS was observed in the tracking mode instead of the snap shot mode.
There are only several ($<$5) beams have \ion{H}{I} spectra with HINSA features.
In the \ion{H}{I}  spectrum of HH25 MMS (Fig. \ref{HI_beamI}),   
besides the dip with velocity similar to those of CO lines ($\sim$10 km s$^{-1}$), there is another 
dip with redder velocity ($\sim$13 km s$^{-1}$) where no corresponding CO emission lines can be seen.
These two dips may be misleadingly produced by a \ion{H}{I} emission peak between them. 
HINSA features tend to be inhibited by star formation activities.

Inhibitions of star formation activities on HINSA features can also be seen in PGCCs. 
Based on the WISE data, \citet{2016MNRAS.458.3479M} presented an all-sky catalog of  133\,980 Class I/II
and 608\,606 Class III young stellar object (YSO) candidates. 
All sources in our sample have WISE YSO associations within the field of views, except for G172.8-14A1 
(see Figs. \ref{HI_CO_fig} and \ref{HI_CO_fig_1}).
The YSOs tend to locate at the margin of the HINSA detection regions, especially 
for sources G174.4-15A1, G175.34-10.8, G192.2-11A2 and G170.88-10.92.
In G165.6-09A1, the map center is associated with 4 YSOs and no HINSA feature is detected.
However, at the north-west margin of G165.6-09A1, there are strong HINSA features  ($\tau_{\rm HINSA} \sim$ 0.1)
but no YSO association.
Since we do not know the velocities of YSOs, the YSOs matched based on angular separations may be not 
truely associated with PGCCs in 3-D space.  The relations between the distribution of YSOs and HINSA confirm that
WISE YSOs  are spatially related to the PGCCs, as pointed out by \citet{2016MNRAS.458.3479M} according to
the YSO surface distributions around PGCCs.

 
Overall, HINSA tends to be not detected in regions associated with warm dust emission and background H$_\alpha$ emission.
In the region associated with young stellar objects or dense gas emission, HINSA feature also tends to be
inhibited. Beside these effects, the varied abundances of \ion{H}{I} in molecular clouds  (Sect. \ref{dis_abhi})
and the different excitation conditions
between HINSA and CO (Sect. \ref{dishinsaorig})
may also lead to the different distributions between the HINSA and CO emission in some sources. 
 
\begin{figure}[t]
\centering
\includegraphics[width=0.8\linewidth]{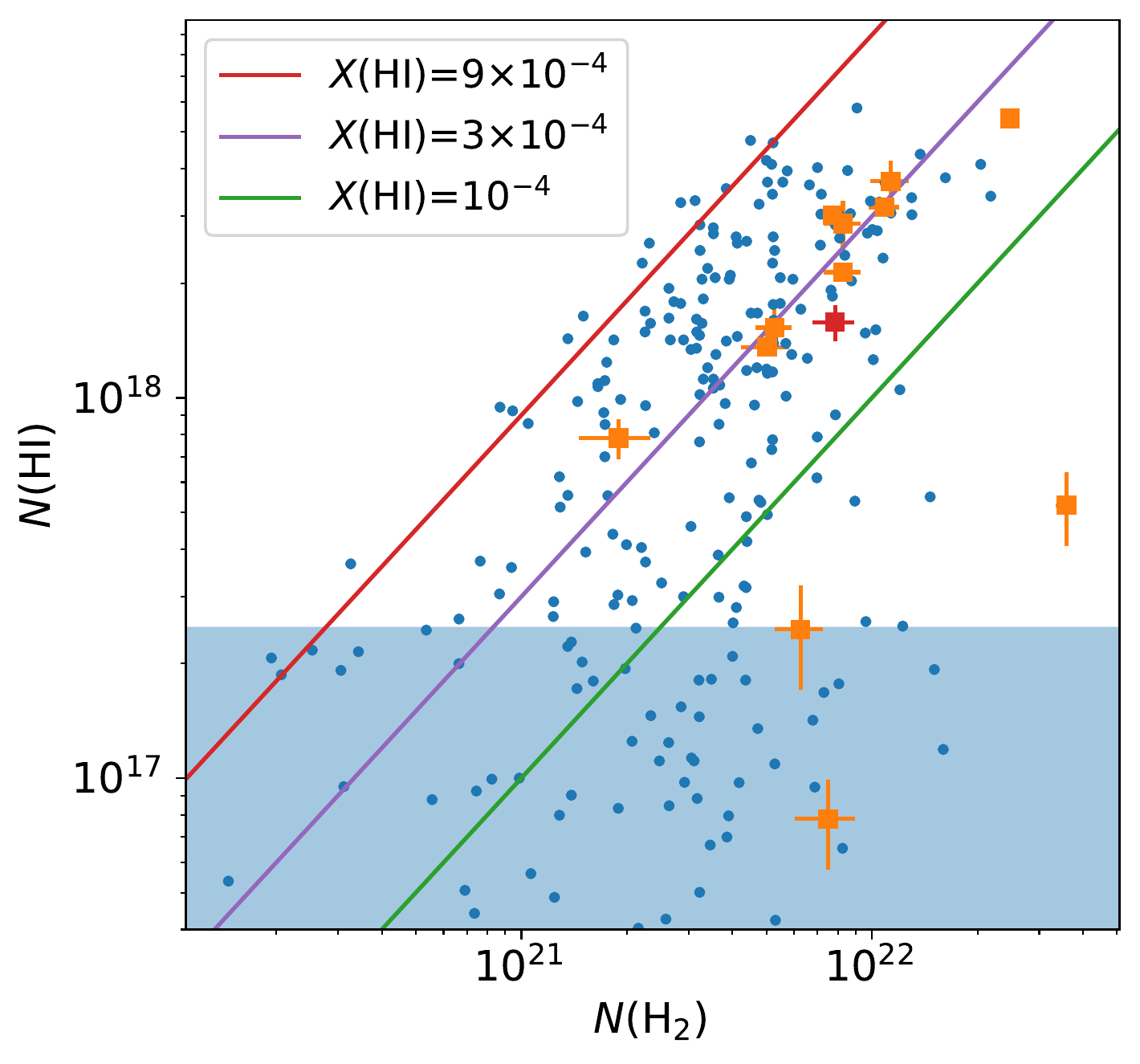}hasing
\caption{
Correlation between the column densities of \ion{H}{I} and those of H$_2$ derived from C$^{18}$O $J$=1-0 emission.
The orange squares represent the data of the central beams.
The blue dots represent the data of other beams.
The red square shows the data at the peak position of the HINSA map of G165.6-09A1. 
\label{corre_HI_H2} }
\end{figure} 
 
\subsection{\ion{H}{I} abundances and evolution status} \label{dis_abhi}
The \ion{H}{I} traced by HINSA is assumed well mixed with molecular gas (this issue will be further discussed
in Sect. \ref{dishinsaorig}).
Fig. \ref{corre_HI_H2} shows the correlation between the column densities of \ion{H}{I} ($N$(\ion{H}{I})) derived 
from HINSA and the column densities of H$_2$ ($N$(H$_2$)) derived from CO spectra.
$N$(\ion{H}{I}) and $N$(H$_2$) are positively correlated, especially for data at the map centers.
The derived abundances of \ion{H}{I}  are approximately $3\times 10^{-4}$, varied by a factor of $\sim$3.
This confirms the similarity between CO and HINSA. 

If a molecular cloud has a typical distance $d=$400 pc, angular diameter $\Theta=5$\arcmin,  and \ion{H}{I} column density 
$N$(\ion{H}{I})=2$\times$10$^{18}$ cm$^{-2}$ (Table \ref{tab:hp}), the typical volume density of \ion{H}{I} can be estimated as
\begin{equation}
n({\ion{H}{I}}) = \frac{N(\ion{H}{I})}{\Theta d} \sim 1.1\ {\rm cm}^{-3} \label{eq_hi_vd}
\end{equation} 
This value is consistent with the value at the steady state ($n^{st}(\ion{H}{I})$) given by \citet{2003ApJ...585..823L}.
If the beam dilution effect and the relative large uncertainty of $N(\ion{H}{I})$ (Fig. \ref{corre_HI_H2})
are considered, the volume density of \ion{H}{I} 
may exceed the $n^{st}$(\ion{H}{I}). In some work such as \citet{2020RAA....20...77T} 
some sources with large \ion{H}{I} abundances ($\sim$10$^{-2}$) were reported, and those sources
were thought to be in transition phase between atomic and molecular states. 

\begin{figure}
\centering
\includegraphics[width=0.8\linewidth]{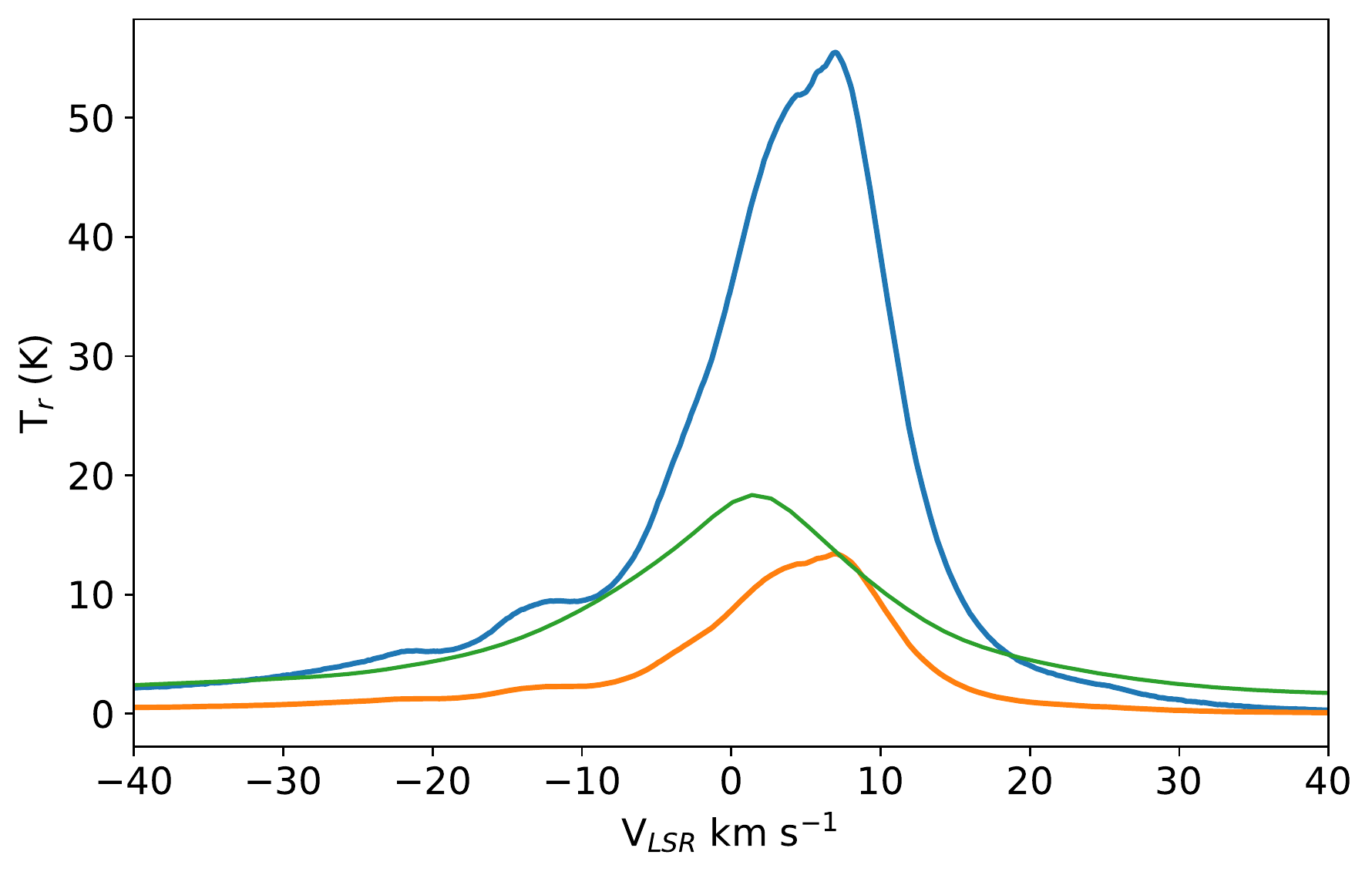}
\caption{The blue line is the average \ion{H}{I} spectrum of the  observed sources in this work. 
The orange line represents the $\overline{T_{\ion{H}{I}}}$ estimated through multiplying the blue line by 
sin(14\degr). The green line shows the $\overline{T_{\ion{H}{I}}}$ evaluated from the 
full-sky \ion{H}{I} survey HI4PI based on EBHIS and GASS \citep{2016A&A...594A.116H}.  \label{fig_ave_spechi} }
\end{figure}

\subsubsection{Production efficiency of \ion{H}{I}}
The steady-state value of \ion{H}{I} abundance reflects the balance between the destruction rate of H$_2$ by cosmic rays and
the formation rate of \ion{H}{I} on the grain surfaces. However, the formation efficiency of H$_2$ relies on the temperature and 
grain types.  When the dust temperature is low (typically several K depending on the binding energies of hydrogen), 
the grain will be covered by a monolayer of \ion{H}{I} or H$_2$, which prevents \ion{H}{I} further
sticking onto the grain surface \citep{1971ApJ...163..155H}. When the dust temperature is higher 
\citep[$>15$ K in astrophysical environments;][]{1999ApJ...522..305K}, the formation efficiency of H$_2$
is small since the depleted \ion{H}{I} will be thermally desorbed before reacting with another \ion{H}{I}.
Thus H$_2$ will only be efficiently produced within a narrow range of temperature \citep{1971ApJ...163..155H,1999ApJ...522..305K}. 
The H$_2$ formation at higher temperature can be efficient if chemisorption site is taken into considered \citep{2002ApJ...575L..29C},
which has a much higher binding energy ($\sim$10$^4$ K) than the value for physisorption site ($\sim$500 K) and provides a shield for 
\ion{H}{I} when dust temperature is high ($>$20 K).

The assumption that a physisorption site occupied by one \ion{H}{I} will not further absorbed another \ion{H}{I}
is not necessary. When dust temperature is low, there may be several \ion{H}{I}  in a same physisorption site. 
The bound energy of H-H is 4.52 eV \citep{1996MNRAS.279..591D}. 
If the reaction heat of a forming H$_2$ is able to peel all H atoms in that site from grain surface,
only the empty site is then truely capable  of absorbing \ion{H}{I} in gas phase.
Under this assumption, the formation efficiency of H$_2$ is not altered, and absorption and synthesis
of other elements and molecules on grain surface at low dust temperature will not be 
weaken by the occupation of sites by hydrogens. 

\begin{figure*}
\centering
\includegraphics[width=0.4\linewidth]{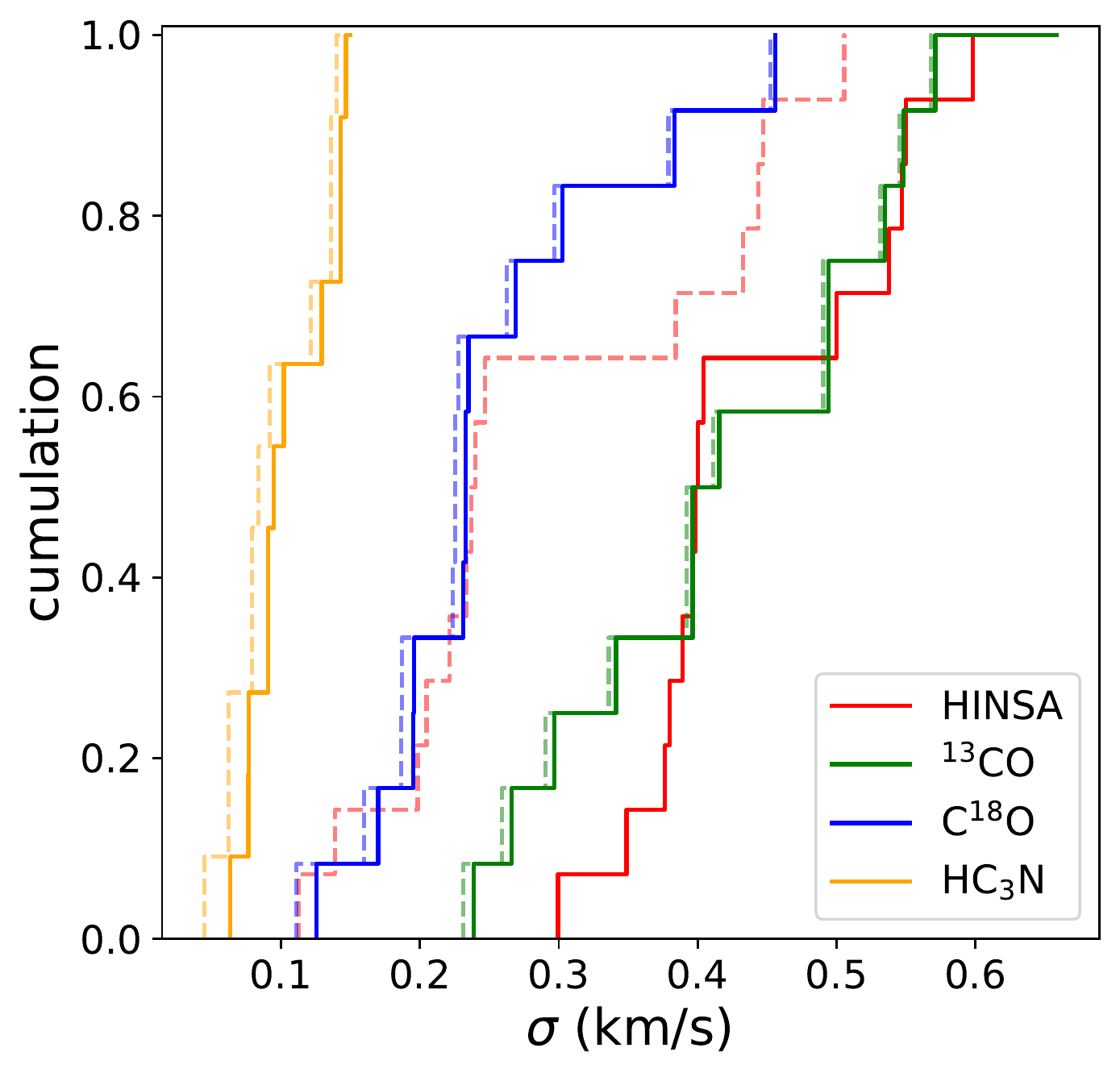}
\includegraphics[width=0.4\linewidth]{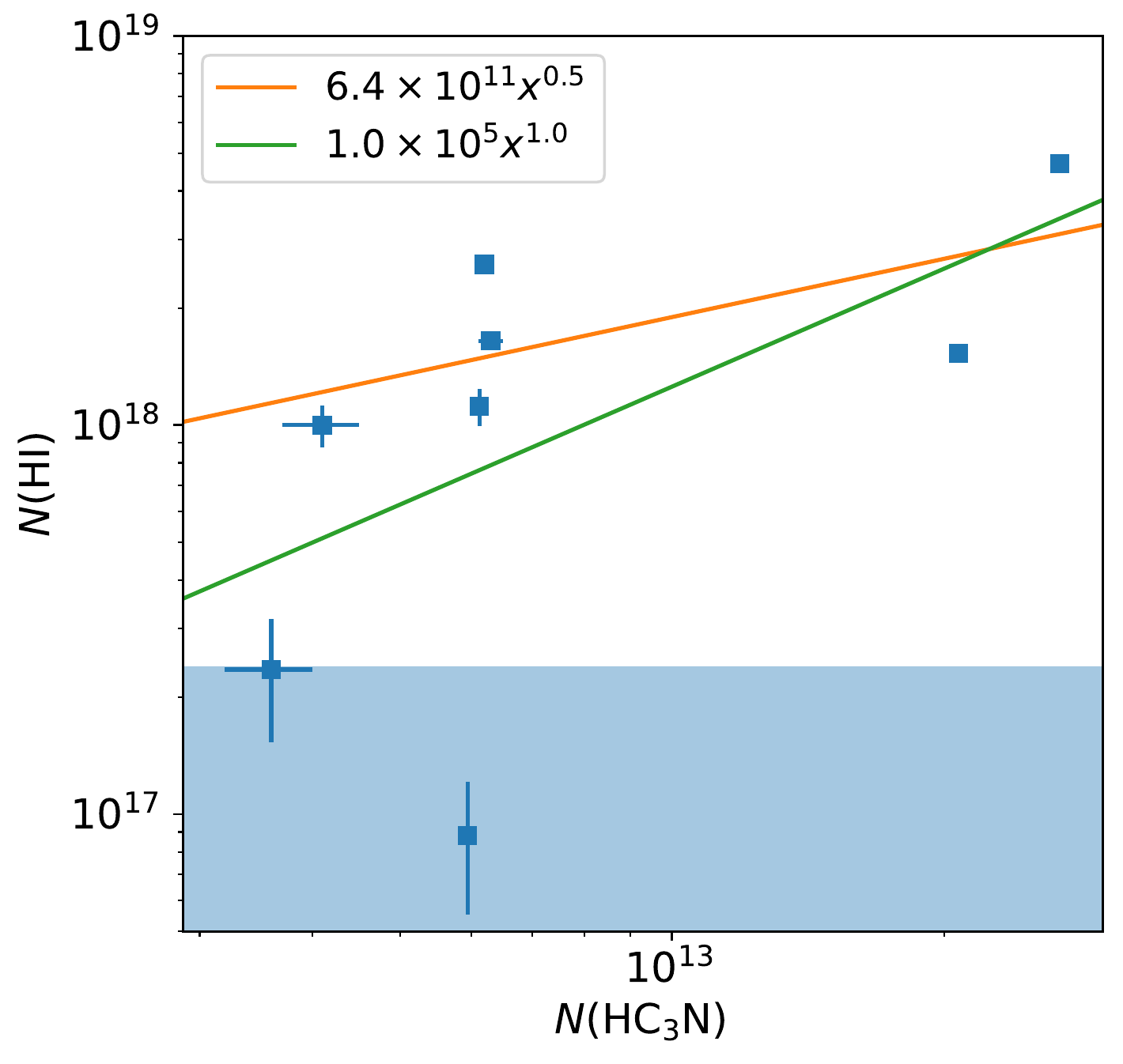}
\caption{Left: Cumulative distributions of velocity dispersions. Soild lines and dashed lines represent 
$\sigma$ and $\sigma^{NT}$, respectively (Sect. \ref{sec_dv}). 
Right: Correlation between the column densities of \ion{H}{I} and those of HC$_3$N.
Green and orange lines represent the linear fitting with and without the two data in blue shadow, respectively. 
For data in the blue shadow,   the $N$(\ion{H}{I}) parameters are not reliable 
with fitted optical depth of HINSA $\tau < 0.01$ (Sect. \ref{sec_lowerlimitforfitting}). 
\label{fig_dv}}
\end{figure*}

\subsubsection{Evolution status}
The sources with relatively higher abundance of \ion{H}{I} may be still in steady state,
and their dust temperatures and grain properities are not suitable for H$_2$ formation.
There is no HINSA in G160.51-17.07, which has a relative high dust temperature (19.2 K) but 
not weak CO emission (Fig. \ref{emission_line_spec}). However, we can not rule out that G160.51-17.07 is in transition phase,
considering its relatively lower  surface density of dust compared with other PGCCs (Table \ref{tab:sources}).


The derived value of $n({\ion{H}{I}})$ for PGCCs may also be underestimated.
It is not sure if the column densities of \ion{H}{I} is evenly distributed within regions shown HINSA features.
The beam dilution effect may be not ignorable, just like the case of CO emission.

\subsection{Does the HINSA we detected trace molecular gas?  } \label{dishinsaorig}
We have assumed that HINSA mainly traces \ion{H}{I}
well mixed with molecular gas in Sect. \ref{dis_abhi}, 
based on the fact that the morphologies of HINSA emissions are overall correlated to those of $^{13}$CO in PGCCs 
(Sect. \ref{sec_dis_me}). However, it is still not  clear if the \ion{H}{I} absorption 
contributed by \ion{H}{I} component around molecular gas cloud is also important.
The spatial resolution of the \ion{H}{I} observations of this present work prevents us to dig deeply into this issue. 
However, from the discussion below about the excitation status of HINSA against background \ion{H}{I} emission, we
try to illustrate that  \ion{H}{I} traced by HINSA should be a gas component of molecular cloud and 
confined within and around  CO emission regions.  

\subsubsection{Collisional excitation of HINSA} 
The H-H spin-exchange collision is the main mechanism of the collisional excitation 
of \ion{H}{I} if the abundance of H atom overwhelm that of electron \citep{1958PIRE...46..240F,2007MNRAS.379..130F}.
The coefficient of H-H spin-exchange collision tabulated in the   column (4) of Table 2 of $\kappa_{10}^{H}$ of \citep{2005ApJ...622.1356Z}
can be empirically fitted as 
\begin{equation}\kappa_{10}^{H}=3.1\times 10^{-11}T_{\rm k}^{0.357}exp(-32/T_{\rm k})\ {\rm cm}^{3}\ {\rm s}^{-1} 
\label{kappa_eq} \end{equation}
for $10<T_k<300$ K. 

In optically thin condition, the excitation temperature of HINSA can be estimated through \citep{1956ApJ...124..542P,2003ApJ...585..823L}
\begin{equation}
T_{\rm ex}(\mbox{HINSA}) = \frac{1}{1+y}{T_{background}}+\frac{y}{1+y}T_{\rm k} \label{ex_hinsa}
\end{equation}  
in which 
\begin{equation}
y=\frac{\kappa_{10}^Hn_{\ion{H}{I}}h\nu}{kT_{k}A_{10}}\ , \label{y_eq_hi}
\end{equation}
$T_{\rm k}$ is the kinematic temperature, $T_{background}$ is the brightness temperature of background emission, 
$n_{\ion{H}{I}}$ is the volume density of hydrogen atom (the major collision partner to excitate the \ion{H}{I} 21 cm line),
and $A_{1,0}$ is the spontaneous decay rate 2.85$\times$10$^{-15}$ s$^{-1}$ \citep{1952ApJ...115..206W}.

If $T_{\rm k}=10$ K is adopted, Eqs. (\ref{kappa_eq}) and (\ref{y_eq_hi}) lead to $y\sim 6.9 n_\ion{H}{I}$. Here,
$n_\ion{H}{I}$ is in unit of cm$^{-3}$. 
For cold \ion{H}{I} within molecular cloud, if we futher adopt $n_\ion{H}{I}$ as the typical volume density of \ion{H}{I} given by
Eq. (\ref{eq_hi_vd}) or the the steady-state value,
the value of $y$ would be larger than 7.
If HINSA is not fully excited, the density of \ion{H}{I} given by Eq. (\ref{eq_hi_vd}) will be underestimated,
which leads to an even larger value of $y$. 
Thus, whihin a molecular cloud, the dnesity of \ion{H}{I} in steady-state should be larger than the critical density of HINSA. 
If a molecular cloud is in transition phase or has grain properities not suitable for H$_2$ formation (Sect. \ref{dis_abhi}),
its density of \ion{H}{I} would be even more higher. 
It is safe to assume that cold \ion{H}{I} within molecular cloud is collisionally excited.


\subsubsection{External excitation of HI 21 cm line}
For diffuse gas component around a molecular cloud with 
high abundance but low density of H atom, 
the \ion{H}{I} 21 cm line may be mainly excited by the  background 
emission. The L-band continumm brightness temperature ($T_{\rm L\ band}$) is averagely smaller than 
4 K (Sect. \ref{sec:method}).
This is consistant with the value estimated adopting the intensity of L-band  standard interstellar radiation field (ISRF) as $\sim$0.8 K
\citep{1980A&A....90..176W,2003ApJ...585..823L} and that of the cosmic microwave background as 2.73 K.
If $T_{\rm L\ band}$ is adopted as the intensity of external emission, the absorption 
feature contributed by  diffuse \ion{H}{I} component may not be ignorable. 

The sources we observed all have $|V_{\rm LSR}| \lesssim 10$ km s$^{-1}$. 
We shall see that the L-band external emission is  mainly contributed by $\overline{T_{\ion{H}{I}}}$, the brightness temperature
of the averaged background 
\ion{H}{I} emission 
\begin{equation} \overline{T_{\ion{H}{I}}} = \frac{1}{4\pi}\int T_{\ion{H}{I}} d\Omega, \end{equation}
The sources we observed have Galactic latitude $|b| \lesssim 20\degr$, with a mean value of $\sim$14$\degr$. 
The lower limit of the intensity of $\overline{T_{\ion{H}{I}}}$
can be estimated through multiplying the average spectrum of observed sources by $\sin(14\degr)$, as
shown in Fig. \ref{fig_ave_spechi}. The peak intensity of $\overline{T_{\ion{H}{I}}}$ is expected to be $\sim$14 K.
This is confirmed by the \ion{H}{I} spectrum averaged over the all-sky  
spectra from the HI4PI survey \citep{2016A&A...594A.116H},
which gives a brightness temperature of $\sim$20 K, $\sim$10 K and $\sim$5 K for $V_{\rm LSR}=$ 0 km s$^{-1}$, 10 km s$^{-1}$  
and 20 km s$^{-1}$, respectively (Fig. \ref{fig_ave_spechi}).  These values are larger than $T_{\rm L\ band}$.

For molecular cloud with short distance to the Earth and thus similar $\overline{T_{\ion{H}{I}}}$ compared with that seen from the Earth,
 $T_{background}$ will be dominated by 
the average brightness temperature of \ion{H}{I} 21 cm spectra if $|V_{\rm LSR}|$ is small ($|V_{\rm LSR}|<20$ km s$^{-1}$). 
If $\overline{T_{\rm \ion{H}{I}}}$ is small, 
the warmer and diffuser \ion{H}{I} around the molecular cloud may be excited by $T_{\rm L\ band}$. 
Then, the \ion{H}{I} in the diffuse gas may also contribute absorption features similar to HINSA and make us overestimate 
the column density of cold \ion{H}{I} associated with molecular gas. If $\overline{{T_{\rm \ion{H}{I}}}}$ is not ignorable, 
the diffuse \ion{H}{I} will not contribute much absorption even if it is not collisionally excited. 
Then, we can safely assume that the HINSA features we extracted are tightly associated with molecular gas.
This effect inhibits the possible absorption contributed by the outter diffuse \ion{H}{I}-rich shell of a molecular cloud,
and confines HINSA within molecular cloud. 


\subsubsection{HINSA and CO emission region}
A molecular cloud may contain  both CO-bright gas and CO-dark gas \citep[e.g.][]{2013ARA&A..51..207B}.
In molecular region with lower gas density, the \ion{H}{I} abundance should be higher
with the volume density of \ion{H}{I} unchanging. There may also be HINSA features produced in CO-dark regions.
However, HINSA features can be restrained because the cloud tends to be warmer in low-density regions.
CO is important for the cooling of molecular clouds \citep{1978ApJ...222..881G}.
For a molecular cloud with $n({\rm H}_2)=10^{3}$ cm$^{-3}$, $T_{\rm k}=10$ K, and abundance of CO $X_{\rm CO}=2\times 10^{-4}$,
  the cooling rate contributed by CO emission is \citep{2018A&A...611A..20W}
\begin{equation}
\frac{\Lambda_{\rm CO}}{\rm erg\ s^{-1}\ cm^{-3}} = 2.16\times10^{-27}  \frac{X_{\rm CO}n^2}{\rm cm^{-6}}  \left(\frac{T_{\rm k}}{\rm K}\right)^{3/2} 
\sim 1.4 \times 10^{-23}
\end{equation}
The heating rate contributed by cosmic rays and X-rays is \citep{1973ApJ...186..859G}
\begin{equation}
\frac{\Gamma_{\rm X}}{\rm erg\ s^{-1}\ cm^{-3}} = 2.7\times10^{-16} \frac{n}{\rm cm^{-3}} {\rm\frac{35eV}{erg}} \sim 1.5\times 10^{-23}  
\end{equation}
If $n({\rm H_2})$ is smaller than 10$^{3}$ cm$^{-3}$, CO cooling is not effective enough, and the gas will be warmer
and  unfavourable to HINSA. The density threshold 10$^{3}$ cm$^{-3}$ is also the lower limit for the excitations of CO transitions,
and this is another reason why HINSA features are confined around  CO emission regions.
PGCCs may be the coldest and earliest samples showing HINSA features. 
 
Overall, the non-small $\overline{T_{\ion{H}{I}}}$ and CO cooling both help to 
confine HINSA features to regions within and around  CO emission kernels.   

\subsection{HINSA and HC$_3$N} \label{sec_dv}
The velocity dispersions ($\sigma$) can be derived from the line widths ($\Delta V$) through
\begin{equation}
\sigma_X = \frac{\Delta V}{\sqrt{8\ln(2)}}
\end{equation}
The non-thermal velocity dispersions ($\sigma^{\rm NT}$) can be calculated through
\begin{equation}
\sigma_X^{\rm NT} =\sqrt{ \sigma_X^2-\frac{kT_{\rm k}}{m_X}}
\end{equation}
where $m_{X}$ represents the mass of molecule $X$.
The left panel of Fig. \ref{fig_dv} shows the cumulative distributions 
of $\sigma$ and $\sigma^{\rm NT}$ for HINSA, $^{13}$CO, C$^{18}$O and HC$_3$N. 
 $\sigma_{\rm HC_3N}$ is similar to $\sigma_{\rm {}^{13}CO}$, and tends to be larger than $\sigma_{\rm C^{18}O}$.
However, the non-thermal velocity dispersion traced by HINSA is comparable with that of C$^{18}$O. 
The broaden effect introduced by non-thin optical depth of $^{13}$CO $J$=1-0 can not fully explain 
the larger $\sigma^{NT}$ traced by $^{13}$CO compared with those traced by C$^{18}$O.
An optical depth of 2 will
contribute to the line width smaller than 30\% \citep{1979ApJ...231..720P}.
$^{13}$CO may trace more extended gas component compared with C$^{18}$O,
because  $^{13}$CO $J$=1-0 has larger optical depth compared with C$^{18}$O $J$=1-0
and is easier to be excitated. HINSA and C$^{18}$O $J$=1-0 are optically thin in molecular clouds,
and originate from similar regions.
However, the spectra of HC$_3$N  are quite narrow, with line widths ($<$0.1 km s$^{-1}$; Fig. \ref{fig_dv})
smaller than those of HINSA and $^{13}$CO.
 
Previous observations shows that HC$_3$N is abundant in some but not all starless cores \citep{1992ApJ...392..551S,2019MNRAS.488..495W}.
In low mass star formation regions,  HC$_3$N will be enhanced again under the heating of YSOs in early
stages \citep{2008ApJ...672..371S,2009ApJ...697..769S}, but may be  destroyed by radiation from YSOs \citep{Liu_2021}. 
In molecular cloud harbouring young stars, it is also correlated with outflows and shocks \citep{2019MNRAS.488..495W,2019MNRAS.489.4497Y}.
The narrow line widths of HC$_3$N $J$=2-1 
imply that the HC$_3$N emission may originate from more compact gas component within PGCCs, instead of the
outer regions dissociated by external radiation fields or shocked by turbulent flows.
Although its CO emission is extended (Sect. \ref{sec_co_dust_emission}), the PGCC could harbor 
dense and compact regions. The narrow line widths of HC$_3$N confirm that these PGCCs are in  quiescent state. 
Carbon chain molecule abundant PGCCs have both cold and condensed structures at the center regions traced by HC$_3$N and  extended
gas components at the outer regions traced by CO. This make them  good targets to observe and study HINSA. 
  
The right panel of Fig. \ref{fig_dv} shows the correlation between the column density of \ion{H}{I}
derived from HINSA and the column densities of HC$_3$N. 
There is a weak positive correlation between $N$(HC$_3$N) and $N$(\ion{H}{I}), and the abundance ratio
$N$(HC$_3$N)/$N$(\ion{H}{I}) is $\sim$ 10$^{-5}$.
The PGCCs with strongest HC$_3$N emission tend to have more remarkable HINSA features.
However, there is not enough evidence to assert that HINSA in sources with weak HC$_3$N emission
will be suppressed. The G178.98-06.7 is such a source with no detection of HC$_3$N but with strong
HINSA. This is not a surprise since not all PGCCs are CCM abundant, and the detection rate of HINSA 
is considerable in PGCCs with no previous observations of CCM emission \citep{2020RAA....20...77T}.
The prior of the presence of HC$_3$N raises the detection rate of HINSA close to $\sim$90\% in PGCCs of this work, although 
the sample size is still limited.

\section{Summary} \label{sec:summary}
Twelve PGCCs, one starless core L1521B and four star forming sources are searched 
for \ion{H}{I} narrow-line self-absorption (HINSA) features, using the  FAST.

Combined the data of \ion{H}{I} 21 cm lines, CO $J$=1-0, HC$_3$N $J$=2-1 and dust continuum,
we studied the morphologies, abundances and excitation conditions of HINSA, CO and HC$_3$N in PGCCs
and star-forming sources. We also investigated  
the relations among these parameters and the environments. 

The main results are as following:
\begin{itemize}
\item[1.] HINSA features are detected in 11 of 12 PGCCs, 
corresponding to a detection rate of 90\%. Eight of the detected ones have HC$_3$N emissions. 

\item[2.] We improve the method of extracting HINSA features (method 2) based on the method of \citet{2008ApJ...689..276K} 
(method 1).  The thresholds of the signal-to-noise ratios (S/Ns)  of these two methods are deduced.
It is shown that method 2 has much lower S/N requirement compared to method 1. 
The improved method performs better when extracting HINSA features with  low signal-to-noise  ratio (S/N),
and large line width resultant from blended velocity components or high velocity resolution. 

\item[3.] Applying the improved method,  HINSA features are extracted in 11 PGCCs, starless core L1251B and 
4 star forming sources.  
The line widths and velocities of CO lines are adopted as the initial values of the fitting parameters
of HINSA.
The optical depths and line widths of HINSA, as well as the column densities of \ion{H}{I} are obtained.
 
\item[4.] HFS fittings are applied to HC$_3$N $J$=2-1. The column densities of HC$_3$N are calculated.
The average abundance of HC$_3$N is 2.2($\pm$0.8)$\times$10$^{-9}$. 
CO emission looks as extended as the Planck dust continuum.  
Part of the CO emission may come from more diffuse regions besides the
regions associated with the cold dust or dense gas.

\item[5.] HINSA tends to be not detected in regions associated with warm dust emission, background ionizing radiations,
stellar objects and dense gas emission.

\item[6.] The abundances of \ion{H}{I} in PGCCs are approximately $3\times 10^{-4}$, varied by a factor of $\sim$3. 
The distribution of HINSA is similar to that of CO emission.
The non-small average itensity of the background \ion{H}{I} emission and CO cooling both helps to 
confine HINSA features to regions within and around  CO emission kernels.


\item[7.] The non-thermal dispersions traced by HINSA are comparable with those of C$^{18}$O,
and larger than those of HC$_3$N. HC$_3$N emissions may originate from the condensed regions within PGCCs. 
Carbon chain molecule abundant PGCCs provide a good sample to study HINSA.
\end{itemize}

Guided by CCM emissions,
HINSA can be a good tool to detect and analyze extended neutral gas components in PGCCs.
The sample size of this work is limited. It would be helpful to search more PGCCs 
in the future, with the supports of the good sensitivity of the FAST and the improved 
HINSA extracting method.

\begin{acknowledgements}
This project was supported by   the National Key R\&D Program of China No. 2017YFA0402600, 
and the NSFC No. 12033005,  11433008, 11725313,   
11373009, 11503035 and 11573036. 
N.-Y. T. is supported by National Key R\&D Program of China No. 2018YFE0202900 and NSFC No. 11803051.
We are grateful to the help of the staff of the FAST during the observations and data conduction.
We show warm thanks to the anonymous referee for providing many constructive 
argues and suggestions which make the content of this work more 
substantial and clear.
\end{acknowledgements}

\bibliographystyle{aa}
\bibliography{ms}

\clearpage
\end{CJK}
\end{document}